\documentclass[aip,reprint,numerical]{revtex4-1}
\usepackage{amsmath}
\usepackage{amssymb}
\usepackage{graphicx}
\usepackage{dcolumn}
\usepackage{bm}
\usepackage{color}
\usepackage{amssymb}
\usepackage{mathtools}

\DeclareMathOperator*{\bigast}{\raisebox{-0.6ex}{\scalebox{2}{$\ast$}}}

\DeclareMathOperator*{\bigblackcircle}{\raisebox{-0.6ex}{\scalebox{2}{$\bullet$}}}

\def\spacce#1{\hskip #1pt}
\def\drawline#1#2{\raise 2.5pt\vbox{\hrule width #1pt height #2pt}}

\def\solid{\drawline{22}{1.0}\nobreak\ }

\def\tdash{\hbox{\drawline{4}{0.8}\spacce{2}}}
\def\dashed{\tdash \tdash \tdash \tdash \nobreak\ }

\def\tdot{\hbox{\drawline{1}{1.0}\spacce{2}}}
\def\dotted{\tdot \tdot \tdot \tdot \tdot \tdot \tdot \nobreak\ }

\def\dashdot{\tdash \tdot \tdash \tdot \tdash \nobreak\ }

\def\tdotlarge{\hbox{\drawline{1}{1.5}\spacce{2}}}
\def\dottedlarge{\tdotlarge \tdotlarge \tdotlarge \tdotlarge \tdotlarge \tdotlarge \tdotlarge \nobreak\ }

\def\ssquare{${\vcenter{\hrule height 0.9pt
       \hbox{\vrule width 0.9pt height 10pt \kern 10pt
       \vrule width 0.9pt}
       \hrule height 0.9pt}}$\nobreak\ }
       
\def\ssquareb{$\Box$\nobreak\ }       
       
\def\blackssquare{$\scriptstyle\blacksquare$\nobreak\ }
       
\def\circle{$\circ$\nobreak\ }

\def\blackcircle{$\bigblackcircle$\nobreak\ }

\def\bigcircle{$\bigcirc$\nobreak\ }

\def\losange{$\Diamond$\nobreak\ }

\def\blacklosange{$\blacklozenge$\nobreak\ }

\def\trianup{\raise 1.25pt\hbox{$\triangle$}\nobreak\ }

\def\blacktrianup{\raise 1.25pt\hbox{$\blacktriangle$}\nobreak\ }

\def\bigtriandown{\raise 1.25pt\hbox{$\bigtriangledown$}\nobreak\ } 

\def\plus{$\bm{+}$ \nobreak}

\def\asterix{$\bigast$ \nobreak}

\def\solidopencircle{\solid \nobreak\spacce{-19.5}\raise
 -0.5pt\hbox{\circle} \nobreak\ }
 
\def\solidblackcircle{\solid \nobreak\spacce{-21}\raise
 -0pt\hbox{\blackcircle} \nobreak\ }

\def\solidopensquare{\solid \nobreak\spacce{-19}\raise
 -1.0pt\hbox{\ssquare} \nobreak\ }
 
\def\solidblacksquare{\solid \nobreak\spacce{-19}\raise
 -1.0pt\hbox{\blackssquare} \nobreak\ }
 
\def\solidopenlosange{\solid \nobreak\spacce{-19}\raise
 -1.0pt\hbox{\losange} \nobreak\ }
 
\def\solidblacklosange{\solid \nobreak\spacce{-21}\raise
 -2.0pt\hbox{\blacklosange} \nobreak\ }

\def\solidopentriangleup{\solid \nobreak\spacce{-19}\raise
 -1.pt\hbox{\trianup} \nobreak\ }
 
\def\solidblacktriangleup{\solid \nobreak\spacce{-21.5}\raise
 -2.0pt\hbox{\blacktrianup} \nobreak\ }

\def\solidopencross{\solid \nobreak\spacce{-22}\raise
 -1.0pt\hbox{\cross} \nobreak\ }
 
\def\solidopenplus{\solid \nobreak\spacce{-21}\raise
 -1.5pt\hbox{\plus} \nobreak\ }
 
\def\solidopenasterix{\solid \nobreak\spacce{-21}\raise
 -1.5pt\hbox{\asterix} \nobreak\ }

\def\dashedopencircle{\dashed \nobreak\spacce{-22}\raise
 -1.0pt\hbox{\circle} \nobreak\ }
 
\def\dashedblackcircle{\dashed \nobreak\spacce{-23}\raise
 -1.4pt\hbox{\blackcircle} \nobreak\ }
 
\def\dashedopensquare{\dashed \nobreak\spacce{-22}\raise
 -1.0pt\hbox{\ssquare} \nobreak\ }
 
\def\dashedblacksquare{\dashed \nobreak\spacce{-23}\raise
 -1.0pt\hbox{\blackssquare} \nobreak\ }
 
\def\dashedopenlosange{\dashed \nobreak\spacce{-22}\raise
 -1.0pt\hbox{\losange} \nobreak\ }
 
\def\dashedblacklosange{\dashed \nobreak\spacce{-23}\raise
 -2.0pt\hbox{\blacklosange} \nobreak\ }
 
\def\dashedopentriangleup{\dashed \nobreak\spacce{-22}\raise
 -1.0pt\hbox{\trianup} \nobreak\ }
 
\def\dashedblacktriangleup{\dashed \nobreak\spacce{-24}\raise
 -2.0pt\hbox{\blacktrianup} \nobreak\ }
 
\def\dashedopencross{\dashed \nobreak\spacce{-24}\raise
 -1.0pt\hbox{\cross} \nobreak\ }
 
\def\dashedopenplus{\dashed \nobreak\spacce{-24}\raise
 -1.0pt\hbox{\plus} \nobreak\ }

\def\dashedopenasterix{\dashed \nobreak\spacce{-22}\raise
 -1.5pt\hbox{\asterix} \nobreak\ }

\def\dottedopencircle{\dotted \nobreak\spacce{-22}\raise
 -1.0pt\hbox{\circle} \nobreak\ }
 
\def\dottedblackcircle{\dotted \nobreak\spacce{-23}\raise
 -1.4pt\hbox{\blackcircle} \nobreak\ }

\def\dottedopensquare{\dotted \nobreak\spacce{-22}\raise
 -1.0pt\hbox{\ssquare} \nobreak\ }
 
\def\dotted\mathbfuare{\dotted \nobreak\spacce{-23}\raise
 -1.0pt\hbox{\blackssquare} \nobreak\ }
 
\def\dottedopenlosange{\dotted \nobreak\spacce{-22}\raise
 -1.0pt\hbox{\losange} \nobreak\ }
 
\def\dottedblacklosange{\dotted \nobreak\spacce{-24}\raise
 -2.0pt\hbox{\blacklosange} \nobreak\ }
 
\def\dottedopentriangleup{\dotted \nobreak\spacce{-22}\raise
 -1.0pt\hbox{\trianup} \nobreak\ }
 
\def\dottedblacktriangleup{\dotted \nobreak\spacce{-25}\raise
 -2.0pt\hbox{\blacktrianup} \nobreak\ }
 
\def\dottedopencross{\dotted \nobreak\spacce{-23.5}\raise
 -1.0pt\hbox{\cross} \nobreak\ }
 
\def\dottedopenplus{\dotted \nobreak\spacce{-23.5}\raise
 -1.0pt\hbox{\plus} \nobreak\ }

\def\dottedopenasterix{\dotted \nobreak\spacce{-23.5}\raise
 -1.0pt\hbox{\asterix} \nobreak\ }

\def\whitehisto{${\vcenter{\color{black} \hrule height 1.0pt
       \hbox{\vrule width 1.0pt height 4pt \kern 8pt
       \vrule width 1.0pt}
       \hrule height 1.0pt}}$\nobreak\ }
       
\def\histosymb#1{${\vcenter{\color{#1} \hrule height 0.0pt
       \hbox{\vrule width 11.0pt height 6pt \kern 0pt 
       \vrule width 0.0pt}
       \hrule height 0.0pt}}$\nobreak\ }

\def\symbol#1#2#3#4#5#6{\color{#5}#1 \nobreak\spacce{-#3}\raise
 -#4pt\hbox{\color{#6}#2}\color{black} \nobreak\ }

\usepackage{bm}
\usepackage{graphicx}
\usepackage[caption=false]{subfig}
\usepackage{geometry}
\usepackage{verbatim}
\usepackage{float}
\usepackage[english]{babel}

\begin{document}

\preprint{AIP/123-QED}

\title{Simulating Brownian suspensions with fluctuating hydrodynamics}

\author{Blaise Delmotte}
\email{blaise.delmotte@gmail.com}
\homepage{https://sites.google.com/site/blaisedelmotte/}
 \affiliation{IMFT - CNRS, UMR 5502 1,  All\'ee du Professeur Camille Soula, Toulouse, France.}
 \affiliation{University of Toulouse, IMFT, Toulouse, France.}
\author{Eric E Keaveny}%
 \email{e.keaveny@imperial.ac.uk}
 \homepage{http://www.imperial.ac.uk/people/e.keaveny}
\affiliation{ 
Department of Mathematics, South Kensington Campus, Imperial College London, LONDON, SW7 2AZ, UK.
}%

\date{\today}

\begin{abstract}
Fluctuating hydrodynamics has been successfully combined with several computational methods to rapidly compute the correlated random velocities of Brownian particles.  In the overdamped limit where both particle and fluid inertia are ignored, one must also account for a Brownian drift term in order to successfully update the particle positions.  In this paper, we present an efficient computational method for the dynamic simulation of Brownian suspensions with fluctuating hydrodynamics that handles both computations and provides a similar approximation as Stokesian Dynamics for dilute and semidilute suspensions.  This advancement relies on combining the fluctuating force-coupling method (FCM) with a new midpoint time-integration scheme we refer to as the drifter-corrector (DC).  The DC resolves the drift term for fluctuating hydrodynamics-based methods at a minimal computational cost when constraints are imposed on the fluid flow to obtain the stresslet corrections to the particle hydrodynamic interactions.  With the DC, this constraint need only be imposed once per time step, reducing the simulation cost to nearly that of a completely deterministic simulation.  By performing a series of simulations, we show that the DC with fluctuating FCM is an effective and versatile approach as it reproduces both the equilibrium distribution and the evolution of particulate suspensions in periodic as well as bounded domains. In addition, we demonstrate that fluctuating FCM coupled with the DC provides an efficient and accurate method for large-scale dynamic simulation of colloidal dispersions and the study of processes such as colloidal gelation.
\end{abstract}

\maketitle
\section{Introduction}
Brownian motion is the random motion exhibited by micron and sub-micron particles immersed in liquid.  It is a fundamental mechanism for material and chemical transport in micron-scale physical and biological systems \cite{Grima2010}, and can play a key role in determining the mechanical response of colloidal suspensions to applied stresses \cite{Batchelor1977,Bossis1989,Foss2000,Banchio2003}.  Brownian motion is also known to affect the aggregation and self-assembly of interacting particles \cite{Anderson2002,Zaccarelli2007,Lu2008}, a fundamental process important in many engineering applications that utilize colloidal particles to tune rheological properties of fluids \cite{Mabille2000,Ten2007} and construct new materials and devices \cite{Whitesides2002,Glotzer2004,Promislow1995}.  

While Brownian motion is clearly important in these situations, the careful study and quantification of its role using simulation remains a computational challenge due to the intimate link between the random motion of the particles and their many-body hydrodynamic interactions. For suspensions of moderate concentration, this requires resolving the hydrodynamic interactions beyond those provided by point forces.  It involves obtaining the second-order corrections to the hydrodynamic interactions that are provided by the stresslets -- the symmetric force moments on the particles -- as well as the rotlets due to torques applied to the particles.  The stresslets are necessary to quantify suspension rheology and accurately resolve particle dynamics in linear flow fields.  

The inclusion of stresslets in Brownian simulations presents several computational challenges and has been addressed directly in Stokesian Dynamics (SD), and in accelerated SD through approximation.  The primary contribution of this work is a complete method to perform dynamic Brownian simulations of dilute and semidilute suspensions that includes the effects of stresslets exactly using fluctuating hydrodynamics.  We do this in a way that incurs only a minimal computational cost over a deterministic simulation of the same system, and in a framework that we show can be extended beyond periodic boundary conditions.

In order to achieve the correct equilibrium distribution, the statistics of the random particle motion must satisfy the fluctuation-dissipation theorem \cite{Kubo1966} which states that the random particle velocity correlations must be proportional to the hydrodynamic mobility matrix. In simulation, this would require the square root of the mobility matrix to be found at each time step to compute the correct particle velocities.  As a result, the inclusion of Brownian motion has often limited simulations to having very small particle numbers, or ignoring completely hydrodynamic interactions between the particles.  Though several methods have been introduced to accelerate the matrix square root computation \cite{Fixman1986,Ando2012,Ando2013}, recent studies have shown this computation can be avoided altogether using fluctuating hydrodynamics.  

Fluctuating hydrodynamics involves generating random fluid flows by including a white-noise fluctuating stress in the equations of fluid motion.  Introduced in the first edition of Landau and Lifshitz \cite{Landau1959}, its effectiveness for yielding the correct random motion of particles was demonstrated in a number of theoretical studies in the 1960s and 1970s\cite{Zwanzig1964,Fox1970,Hauge1973,Chow1973,Bedeaux1974,Velarde1974,Hills1975,Hinch1975,Bedeaux1977}.  As a result of this fundamental work, fluctuating hydrodynamics has found success in numerical simulations of micron-scale fluid-structure interactions in methods such as Lattice-Boltzmann \cite{Ladd1993, Ladd1994, Ladd1994b}, hybrid Eulerian-Lagrangian approaches for point particles \cite{Usabiaga2013,Usabiaga2013b}, distributed Lagrange-multiplier method \cite{Sharma2004}, finite-element simulation \cite{Plunkett2014}, the stochastic and fluctuating immersed-boundary methods (IBM) \cite{Atzberger2007, Atzberger2011, Usabiaga2014, Delong2014}, and the fluctuating force-coupling method (FCM) \cite{Keaveny2014}.  

IBM and FCM are similar in that they both use projection and volume averaging operations to first transfer the forces experienced by the particles to the fluid and subsequently, extract the motion of the particle phase from the motion of the fluid.  With IBM, these operators are grid-dependent and are designed to provide minimal resolution of the particles.  The resulting flow due to the particles is typically computed using finite-difference or finite volume methods.  With FCM, the operators are built around Gaussian functions which require more resolution than their IBM counterparts, but due to their smoothness, can be used in conjunction with very accurate spectral solvers for the fluid flow.  In addition, FCM aims to provide a higher resolution of particle finite size by including the operators that correspond to the second order effects of the rotlet and stresslet.  The particle hydrodynamic interactions provided by FCM have been studied extensively\cite{Lomholt2003,Dance2003,Yeo2010} and in many cases a successful comparison with SD results is found.

To resolve Brownian motion with these methods, the volume averaging operators can be used to obtain the random motion of the particles from the flow field generated by the fluctuating stress.  As the fluctuating stress is based on spatially uncorrelated white-noise, the matrix square root computation does not have to be performed.  For the stochastic and fluctuating IBMs and fluctuating FCM, it has been shown\cite{Atzberger2007,Atzberger2011,Keaveny2014,Delong2014} explicitly that the resulting particle velocity correlations satisfy the fluctuation-dissipation theorem.

While the usage of fluctuating hydrodynamics has accelerated the computation of the random particle velocities, in the overdamped, or Brownian dynamics limit where one can ignore both fluid and particle inertia, a seemingly-problematic Brownian drift term proportional to the divergence of the particle mobility matrix also needs to be accounted for even to achieve the correct equilibrium distribution of the suspension.  

Even though the drift term is identically zero for periodic suspensions when the point force solution is used to provide the mobility matrix, this is not the case when second-order corrections provided by the stresslets are included.  This presented particular difficulty in developing fluctuating FCM\cite{Keaveny2014}.  While we were able to successfully and rapidly compute the correct random displacements with fluctuating FCM, the Brownian drift arising because of the stresslets limited any dynamic simulations we performed to only $O(10^2)$ particles.  In this work, we address this challenge directly and present a complete methodology for dynamic simulation with fluctuating FCM that accounts for the drift term due to the stresslets at a minimal computational cost.

Our approach relies on the construction of a time integration scheme that naturally accounts for the drift term.  Fixman\cite{Fixman1978,Grassia1995} showed that the effects of the drift term can be recovered without a direct computation by using a specific mid-point time integration scheme.  This approach, however, relies on the usage of random forces and torques, rather than random velocities and angular velocities, making its implementation with the stochastic and fluctuating IBMs and fluctuating FCM rather cumbersome and computationally expensive\cite{Keaveny2014}.  To overcome this difficulty, Delong \emph{et al.} \cite{Delong2014} proposed an integration scheme for the IBM using what they termed as random finite differencing (RFD).  By randomly displacing and forcing the particles in a particular way, they showed that divergence of the mobility matrix due to point force hydrodynamic interactions can be recovered without ever needing to perform the onerous computations involved with Fixman's method. 

In this study, we construct a time integration for fluctuating hydrodynamics-based simulations of Brownian particles that include the second-order stresslet and rotlet corrections to the particle hydrodynamic interactions, for which RFD still incurs a significant computational cost.  We develop a midpoint time integration scheme that we refer to as the drifter-corrector (DC) that requires the stresslet computation be performed only once per timestep, leading to the desired reduction in computational cost.  

The DC works by first advecting the particle positions a half timestep by the unconstrained fluctuating flow field.  At the midpoint, the complete calculation to determine the partice motion is performed, including the computation of the stresslets.  This time integration scheme with fluctuating FCM forms an efficient and complete method for the dynamic simulation of Brownian suspensions of moderate concentration.  In fact, we show that when the DC is used in conjunction with fluctuating FCM, a Brownian simulation requires just a single additional unconstrained Stokes solve per timestep as compared to a deterministic FCM simulation of the same system.  

We provide an extensive validation of the DC by performing dynamic fluctuating FCM simulations of periodic suspensions and of a single particle between two slip surfaces.  We show how to impose these boundary conditions with fluctuating hydrodynamics by modifying the spatial correlations of the fluctuating stress, rather than imposing it directly through the Stokes solver.  This additional contribution provides a framework to study Brownian particles in thin films, allowing for direct comparison with experimental results such as those on active particles \cite{Leptos2009,Kurtuldu2011}.  In addition, these simulations confirm our theoretical analysis of the scheme and shows that the DC is able to yield both the correct equilibrium distribution and the correct distribution dynamics as described by the Smoluchowski equation.  

We consider the canonical problem of a collapsing cluster of interacting colloidal particles, allowing for comparison with results given by other methodologies and the demonstration that the higher-order stresslet corrections resolved in fluctuating FCM yield quantitative differences in the results. We show that, by avoiding additional stresslet iterations, the DC has a computational cost half that of forward RFD. 

Finally, using the DC with fluctuating FCM, we also perform simulations of suspensions of interacting Brownian particles to explore colloidal gelation and percolated network formation.  We show that in addition to reproducing the results of other simulation techniques such as SD, fluctuating FCM with the DC allows for large-scale simulation of hydrodynamically interacting Brownian particles at a low computational cost.
\section{Equations of motion}
In this study, we will be considering a suspension of $N_p$ spherical Brownian particles, each having radius $a$.  The position of particle $n$ is denoted by $\mathbf{Y}^n$.  Each particle may also be subject to non-hydrodynamic forces, $\mathbf{F}^n$, and torques, $\bm{\tau}^n$.  In the overdamped or Brownian dynamics limit, where both fluid and particle inertia are negligible \cite{Ermak1978,Brady1988}, the dynamics of the particles is described by the system of stochastic differential equations
\begin{eqnarray}
d\mathcal{Y} & = & \left(\mathcal{M}^{\mathcal{VF}}\mathcal{F}+ \mathcal{M}^{\mathcal{VT}}\mathcal{T}\right)dt \nonumber \\
& &   + k_BT\bm\nabla_{\mathcal{Y}}\cdot \mathcal{M}^{\mathcal{VF}}dt + d\tilde{\mathcal{V}}
\label{eq:SDE}
\end{eqnarray}
where $\mathcal{Y}$ is the $3N_p \times 1$ vector that contains the position information for all particles, $\mathcal{F}$ is the $3N_p \times 1$ vector of forces on all of the particles, and $\mathcal{T}$ is the similar vector for the torques.  We note that because of their isotropic shape, there is no need to keep track of orientations for passive spherical particles and Eq. \eqref{eq:SDE} is sufficient to describe the dynamics of the suspension.  For active or Janus spherical particles that possess an inherent swimming direction, one must also keep track of orientation and an additional equation similar to Eq. \eqref{eq:SDE} for particle orientation must also be considered.  The vector $d\tilde{\mathcal{V}}$ is the incremental random velocity which, along with the incremental angular velocity, $d\tilde{\mathcal{W}}$, is related to the incremental $6N_p\times 1$ Wiener process, $d\mathcal{B}$, through
\begin{equation}
\left[\begin{array}{c}
 d\tilde{\mathcal{V}} \\
 d\tilde{\mathcal{W}}
\end{array}\right]=\sqrt{2k_BT}\mathcal{M}^{1/2}d\mathcal{B}
\label{eq:FDT}
\end{equation}
where $k_B$ is Boltzmann's constant and $T$ is the temperature.  Appearing also in Eqs. (\ref{eq:SDE}) and (\ref{eq:FDT}) is the $6N_p \times 6N_p$ hydrodynamic mobility matrix
\begin{equation}
\mathcal{M} = 
\left[\begin{array}{cc}
 \mathcal{M}^{\mathcal{VF}} & \mathcal{M}^{\mathcal{VT}}\\
 \mathcal{M}^{\mathcal{WF}} & \mathcal{M}^{\mathcal{WT}}
\end{array}\right],
\end{equation}
as well as its submatrices $\mathcal{M}^{\mathcal{VF}}$ and $\mathcal{M}^{\mathcal{VT}}$.  The mobility matrix, which in general depends on the particle positions, provides the linear relationship between the forces and torques on the particles and their resulting velocities and angular velocities.  All the information about how the particles interact through the fluid is contained in the mobility matrix.  The mobility matrix is determined by solving the Stokes equations,
\begin{eqnarray}
\bm{\nabla} p - \eta \nabla^2 \mathbf{u} &=& \mathbf{0}\nonumber\\
\bm{\nabla}\cdot \mathbf{u} &=& 0
\end{eqnarray}
that govern the flow induced in the surrounding fluid as the particles move through it.  \\

For methods such as Brownian dynamics or SD, the mobility matrix is constructed using the flow generated by a force multipole expansion in the Stokes equations and Fax\'en laws\cite{Happel2012,Kim1991} to extract the particle motion from the fluid velocity.  In this paper, we adopt an alternative, but parallel approach known as the force-coupling method\cite{Maxey2001,Lomholt2003} (FCM) that utilizes regularized, rather than singular, force distributions in the Stokes equations and replaces the Fax\'en laws by volume averaging operators.  As noted by Delong \emph{et al.}\cite{Delong2014}, this approach and the IBM share similar features and have distinct advantages over methods based on singular force distributions, particularly when Brownian motion is included in the simulation.

In the absence of Brownian motion, only the first term in Eq. (\ref{eq:SDE}), the mobility matrix multiplied by the forces and torques, will be non-zero.  The inclusion of Brownian motion gives rise to the two additional terms -- the incremental random velocity, $d\tilde{\mathcal{V}}$, as well as an additional drift known as Brownian drift.  To satisfy the fluctuation-dissipation theorem\cite{Kubo1966},  $d\tilde{\mathcal{V}}$ depends on the square root of the mobility matrix through Eq. (\ref{eq:FDT}).  This links the random motion of the particles with how they interact through the fluid.  As we discuss in the next section, we can compute this term rather efficiently by volume averaging and constraining the fluid flow generated by a white-noise fluctuating stress.  

The Brownian drift term has a more subtle origin.  It arises as a result of taking the overdamped limit of the Langevin dynamics where particle inertia is present \cite{Ermak1978,Atzberger2011}.  In order to successfully ignore the suspension dynamics during the short inertial relaxation timescale, one must include the Brownian drift term to obtain an SDE that is consistent with Smoluchowski's equation \cite{Ermak1978}.  Note that since the particles considered here are spherical, the Brownian drift term does not involve the divergence of the mobility matrix with respect to particle orientations, only particle positions.  The purpose of this work is to introduce a new and study existing\cite{Fixman1978,Grassia1995,Delong2014} time integration schemes that automatically account for Brownian drift.  We pay particular attention to the case where the second-order correction to the particle mobility matrix is obtained by imposing a local rate-of-strain constraint on the flow.  We show that for methods employing fluctuating hydrodynamics, the Brownian drift can be accounted for at the cost of a single additional unconstrained Stokes solve per time-step by using an appropriately designed time integration scheme.
\section{Fluctuating FCM}
To compute the terms in Eq. (\ref{eq:SDE}) that depend on the mobility matrix, we utilize fluctuating FCM that is described and analyzed in detail in our previous work \cite{Keaveny2014}.  Fluctuating FCM combines FCM\cite{Maxey2001,Lomholt2003} with fluctuating hydrodynamics resulting in an efficient methodology to determine particle hydrodynamic interactions while simultaneously yielding the correct random particle velocities that satisfy the fluctuation-dissipation theorem.  In fluctuating FCM, the fluid velocity is given by the Stokes equations driven by a white-noise fluctuating stress, and a low-order, regularized multipole expansion representing the force the particles exert on the fluid.  The motion of the particles is then found by taking volume averages of the fluid velocity.  This process is similar to that of the stochastic and fluctuating IBMs, and as we demonstrate below, fluctuating FCM can be expressed in terms of projection, or spreading, operators and volume averaging, or interpolation, operators commonly used to describe IBM.  We present fluctuating FCM using this framework to emphasize the connection between the methodologies, indicating that the time integration schemes that we explore using fluctuating FCM in subsequent sections could also be used more widely.

For fluctuating FCM, the fluid velocity is given by the following Stokes flow
\begin{eqnarray}
\bm{\nabla} p - \eta \nabla^2 \mathbf{u} &=& \bm{\nabla} \cdot \mathbf{P}+ \mathcal{J}^\dagger[\mathcal{F}]\nonumber\\
&& + \mathcal{N}^\dagger[\mathcal{T}] + \mathcal{K}^\dagger[\mathcal{S}] \nonumber \\
\bm{\nabla}\cdot \mathbf{u} &=& 0,\nonumber\\
&&\label{eq:Stokes}\\
\mathcal{K}[\mathbf{u}] &=& \mathbf{0},\label{eq:Econ}
\end{eqnarray}
where the fluctuating stress, $\mathbf{P}$, has the following statistics
\begin{eqnarray}
\langle P_{ij}(\mathbf{x})\rangle &=& 0\\
\langle P_{ij}(\mathbf{x})P_{kl}(\mathbf{y})\rangle &=& 2k_BT(\delta_{ik}\delta_{jl} + \delta_{il}\delta_{jk})\delta(\mathbf{x} - \mathbf{y})\nonumber\\
& & 
\end{eqnarray}
with $\langle \cdot \rangle$ denoting the ensemble average of a quantity. 

The non-hydrodynamic forces, $\mathcal{F}$, and torques, $\mathcal{T}$, on the particles, as well as the particle stresslets, $\mathcal{S}$, the symmetric force-moment on each particle, are projected onto the fluid using the linear operators $\mathcal{J}^\dagger$, $\mathcal{N}^\dagger$, and $\mathcal{K}^\dagger$ which are given by
\begin{eqnarray}
\mathcal{J}^\dagger[\mathcal{F}] &=& \sum_{n} \mathbf{F}^n\Delta_n(\mathbf{x})\\
\mathcal{N}^\dagger[\mathcal{T}] &=& -\frac{1}{2}\sum_{n} \bm{\tau}^n\times \bm{\nabla}\Theta_n(\mathbf{x})\\
\mathcal{K}^\dagger[\mathcal{S}] &=& \frac{1}{2}\sum_{n} \mathbf{S}^n\cdot\left( \bm{\nabla}\Theta_n(\mathbf{x})+\left(\bm{\nabla}\Theta_n(\mathbf{x})\right)^{T}\right).\nonumber\\
\end{eqnarray}
Appearing in these expressions are the two Gaussian envelopes, or spreading functions,
\begin{eqnarray}
\Delta_n(\mathbf{x})&=&(2\pi\sigma_{\Delta}^2)^{-3/2}\textrm{e}^{-|\mathbf{x} - \mathbf{Y}^n|^2/2\sigma_{\Delta}^2} \label{eq:monoGauss}\\
\Theta_n(\mathbf{x})&=&(2\pi\sigma_{\Theta}^2)^{-3/2}\textrm{e}^{-|\mathbf{x} - \mathbf{Y}^n|^2/2\sigma_{\Theta}^2} \label{eq:dipGauss}
\end{eqnarray} 
where $\sigma_{\Delta}$ and $\sigma_{\Theta}$ are related to the radius of the particles through $\sigma_{\Delta} = a/\sqrt{\pi}$ and $\sigma_{\Theta} = a/\left(6\sqrt{\pi}\right)^{1/3}$.  Unlike the forces and torques which are typically set by external or inter-particle potentials, the stresslets arise as a result of the constraint on the flow given by Eq. (\ref{eq:Econ}) and, consequently, need to be solved for as part of the general flow problem.
The projection operators are the adjoints of the volume averaging operators $\mathcal{J}$, $\mathcal{N}$, and $\mathcal{K}$ that are used to extract the particle velocities, $\mathcal{V}$, angular velocities, $\mathcal{W}$, and local rates-of-strain, $\mathcal{E}$, from the fluid velocity and its derivatives, 
\begin{eqnarray}
\mathcal{V} &=& \mathcal{J}[\mathbf{u}]\label{eq:parV} \\
\mathcal{W} &=& \mathcal{N}[\mathbf{u}]  \\
\mathcal{E} &=& -\mathcal{K}[\mathbf{u}]. 
\end{eqnarray}
The expressions for these operators are most clearly expressed when they are restricted to particle $n$,
\begin{eqnarray}
\mathbf{V}^n &=& (\mathcal{J}[\mathbf{u}])_n = \int \mathbf{u}\Delta_n(\mathbf{x})d^3\mathbf{x} \label{eq:FCM_Ju}\\
\bm{\Omega}^n &=& (\mathcal{N}[\mathbf{u}])_n = \frac{1}{2}\int \mathbf{u}\times\bm{\nabla} \Theta_n(\mathbf{x})d^3\mathbf{x} \label{eq:FCM_Qu} \\
\mathbf{E}^n &=& -(\mathcal{K}[\mathbf{u}])_n\nonumber\\
&=& -\frac{1}{2}\int \left[\mathbf{u}\left(\bm{\nabla}\Theta_n(\mathbf{x})\right)^T +\bm{\nabla}\Theta_n(\mathbf{x})\mathbf{u}^T \right]d^3\mathbf{x} \nonumber \label{eq:FCM_Ku}. \\
\end{eqnarray}
It is important to note that these definitions for the particle angular velocity and local rate of strain depart slightly from the standard FCM definitions provided by Lomholt \& Maxey\cite{Lomholt2003}.  As we show in Appendix \ref{sec:VisDis}, using Eqs. (\ref{eq:FCM_Qu}) and (\ref{eq:FCM_Ku}) ensures that, in general, the rate of work done by the particles on the fluid is correctly balanced by the viscous dissipation.  These definitions also ensure that the stresslets do no work on the fluid.  If periodic or no-slip conditions are imposed on the bounding surfaces, integration by parts of Eqs. (\ref{eq:FCM_Qu}) and (\ref{eq:FCM_Ku}) reveals that the particle angular velocities and local rates-of-strain reduce to the standard FCM definitions for these quantities which are the local volume averages of the fluid vorticity and rate-of-strain, respectively.  With this in mind, the constraint Eq. (\ref{eq:Econ}) insists the local rate-of-strain must be zero as is the case for a rigid particle, and the stresslets can be viewed as the Lagrange multipliers included to enforce this constraint.  

It is also important to note the dependence of the Gaussian envelopes, Eqs. \eqref{eq:monoGauss} and \eqref{eq:dipGauss}, on particle position, $\mathbf{Y}^n$.  For a fixed flow field $\mathbf{u}$, the values of $\mathbf{V}^n$,  $\bm{\Omega}^n$, and $\mathbf{E}^n$ are continuous functions of $\mathbf{Y}^n$ and can be differentiated with respect to this quantity.  We will see that this dependence is quite important and needs to be considered to design schemes that correctly update the particle positions when Brownian motion is included in simulation.

At this stage, it is useful to again note that the fluctuating and stochastic IBM share this framework with fluctuating FCM.  The main differences between IBM and FCM are the choice of spreading function, and the only operators typically used with IBM are $\mathcal{J}$ and its adjoint.  The higher-order correction to the hydrodynamic interactions due to torques or stresslets resolved in FCM are not typically included with IBM.  

For FCM, the stresslets can be obtained using the conjugate gradient procedure describe by Yeo and Maxey\cite{Yeo2010}, and once found, the solution to Eq. (\ref{eq:Stokes}) that satisfies Eq. (\ref{eq:Econ}) is determined.  For the simulations performed in the proceeding sections, approximately 10 conjugate gradient iterations, each needing one Stokes solve, are required to reach a residual of
$\|\mathcal{E}\|=7\cdot10^{-5}D_0/a^2$, where
\begin{equation}
\|\mathcal{E}\| =\underset{n\in[1,N_p]}{\max}\|\mathbf{E}^n\|_2,
\end{equation}
and $D_0 = k_BT/(6\pi\eta a)$ is the Stokes-Einstein diffusion coefficient for an isolated sphere.

In designing, exploring, and constructing the time integration scheme that provides the Brownian drift term, we are particularly mindful of the computational cost associated with the stresslet computation, and in fact, seek to limit this computation to once per timestep. 
\subsection{Mobility matrices and the fluctuation dissipation theorem}
Fluctuating FCM yields the deterministic and random velocities, corresponding to the first and third terms in Eqs (\ref{eq:SDE}) without ever directly computing the mobility matrix.  Although they are never computed explicitly, expressions for the mobility matrices can be determined \cite{Keaveny2014}.  For FCM, the mobility matrices provide the linear relationship between the deterministic particle velocities, angular velocities, and local rates-of-strain and the forces, torques, and stresslets associated with the particles,
\begin{equation}
\left[\begin{array}{c}
\mathcal{V} \\
\mathcal{W} \\
\mathcal{E}
\end{array}\right]=
\left[\begin{array}{ccc}
\mathcal{M}^{\mathcal{V}\mathcal{F}}_{FCM} & \mathcal{M}^{\mathcal{V}\mathcal{T}}_{FCM} & \mathcal{M}^{\mathcal{V}\mathcal{S}}_{FCM} \\
\mathcal{M}^{\mathcal{W}\mathcal{F}}_{FCM} & \mathcal{M}^{\mathcal{W}\mathcal{T}}_{FCM} & \mathcal{M}^{\mathcal{W}\mathcal{S}}_{FCM} \\
\mathcal{M}^{\mathcal{E}\mathcal{F}}_{FCM} & \mathcal{M}^{\mathcal{E}\mathcal{T}}_{FCM} & \mathcal{M}^{\mathcal{E}\mathcal{S}}_{FCM}
\end{array}\right]
\left[\begin{array}{c}
\mathcal{F} \\
\mathcal{T} \\
\mathcal{S}
\end{array}\right].
\label{eq:GMM}
\end{equation}
As shown in our previous work\cite{Keaveny2014}, the submatrices can be expressed using the Gaussian spreading functions, Eqs. (\ref{eq:monoGauss}) and (\ref{eq:dipGauss}) and the Green's function $\mathbf{G}(\mathbf{x},\mathbf{y})$ for the Stokes equations.  For example, the entries of the matrix $\mathcal{M}^{\mathcal{V}\mathcal{F}}_{FCM}$ linking particles $n$ and $m$ are given by
\begin{equation}
\mathcal{M}^{\mathcal{V}\mathcal{F};nm}_{FCM} = \int \int \Delta_n(\mathbf{x}) \mathbf{G}(\mathbf{x},\mathbf{y}) \Delta_m(\mathbf{y})d^3\mathbf{x}d^3\mathbf{y}.
\label{eq:MVF}
\end{equation}
Similar expressions can be found for the other matrices in the appendix of our previous work\cite{Keaveny2014}.  As described by Delong \emph{et al.}\cite{Delong2014}, these matrices can also be expressed using the projection, volume averaging, and Stokes operators.  For $\mathcal{M}^{\mathcal{V}\mathcal{F}}_{FCM}$, the expression is
\begin{equation}
\mathcal{M}_{FCM}^{\mathcal{V}\mathcal{F}} = \mathcal{J}[\mathcal{L}^{-1}[\mathcal{J}^\dagger[\cdot]]]
\end{equation}
where $\mathcal{L}^{-1}$ represents the inverse Stokes operator.

In the absence of the stresslets, the mobility matrix for FCM reduces to
\begin{equation}
\mathcal{M}_{FCM}=
\left[\begin{array}{cc}
\mathcal{M}^{\mathcal{V}\mathcal{F}}_{FCM} & \mathcal{M}^{\mathcal{V}\mathcal{T}}_{FCM} \\
\mathcal{M}^{\mathcal{W}\mathcal{F}}_{FCM} & \mathcal{M}^{\mathcal{W}\mathcal{T}}_{FCM} 
\end{array}\right]
\left[\begin{array}{c}
\mathcal{F} \\
\mathcal{T} 
\end{array}\right].
\label{eq:FCMmonomat}
\end{equation}
If the stresslets are included, they can be found using Eq. (\ref{eq:GMM}).  Since for rigid particles the local rates-of-strain are zero, we have $\mathcal{E}=0$ and, in terms of the forces and torques, the stresslets will be given by
\begin{equation}
\mathcal{S} = -\mathcal{R}_{FCM}^{\mathcal{E}\mathcal{S}}\left(\mathcal{M}_{FCM}^{\mathcal{E}\mathcal{F}}\mathcal{F} + \mathcal{M}_{FCM}^{\mathcal{E}\mathcal{T}}\mathcal{T}\right).
\end{equation}
where we have written $\mathcal{R}_{FCM}^{\mathcal{E}\mathcal{S}} = (\mathcal{M}_{FCM}^{\mathcal{E}\mathcal{S}})^{-1}$.  From this expression for $\mathcal{S}$, we find the stresslet-corrected mobility matrix is
\begin{equation}
\mathcal{M}_{FCM-S} = 
\left[\begin{array}{cc}
\mathcal{M}^{\mathcal{V}\mathcal{F}}_{FCM-S} & \mathcal{M}^{\mathcal{V}\mathcal{T}}_{FCM-S}  \\
\mathcal{M}^{\mathcal{W}\mathcal{F}}_{FCM-S} & \mathcal{M}^{\mathcal{W}\mathcal{T}}_{FCM-S}
\end{array}\right]
\end{equation}
where
\begin{eqnarray}
\mathcal{M}^{\mathcal{V}\mathcal{F}}_{FCM-S} &=& \mathcal{M}^{\mathcal{V}\mathcal{F}}_{FCM}-\mathcal{M}^{\mathcal{V}\mathcal{S}}_{FCM}\mathcal{R}^{\mathcal{E}\mathcal{S}}_{FCM}\mathcal{M}^{\mathcal{E}\mathcal{F}}_{FCM}, \nonumber \\
\mathcal{M}^{\mathcal{V}\mathcal{T}}_{FCM-S} &=& \mathcal{M}^{\mathcal{V}\mathcal{T}}_{FCM}-\mathcal{M}^{\mathcal{V}\mathcal{S}}_{FCM}\mathcal{R}^{\mathcal{E}\mathcal{S}}_{FCM}\mathcal{M}^{\mathcal{E}\mathcal{T}}_{FCM}, \nonumber\\
\mathcal{M}^{\mathcal{W}\mathcal{F}}_{FCM-S} &=& \mathcal{M}^{\mathcal{W}\mathcal{F}}_{FCM}-\mathcal{M}^{\mathcal{W}\mathcal{S}}_{FCM}\mathcal{R}^{\mathcal{E}\mathcal{S}}_{FCM}\mathcal{M}^{\mathcal{E}\mathcal{F}}_{FCM}, \nonumber\\
\mathcal{M}^{\mathcal{W}\mathcal{T}}_{FCM-S} &=& \mathcal{M}^{\mathcal{W}\mathcal{T}}_{FCM}-\mathcal{M}^{\mathcal{W}\mathcal{S}}_{FCM}\mathcal{R}^{\mathcal{E}\mathcal{S}}_{FCM}\mathcal{M}^{\mathcal{E}\mathcal{T}}_{FCM}.  \nonumber\\
\label{eq:FCM-S}
\end{eqnarray}

Again, while these matrices are never computed explicitly, we know that they exist.  This allows us to draw a parallel between fluctuating FCM and traditional methods for suspended particles such as Brownian dynamics\cite{Ermak1978} and SD\cite{Brady1988}.  In addition, the expressions for the mobility matrices will prove useful in the analysis that demonstrates the time-integration scheme we propose in this study yields the correct first and second moments of the incremental change in the particle positions to first order in time.

The expressions for the mobility matrices can also be used to demonstrate that the random motion of the particles obtained using fluctuating FCM complies with the fluctuation-dissipation theorem\cite{Kubo1966}.  If we consider the flow, 
\begin{eqnarray}
\bm{\nabla} p - \eta \nabla^2 \mathbf{\tilde u} &=& \bm{\nabla} \cdot \mathbf{P}  \nonumber\\
\bm{\nabla}\cdot \mathbf{\tilde{u}} &=& 0 \label{eq:FlucStokes}\\
\mathcal{K}[\mathbf{\tilde{u}}] &=& \mathbf{0}, \label{eq:FlucEcon}
\end{eqnarray}
the resulting particle velocities are then
\begin{equation}
\tilde{\mathcal{V}}=\mathcal{J}[\mathbf{\tilde{u}}]. \label{eq:RandVel}
\end{equation}
By examining the velocity corrections explicitly, one can show \cite{Keaveny2014} that the particle velocities given by Eq. (\ref{eq:RandVel}) satisfy the fluctuation-dissipation theorem \cite{Kubo1966}.  For fluctuating FCM, we have that
\begin{equation}
\langle \tilde{\mathcal{V}}\tilde{\mathcal{V}}^{T}\rangle = 2k_BT \mathcal{M}^{\mathcal{VT}}_{FCM}
\label{eq:fluctu_dissip_no_stress}
\end{equation}
when Eq. (\ref{eq:FlucEcon}) is not enforced (i.e. $\mathcal{S} = 0$) and
\begin{equation}
\langle \tilde{\mathcal{V}}\tilde{\mathcal{V}}^{T}\rangle = 2k_BT \mathcal{M}^{\mathcal{VT}}_{FCM-S}
\end{equation}
when the constraint Eq. (\ref{eq:FlucEcon}) is enforced.  A similar result has been shown in the context of the stochastic and fluctuating IBMs\cite{Atzberger2007,Delong2014}.  Thus, by simply applying the volume averaging operators to the fluid flow induced by the white-noise fluctuating stress, we can obtain random particle velocities consistent with the noise terms in the stochastic equations of motion.  This is done without ever needing to explicitly form and decompose the mobility matrix, leading to a great reduction in the computational effort needed to compute these terms.
\section{Time integration}

While fluctuating FCM provides the deterministic and random particle velocities corresponding to the first and last terms in the equations of motion Eq. (\ref{eq:SDE}), the Brownian drift, $k_BT\bm\nabla_{\mathcal{Y}} \cdot \mathcal{M}^{\mathcal{VF}}_{FCM-S}$, would also need to be accounted for to advance the particle positions in time. 
Therefore, simply by applying the Euler-Maruyama (EM) scheme\cite{Kloeden1992} to fluctuating FCM where one first solves the Stokes equations
\begin{eqnarray}
\bm{\nabla} p^k - \eta \nabla^2 \mathbf{u}^k &=& (\Delta t)^{-\frac{1}{2}}\bm{\nabla} \cdot \mathbf{P}^k\nonumber\\
&&+  \mathcal{J}^{\dagger;k}[\mathcal{F}^k] \nonumber\\
&&+ \mathcal{N}^{\dagger;k}[\mathcal{T}^k]+ \mathcal{K}^{\dagger;k}[\mathcal{S}^k]\nonumber\\
\bm{\nabla}\cdot \mathbf{u}^k &=& 0,\nonumber\\
\mathcal{K}^k[\mathbf{u}^k] &=& \mathbf{0},\nonumber
\end{eqnarray}
and then updates the particle positions using
\begin{eqnarray}
\mathcal{Y}^{k+1} &=&  \mathcal{Y}^{k} + \Delta t\mathcal{J}^{k}[\mathbf{u}^{k}],\nonumber
\label{eq:EMFCM}
\end{eqnarray}
one does not account for $k_BT\bm\nabla_{\mathcal{Y}} \cdot \mathcal{M}^{\mathcal{VF}}_{FCM-S}$ and will not achieve the correct suspension dynamics.  Rather than resorting to computing the drift term explicitly, however, its effects can be successfully incorporated into a simulation by using an appropriately designed time integration scheme\cite{Fixman1978,Grassia1995,Delong2014}.  In this section, we discuss these schemes, and building from the ideas used in their construction, we introduce a new scheme called the drifter-corrector (DC).  The DC has the particular advantage that when constraints are imposed on the flow to recover stresslet corrections for the particle hydrodynamic interactions, they need only to be imposed once per timestep.  This leads to a non-negligible reduction in computational cost with respect to other schemes.   For the DC, the total additional computational cost per timestep for fluctuating FCM with respect to its deterministic counterpart is a single Stokes solve per timestep.
\subsection{Fixman's method}
In the late 1970's, Fixman \cite{Fixman1978,Grassia1995} developed a midpoint integration scheme to account for the drift term.  The key to recovering the drift is that the scheme employs the same random forces at time levels $t_k$ and $t_{k+1/2}$ while using updated values for the particle positions at level $t_{k+1/2}$.  Applying Fixman's method to fluctuating FCM yields the following scheme:
\begin{enumerate}
\item Compute the random velocities, $\tilde{\mathcal{V}}^k$ and angular velocities, $\tilde{\mathcal{W}}^k$
\begin{eqnarray}
\bm{\nabla} \tilde{p}^k - \eta \nabla^2 \mathbf{\tilde{u}}^k &=& (\Delta t)^{-\frac{1}{2}}\bm{\nabla} \cdot \mathbf{P}^k\nonumber\\
&&+ \mathcal{K}^{\dagger;k}[\mathcal{\tilde{S}}^k]\nonumber\\
\bm{\nabla}\cdot \mathbf{\tilde{u}}^k &=& 0,\nonumber\\
\mathcal{K}^k[\mathbf{\tilde{u}}^k] &=& \mathbf{0},\nonumber\\
\tilde{\mathcal{V}}^k &=& \mathcal{J}[\mathbf{\tilde{u}}^k]\nonumber \\
\tilde{\mathcal{W}}^k &=& \mathcal{N}[\mathbf{\tilde{u}}^k]\nonumber 
\end{eqnarray}
\item Compute random forces, $\tilde{\mathcal{F}}^k$, and torques, $\tilde{\mathcal{T}}^k$, from $\tilde{\mathcal{V}}^k$ and $\tilde{\mathcal{W}}^k$
\begin{equation}
\left[\begin{array}{c}
\tilde{\mathcal{F}}^k \\
\tilde{\mathcal{T}}^k
\end{array}\right]=
\mathcal{M}^{-1}_{FCM-S}
\left[\begin{array}{c}
\tilde{\mathcal{V}}^k \\
\tilde{\mathcal{W}}^k
\end{array}\right].
\label{eq:BD}
\end{equation}
\item Solve the predictor constrained Stokes problem
\begin{eqnarray}
\bm{\nabla} p^k - \eta \nabla^2 \mathbf{u}^k &=& \mathcal{J}^{\dagger;k}[\mathcal{F}^k + \tilde{\mathcal{F}}^k] \nonumber\\
&&+ \mathcal{N}^{\dagger;k}[\mathcal{T}^k + \tilde{\mathcal{T}}^k]\nonumber\\
&&+ \mathcal{K}^{\dagger;k}[\mathcal{S}^k]\nonumber\\
\bm{\nabla}\cdot \mathbf{u}^k &=& 0\nonumber\\
\mathcal{K}^k[\mathbf{u}^k] &=& \mathbf{0}\nonumber\\
\end{eqnarray}
\item Move particles to the midpoint,
\begin{eqnarray}
\mathcal{Y}^{k+1/2} &=& \mathcal{Y}^{k} + \frac{\Delta t}{2}\mathcal{J}^k[\mathbf{u}^k]
\end{eqnarray}
\item Solve the constrained Stokes problem at the midpoint,
\begin{eqnarray}
\bm{\nabla} p^{k+1/2} - \eta \nabla^2 \mathbf{u}^{k+1/2} &=& \mathcal{J}^{\dagger;k+1/2}[\mathcal{F}^k + \tilde{\mathcal{F}}^k]\nonumber\\
&&+ \mathcal{N}^{\dagger;k+1/2}[\mathcal{T}^k + \tilde{\mathcal{T}}^k]\nonumber\\
&&+ \mathcal{K}^{\dagger;k+1/2}[\mathcal{S}^{k+1/2}]\nonumber\\
\bm{\nabla}\cdot \mathbf{u}^{k+1/2} &=& 0\nonumber\\
\mathcal{K}^{k+1/2}[\mathbf{u}^{k+1/2}] &=& \mathbf{0}\nonumber\\
\end{eqnarray}
\item Update the particle positions
\begin{eqnarray}
\mathcal{Y}^{k+1} &=&  \mathcal{Y}^{k} + \Delta t \mathcal{J}^{k+1/2}[\mathbf{u}^{k+1/2}]\nonumber\\
\end{eqnarray}
\end{enumerate}
The subscripts $k$ and $k+1/2$ indicate whether the operators are evaluated at the positions $\mathcal{Y}^k$, or $\mathcal{Y}^{k+1/2}$.  This scheme provides the first and second moments of the increment up to first order in $\Delta t$.  We note that the approximation of the deterministic aspects of the motion can be improved to second order by using $\mathcal{F}^{k+1/2}$ and $\mathcal{T}^{k+1/2}$ instead of $\mathcal{F}^{k}$ and $\mathcal{T}^{k}$ in Step 5.

While Fixman's scheme does avoid a direct calculation of the Brownian drift term, it does introduce the need to solve a large linear system in Step 2 to find the random forces and torques.  This system can be solved using conjugate gradient methods, with each iteration requiring the Stokes equations to be solved.  Simulations with $N = 183$ particles corresponding to a volume fraction of $\phi = 0.1$ required approximately 25 iterations to find the random forces and torques with the residual reduced to 1e-4 times its original value \cite{Keaveny2014}.  This computation, along with having to compute the stresslets twice per timestep, severely limit the scale at which Brownian simulations could be performed using the fluctuating FCM.
 \subsection{Random Finite Differencing}
To help overcome this challenge, Delong \emph{et al.}\cite{Delong2014} introduced the random finite difference (RFD).  RFD takes advantage of the fact that the random forces and torques and the resulting displacements used in Fixman's method are in fact one of many possible choices that could be used to account for Brownian drift.  Specifically, they show that using random displacements $\delta \Delta \mathcal{Y}$, as well as randomly particle forcing $\Delta \mathcal{F}/\delta $, the divergence of a mobility matrix, $\mathcal{M}$, can be approximated from
\begin{eqnarray}
\frac{1}{\delta}\langle \mathcal{M}\left(\mathcal{Y} + \delta\Delta \mathcal{Y}\right)\Delta \mathcal{F} -\mathcal{M}(\mathcal{Y})\Delta \mathcal{F}\rangle &=& \nabla_{\mathcal{Y}}\cdot \mathcal{M}\nonumber\\
&&+ O(\delta)\nonumber \\
\end{eqnarray}
provided that $\langle \Delta \mathcal{Y} \Delta \mathcal{F}\rangle = \mathcal{I}$.  Therefore, one can simply choose $\Delta \mathcal{Y}=\Delta \mathcal{F}=\bm{\xi}$, where $\bm{\xi}$ is a $3N_p \times 1$ vector of independent Gaussian random variables with zero mean and unit variance.  This eliminates the need to solve any linear system. The concept of random finite differencing can also be extended to higher-order accuracy.  For example, the central random finite difference is given by
\begin{eqnarray}
\frac{1}{\delta}\langle \mathcal{M}(\mathcal{Y} + \delta \bm{\xi}/2)\bm{\xi} -\mathcal{M}(\mathcal{Y} - \delta \bm{\xi}/2)\bm{\xi}\rangle &\nonumber\\
= \nabla_{\mathcal{Y}}\cdot \mathcal{M} + O(\delta^2)\nonumber \\
\end{eqnarray}
Thus, in general, a weakly first-order accurate scheme that accounts for the Brownian drift term can be constructed by adding an RFD to the Euler-Maruyama scheme.  Applying this to fluctuating FCM, we have:
\begin{enumerate}
\item Solve the constrained Stokes problem
\begin{eqnarray}
\bm{\nabla} p^k - \eta \nabla^2 \mathbf{u}^k &=& (\Delta t)^{-\frac{1}{2}}\bm{\nabla} \cdot \mathbf{P}^k\nonumber\\
&&+  \mathcal{J}^{\dagger;k}[\mathcal{F}^k - \bm{\xi}/\delta] \nonumber\\
&&+ \mathcal{N}^{\dagger;k}[\mathcal{T}^k]+ \mathcal{K}^{\dagger;k}[\mathcal{S}^k]\nonumber\\
\bm{\nabla}\cdot \mathbf{u}^k &=& 0\nonumber\\
\mathcal{K}^k[\mathbf{u}^k] &=& \mathbf{0}
\end{eqnarray}
\item Solve the constrained Stokes problem with disturbed particle positions but without the fluctuating stress tensor, 
\begin{eqnarray}
\bm{\nabla} p^{k+\delta} - \eta \nabla^2 \mathbf{u}^{k+\delta} &=& \mathcal{J}^{\dagger;k+\delta}[\bm{\xi}/\delta]\nonumber\\
&&+ \mathcal{K}^{\dagger;k+\delta}[\mathcal{S}^{k+\delta}]\nonumber\\
\bm{\nabla}\cdot \mathbf{u}^{k+\delta} &=& 0\nonumber\\
\mathcal{K}^{k+\delta}[\mathbf{u}^{k+\delta}] &=& \mathbf{0}
\end{eqnarray}
\item Add the corresponding particle velocities and move particles to the next timestep, 
\begin{eqnarray}
\mathcal{Y}^{k+1} &=&  \mathcal{Y}^{k} + \Delta t\mathcal{J}^{k}[\mathbf{u}^{k}] \nonumber\\
&& +\Delta t \mathcal{J}^{k+\delta}[\mathbf{u}^{k+\delta}]
\end{eqnarray}
\end{enumerate}
where the superscript $k+\delta$ indicates that the operators are evaluated at $\mathcal{Y}^k + \delta \bm{\xi}$.  We emphasize that for passive spherical particles, we only need the divergence of the velocity-force mobility matrix.  We, therefore, only need to consider random displacements and forces with the RFD.  A similar scheme can be constructed using central RFD to account for the Brownian drift.  Comparing the resulting scheme with Fixman's method, we immediately see that using RFD eliminates the need to compute any random forces or torques, making this approach much more suited for fluctuating FCM.  

\subsection{Drifter-Corrector}
While the usage of RFD provides a clear advantage over Fixman's method for fluctuating hydrodynamics-based methods, we see that it does require that the stresslets be determined twice per timestep (three times per timestep if using central RFD).  The cost of performing a Brownian simulation is then at least double that of a deterministic simulation of the same system.  For fluctuating FCM where the conjugate gradient method is used to find the stresslets, the additional cost of including Brownian motion would then be approximately $10$ Stokes solves per timestep.  

In an effort to mitigate this computational cost as much as possible, we build on the ideas introduced by Fixman\cite{Fixman1978,Grassia1995} and Delong \emph{et al.}\cite{Delong2014} and construct a mid-point scheme that accounts for Brownian drift at the cost of one Stokes solve per timestep. We will refer to this scheme as the \emph{drifter-corrector} (DC).  Specifically, the DC is:
\begin{enumerate}
\item Solve the unconstrained Stokes problem
\begin{eqnarray}
\bm{\nabla} \tilde{p}^k - \eta \nabla^2 \mathbf{\tilde{u}}^k &=&(\Delta t)^{-\frac{1}{2}}\bm{\nabla} \cdot \mathbf{P}^k \nonumber \\
\bm{\nabla}\cdot \mathbf{\tilde{u}}^k &=& 0 \label{eq:DC1}
\end{eqnarray}
\item Move particles to the midpoint,
\begin{eqnarray}
\mathcal{Y}^{k+1/2} &=& \mathcal{Y}^{k} + \frac{\Delta t}{2}\mathcal{J}^k[\mathbf{\tilde{u}}^k] 
\end{eqnarray}
\item Solve the constrained Stokes problem, 
\begin{eqnarray}
\bm{\nabla} p^{k+1/2} - \eta \nabla^2 \mathbf{u}^{k+1/2} &=& (\Delta t)^{-\frac{1}{2}}\bm{\nabla} \cdot \mathbf{P}^k \nonumber\\
&&+  \mathcal{J}^{\dagger;k+1/2}[\mathcal{F}^{k+1/2}] \nonumber\\
&&+ \mathcal{N}^{\dagger;k+1/2}[\mathcal{T}^{k+1/2}]\nonumber\\
&&+ \mathcal{K}^{\dagger;k+1/2}[\mathcal{S}^{k+1/2}]\nonumber\\
\bm{\nabla}\cdot \mathbf{u}^{k+1/2} &=& 0\nonumber\\
\mathcal{K}^{k+1/2}[\mathbf{u}^{k+1/2}] &=& \mathbf{0}
\end{eqnarray}
\item Update the particle positions 
\begin{eqnarray}
\mathcal{Y}^{k+1} &=&  \mathcal{Y}^{k} \nonumber\\
&&+ \Delta t(1 + v^k) \nonumber\\
&&\times \mathcal{J}^{k+1/2}[\mathbf{u}^{k+1/2}]. \label{eq:Move_with_factor}
\end{eqnarray}
\end{enumerate}
The scalar factor, $v^k$, is related to the divergence of the unconstrained particle velocities, $\mathbf{\tilde{U}}^{k}_n = (\mathcal{J}^k[\mathbf{\tilde{u}}^k])_n$, through
\begin{eqnarray}
v^k &=& \frac{\Delta t}{2} \sum_{n} \bm{\nabla}_{\mathbf{Y}^n} \cdot\mathbf{\tilde{U}}^{k}_n \label{eq:vkexp}
\end{eqnarray}
For fluctuating FCM, we compute $v^k$ directly by evaluating
\begin{eqnarray}
v^k &=& \frac{\Delta t}{2\sigma_{\Delta}^2} \sum_{n} \int (\mathbf{x} - \mathbf{Y}^n)\cdot\mathbf{\tilde{u}}^k  \Delta_n(\mathbf{x}) d^3\mathbf{x}, \label{eq:vkdir}
\end{eqnarray}
but it can also be computed using finite-differencing or even RFD.  

We show explicitly in Appendix \ref{sec:AppB} that the expansions of the DC for the first and second moments of the increment $\Delta \mathcal{Y}^k = \mathcal{Y}^{k+1} - \mathcal{Y}^k$ are 
\begin{eqnarray}
\langle \Delta \mathcal{Y}^k \rangle &=& \Delta t\mathcal{M}^{\mathcal{VF};k}_{FCM-S} \mathcal{F}^k + \mathcal{M}^{\mathcal{VT};k}_{FCM-S} \mathcal{T}^k   \nonumber\\
&& + \Delta t k_BT\nabla_{\mathcal{Y}}\cdot\mathcal{M}^{\mathcal{VF};k}_{FCM-S} + O(\Delta t^2), \nonumber\\
\end{eqnarray}
and 
\begin{eqnarray}
\langle \Delta \mathcal{Y}^k (\Delta \mathcal{Y}^{k})^T \rangle &=& 2k_BT \Delta t \mathcal{M}^{\mathcal{VF};k}_{FCM-S} \nonumber\\
&& + O(\Delta t^2),
\end{eqnarray}
respectively.

The DC, therefore, first moves the particles a half timestep using the particle velocities obtained from the unconstrained fluctuating flow field.  Then, using the updated positions, but the same realization of the fluctuating stress, the particle forces and torques are projected onto the fluid, and the local rate-of-strain constraint is imposed.  The particle positions are then updated using the resulting particle velocities.  Thus, for the DC, the stresslets need to be computed only once per timestep, and the only additional cost is the single Stokes solve to determine the unconstrained fluctuating flow field.  A second-order approximation of the deterministic terms may be achieved by also including particle forcing (torques, forces, and stresslets) in Eq. (\ref{eq:DC1}), but this would come at the cost of an additional stresslet iteration per timestep.  

We note that similar ideas were employed by Delong \emph{et al.}\cite{Delong2014} to construct a midpoint scheme using RFD.  In their construction, however, they relied on a specific decomposition of the mobility matrix that is not applicable in the case where local rate-of-strain constraints are enforced.  Nevertheless, our analysis reveals that this similar technique extends to the case where the stresslets are present.  

If we have periodic boundary conditions, or if $\mathbf{u}\cdot \mathbf{\hat{n}} = 0$ pointwise on the boundary, where $\mathbf{\hat{n}}$ is the unit normal to the boundary, an analysis of the scheme using continuous spatial operators reveals that the correct first and second moments can be achieved with $v^k = 0$.  This condition holds for no-slip or no shear on the fluid boundaries.  As shown in Appendix  \ref{sec:AppB}, this is derived from the condition that $\bm{\nabla}_{\mathbf{Y}^n}\cdot \mathbf{\tilde{U}}_n = 0$ for all particles.  In this case, Eq. \eqref{eq:Move_with_factor} simplifies to become
\begin{eqnarray}
\mathcal{Y}^{k+1} &=&  \mathcal{Y}^{k} + \Delta t \mathcal{J}^{k+1/2}[\mathbf{u}^{k+1/2}].
\end{eqnarray}

It should be emphasized that these arguments based on continuous spatial operators may not hold in practice when the Stokes equations and the spreading and volume averaging operators are discretized.  We may not, therefore, necessarily set $v^k=0$ without incurring an additional small error due to spatial discretization.  Delong \emph{et al.}\cite{Delong2014} have shown for the fluctuating IBM that even in the case of periodic boundary conditions, spatial discretization does give rise to this small error, but it can be eliminated by including an RFD of the spreading operator as part of the time integration.  For the DC, this would correspond to including the $v^k$ term.  In Appendix \ref{sec:AppB} and Section \ref{sec:Periodic}, we explore these errors for fluctuating FCM.  We find that for periodic boundary conditions, the spatial discretization does not introduce any additional errors for fluctuating FCM and we may use the scheme with $v^k=0$.  We do find, however, that for no-shear conditions on the boundaries, the numerical integration of the Gaussian functions introduces this small, but non-negligible, error.  Thus, we retain the $v^k$ terms in our simulations where we apply no-shear boundary conditions.

\section{Numerical studies}

To demonstrate the performance of the DC with fluctuating FCM, we perform simulations of particulate suspensions under both dynamic and equilibrium conditions.  These simulations confirm the results of our theoretical analysis of the DC, and show, in practice, that it is able to produce the correct microstructure, dynamics and final equilibrium states for distributions of particles.  Our results also show that with the DC, fluctuating FCM can be used for large-scale simulation of Brownian suspensions even when higher order corrections to the hydrodynamic interactions are included.
\subsection{Spatial discretization}
\label{sec:SpatDisc}
In our simulations, we mainly consider periodic boundary conditions and use a Fourier spectral method with fast Fourier transforms to solve the Stokes equations, taking advantage of the highly scalable MPI library P3DFFT \cite{Pekurovsky2012}. We give here a summary of the main steps of the discretization scheme as a detailed description is provided in our previous work\cite{Keaveny2014}.

For the case where each side of the domain has length $L$, we use $M$ grid points in each direction.  The total number of grid points is then $N_g = M^3$ and the grid spacing is given by $\Delta x = L/M$.  The position of a grid point in a given direction is then $x_\alpha = \alpha \Delta x$ for $\alpha = 0, \dots, M-1$.  At each grid point, the entries of the fluctuating stress are independent Gaussian random variables.  Since the fluctuating stress $\mathbf{P}$ is symmetric, at each grid point we generate six random numbers based on a Gaussian distribution with zero mean and unit variance.  We then multiply the off-diagonal entries by $\sqrt{2k_BT/(\Delta x^3)}$ and the diagonal ones by $2\sqrt{k_BT/(\Delta x^3)}$.

To the fluctuating stress, we add the FCM force distribution, 
\begin{eqnarray}
\mathbf{f}_{FCM}(\mathbf{x}) &=& \mathcal{J}^\dagger[\mathcal{F}](\mathbf{x})\nonumber\\
&& + \mathcal{N}^\dagger[\mathcal{T}](\mathbf{x}) + \mathcal{K}^\dagger[\mathcal{S}](\mathbf{x})
\end{eqnarray}
evaluated at the grid points, $\mathbf{x} = [\begin{array}{ccc}x_{\alpha} & x_{\beta} & x_{\gamma}\end{array}]^{T}$, taking also $\Delta_n(\mathbf{x}) = 0$ and $\Theta_n(\mathbf{x}) = 0$ for $|\mathbf{x} - \mathbf{Y}^n| > 3a$.  To ensure sufficient resolution of $\mathbf{f}_{FCM}$, we set $\sigma_{\Theta}/\Delta x = 1.5$ and $\sigma_{\Delta}/\Delta x = 1.86$.  After taking the discrete Fourier transform (DFT) of the total force distribution, we compute the fluid velocity in Fourier space 
\begin{eqnarray}
\hat{\mathbf{u}}(k_{\alpha}, k_{\beta}, k_{\gamma})&=& \frac{1}{\eta |\mathbf{k}|^2} \left(\mathbf{I} - \frac{\mathbf{k}\mathbf{k}}{|\mathbf{k}|^2} \right)\nonumber\\
& &\times\left(i\mathbf{k} \cdot (\Delta t)^{-1/2}\hat{\mathbf{P}}(k_{\alpha}, k_{\beta}, k_{\gamma})\right. \nonumber\\
& & \left. + \hat{\mathbf{f}}_{FCM}(k_{\alpha}, k_{\beta}, k_{\gamma})\right), \nonumber \\
\label{eq:DFTu}
\end{eqnarray}
where $\mathbf{k} = [\begin{array}{ccc}k_{\alpha} & k_{\beta} & k_{\gamma}\end{array}]^{T}$.  We set $\hat{\mathbf{u}}(\mathbf{k}) = 0$ if $|\mathbf{k}| = 0$ or if $\mathbf{k}\cdot \mathbf{\hat{e}}^{(i)} = M\pi/L$ for $i = 1, 2, \mbox{or } 3$.  After taking the inverse DFT to obtain the fluid velocity at the grid points,  the velocity, angular velocity, and local rate-of-strain for each particle is computed by applying the spectrally accurate trapezoidal rule to equations Eqs. (\ref{eq:FCM_Ju} - \ref{eq:FCM_Ku}), where we again set $\Delta_n(\mathbf{x}) = 0$ and $ \Theta_n(\mathbf{x}) = 0$ for $|\mathbf{x} - \mathbf{Y}^n| > 3a$.  

For the case where the stresslets are ignored, we have $\mathcal{S} = 0$ and the Stokes equations need only to be solved once per time-step to obtain the particle velocities.  If the stresslets are included, however, they must be solved for, and to do this, we employ the iterative conjugate gradient approach described in Yeo and Maxey\cite{Yeo2010}.  

\subsection{Brownian particles in a periodic domain}\label{sec:Periodic}
In this section, we perform two simulations to test the DC with fluctuating FCM in periodic domains.  We first simulate the free diffusion of non-interacting particles to quantify the effect of spatial discretization of the operators.  This same test was performed by Delong \emph{et al.}\cite{Delong2014} for IBM.  We then simulate a hard-sphere suspension to demonstrate that the DC successfully accounts for the Brownian drift term that arises due to the stresslets.  We show that with the DC, the correct radial distribution function is reproduced, while neglecting the Brownian drift term by using the EM scheme does not provide the correct suspension microstructure.  We note here that as we have periodic boundary conditions, we take $v^k = 0$ in our DC simulations.

\subsubsection{Free diffusion}
For IBM and FCM without stresslets, the translational invariance associated periodic boundary conditions indicates that $\nabla_{\mathcal{Y}}\cdot\mathcal{M}^{\mathcal{V}\mathcal{F}} = 0$.  Delong \emph{et al.}\cite{Delong2014}, however, showed that while this holds true for the spatially continuous formulations of these methods, the spatially discrete operators used in simulations may not preserve translational invariance, and hence the divergence of the mobility matrix may be non-zero.  Indeed for the IBM\cite{Delong2014} , spatial discretization leads to a non-zero divergence and when EM is used to integrate the equation of motion, variations in the particle distribution are found at the level of the grid spacing even though the distribution should be uniform.

\begin{figure*}
\centering
\subfloat[]{\label{fig:PDF_1cell_EM} \includegraphics[width=7.25cm]{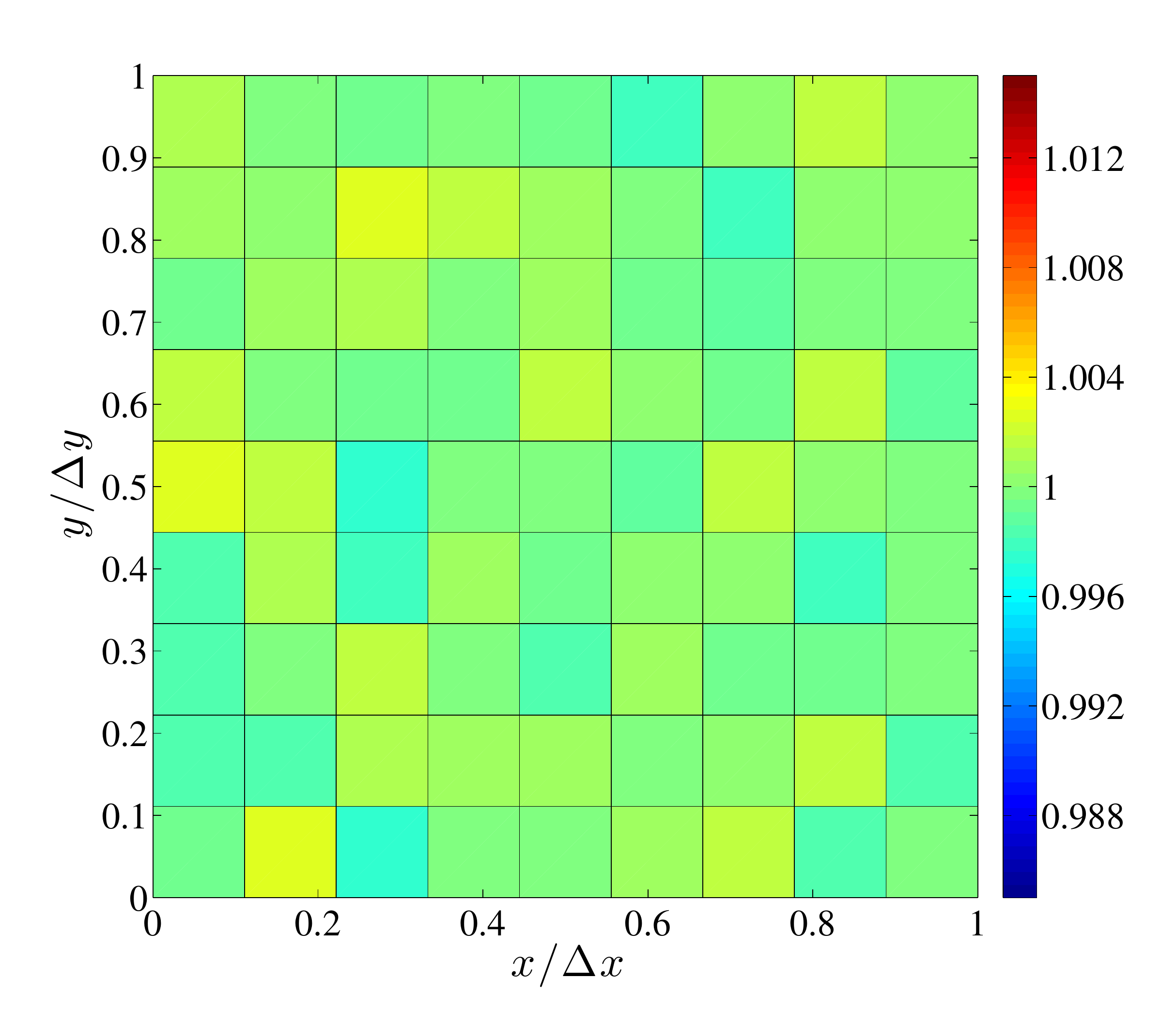}}
\subfloat[]{\label{fig:PDF_1cell_DC} \includegraphics[width=7.25cm]{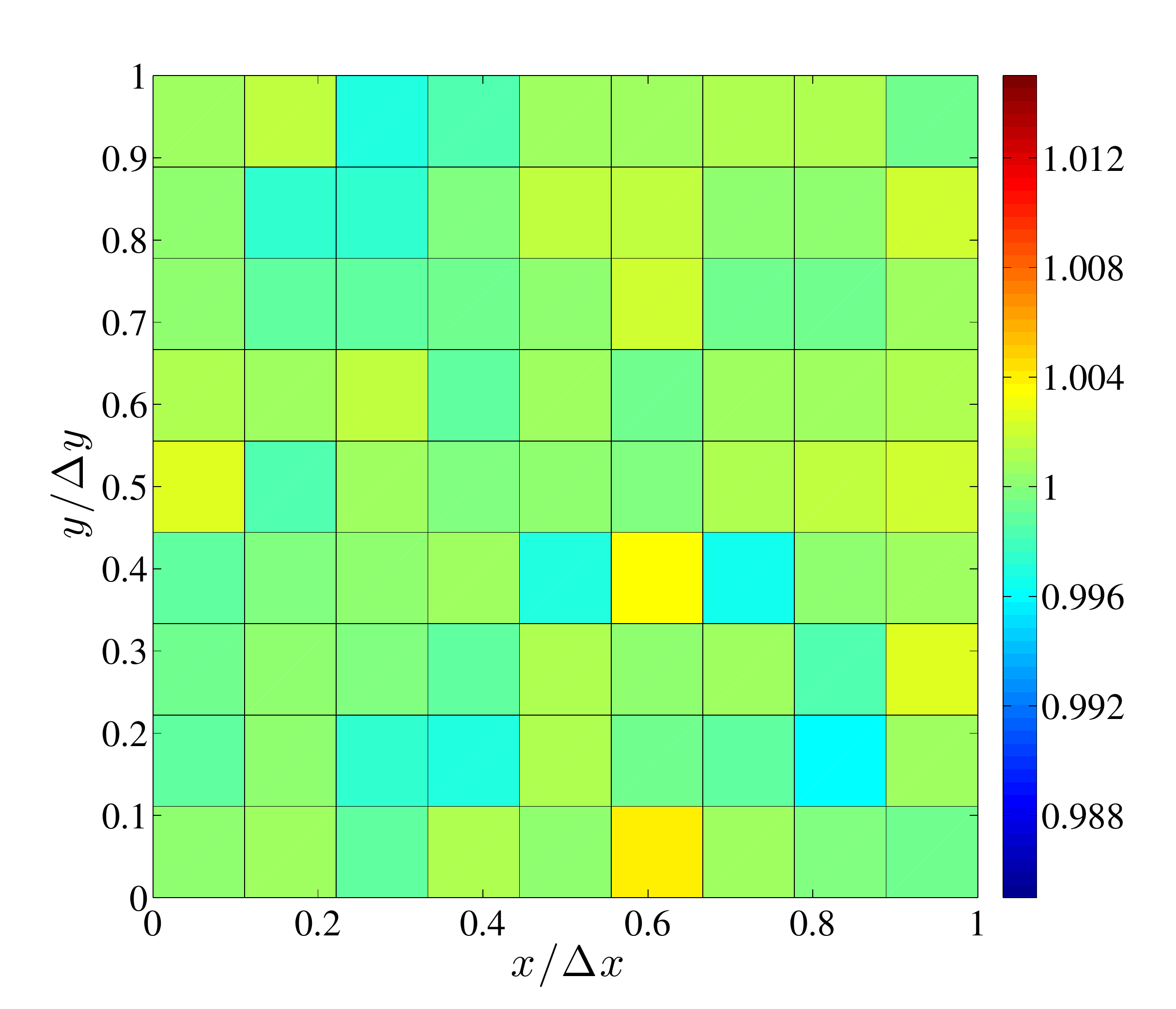}}
\caption{Normalized depth-averaged equilibrium distribution in a grid cell for simulations using  
(a) EM, 
(b) the DC}
\label{fig:Distrib1cell}
\end{figure*}

To quantify this effect for FCM, we let $522$ non-interacting particles diffuse in a 3D periodic system for $t/t_{D_a} = 170$ Brownian times where $t_{D_a} = a^2/D_0$ is the characteristic time for a particle to diffuse one radius and $D_0$ is the Stokes-Einstein diffusion coefficient. The stresslets are not included in these simulations, making this test case identical to that performed by Delong \emph{et al.}\cite{Delong2014}.  We perform $20$ independent simulations, each run using the very small time-step $\Delta t = 4.1\textrm{e-}4(\Delta x^2/D_0)$ where $\Delta x$ is the grid spacing.  This small value of $\Delta t$ is chosen to ensure that any systematic variations in the distribution are due to the spatial discretization of the operators.  Fig. \ref{fig:Distrib1cell} shows the depth-averaged normalized particle distribution over a single grid cell for FCM simulations using both EM and the DC.  To allow for a quantitative comparison with Delong \emph{et al.}\cite{Delong2014}, we have set the limits of the color axis exactly as it is in Fig. 3 of their work.  Apart from infinitesimal statistical fluctuations, there is no clear bias in the distributions for either integration scheme.  This means that our spatial discretization preserves translational invariance within a tight tolerance.  This property is likely due to the smoothness of the Gaussian kernels used with FCM, as well as the spectral accuracy of the Fourier method used to solve the Stokes equations. 

\subsubsection{Hard-sphere suspension simulations}
\begin{figure}
\centering
\includegraphics[width=7.5cm]{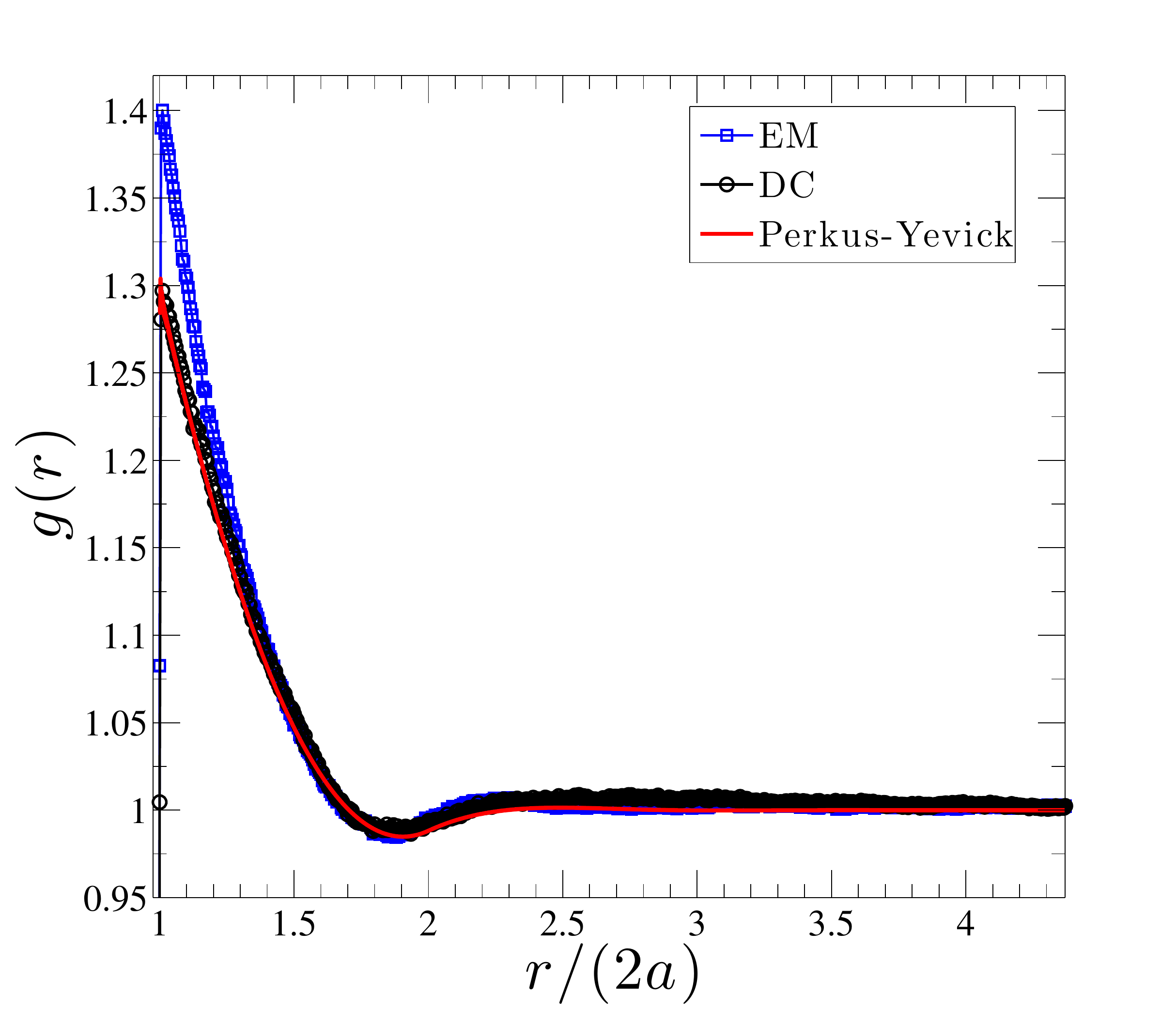}
\caption{Radial distribution function for a semi-dilute suspension ($\phi=0.1$) of hard-spheres in a periodic box. The blue line with square markers corresponds to simulations with EM time integration, the black line with circles corresponds to the DC, and the red line is given by the Percus-Yevick approximation\cite{Perkus1958}.}
\label{fig:RDF_HS}
\end{figure}
When stresslet corrections to the particle hydrodynamics interactions are included, the divergence of the mobility matrix is no longer zero for periodic systems.  We show here that time integration with the DC recovers the drift term due to the stresslet corrections in FCM and show that neglecting the drift term by using EM leads to an erroneous characterization of the suspension microstructure. 

To do this, we simulate a suspension of $174$ particles in a periodic domain that interact via a hard-sphere potential.  The hard-sphere potential is included by rejecting timesteps where the particles are observed to overlap.  The volume fraction is taken to be $\phi = 0.1$.  We perform these simulations using both EM and the DC to advance the particle positions in time.  For each time integration scheme, we perform 20 independent simulations, taking each simulation to a final time of $t/t_{D_a} = 125$.  Fig. \ref{fig:RDF_HS} shows the radial distribution function $g(r)$ obtained by averaging over the EM simulations and the DC simulations, as well the values given by the Percus-Yevick approximation\cite{Perkus1958}.  The DC simulations are in very good agreement with the Percus-Yevick approximation for all values of $r$, showing that the drift term is accounted for in these simulations.  The EM results, however, deviate from the Percus-Yevick approximation for $r < 2.6a$ and clearly overestimate $g(r)$ at contact.  The separations $r < 2.6a$ correspond to the range over which the stresslets strongly modify the mobility matrix.  These results clearly indicate that even for periodic boundary conditions, stresslet corrections to the hydrodynamic mobility yield a non-negligible drift term, which is correctly accounted for by the DC.

\subsection{Brownian particles between two slip surfaces}
We examine the Brownian dynamics of a spherical particle between two slip surfaces at $z = 0$ and $z = L/2$.  On these surfaces, the boundary conditions for the flow are given by $\mathbf{u}\cdot \mathbf{\hat z} = 0$ and $(\mathbf{I} - \mathbf{\hat z}\mathbf{\hat z}^T)\bm{\nabla}\mathbf{u} = \mathbf{0}$.    The divergence of the mobility matrix is non-zero for these conditions and the drift term must be properly accounted for in simulation.  Even though we impose periodicity or have $\mathbf{u}\cdot \mathbf{\hat{n}} = 0$ on the boundaries, it is important to include $v^k$ in the time integration to eliminate errors that arise when the Gaussian functions overlap the boundaries (see Appendix \ref{sec:AppB}).

\subsubsection{Image system and fluctuating stress for slip surfaces}
Rather than enforcing these boundary conditions directly through the numerical solver, we can create the slip surfaces by introducing the appropriate image system for both the FCM particle force distribution, $\mathbf{f}^{FCM}$, and the fluctuating stress, $\mathbf{P}$.  The flow due to the force distribution and its image can then be found using the Fourier spectral method described in Section \ref{sec:SpatDisc}.

In our simulations, the fluid domain is given by $\Omega = [0, L]^2\times[0,L_z]$.  We impose the slip conditions at $z = 0$ and $z = L_z$ and periodic boundary conditions on the other boundaries such that $\mathbf{u}(0,y,z) = \mathbf{u}(L,y,z)$ and $\mathbf{u}(x,0,z) = \mathbf{u}(x,L,z)$.  The force distribution due to particle forces is given by Eq. \eqref{eq:force_system}.  To enforce the slip conditions at $z = 0$ and $z = L_z$, we first extend the fluid domain to $\Omega^\star = [0, L]^2\times[0,2L_z]$ and utilize the modified FCM force distribution and its image,
\begin{widetext}
\begin{equation}
\mathbf{f}^{FCM}(\mathbf{x})=\left\{ \begin{array}{ll}
\mathcal{J}^\dagger[\mathcal{F}](\mathbf{x}) + \mathcal{Q}^\dagger[\mathcal{T}](\mathbf{x}) + \mathcal{K}^\dagger[\mathcal{S}](\mathbf{x}) & ,\,\mathbf{x}\in\ensuremath{[0,L]^{2}\times[0,L_{z}]}\\
\mathbf{0} & ,\,\mathbf{x}\in\ensuremath{[0,L]^{2}\times]L_{z},2L_{z}[}
\end{array}\right.
\label{eq:force_system}
\end{equation}
\begin{equation}
\mathbf{f}^{FCM,im}(\mathbf{x})=\left\{ \begin{array}{ll}
\mathbf{0} & ,\,\mathbf{x}\in\ensuremath{[0,L]^{2}\times]0,L_{z}[}\\
\left(\bm{I}-2\bm{\hat{z}}\bm{\hat{z}}^T\right)\mathbf{f}^{FCM}(\mathbf{X}) & ,\,\mathbf{x}\in\ensuremath{[0,L]^{2}\times[L_{z},2L_{z}]}
\end{array}\right. .
\label{eq:force_im_system}
\end{equation}
\end{widetext}
where $\mathbf{X} = \mathbf{x} - 2(\mathbf{x}\cdot\mathbf{\hat{z}} - L_z)\mathbf{\hat{z}}$. The total forcing term that will now appear in the right hand side of Eq. \eqref{eq:DFTu} is given by the summation $\mathbf{f}^{FCM}+ \mathbf{f}^{FCM,im}$.  The flow is then determined over the domain $\Omega^\star$ with the periodic boundary conditions in all directions.  The combined effects of the image system and periodicity in the $z$-direction yields the desired slip conditions at $z = 0$ and $z = L_z$.  Once the fluid flow is determined, the particle velocities, angular velocities, and local rates-of-strain are computed using the resulting flow restricted to the domain $\Omega$.

In introducing the slip surfaces, one must make a choice as how to modify the Gaussian functions when the particles get close to the boundaries.  In this work, we simply truncate any part of the Gaussian functions that extends beyond the domain $\Omega$.  This is already reflected in our definition of the force distribution, Eq. \eqref{eq:force_system} and its image, Eq. \eqref{eq:force_im_system}.  We use the same truncated Gaussians to compute the particle velocities, angular velocities, and local rates-of-strain, thereby preserving the adjoint properties of the operators.  For particles in contact with the boundary, the truncated volume in Eq. \eqref{eq:FCM_Ju} is approximately $3.8\%$ of the total and for Eq. \eqref{eq:FCM_Ku} it is only $1.4 \%$.  We note, that more sophisticated ways to modify the Gaussian envelopes have been explored and are implemented in FCM elsewhere \cite{Yeo2010jfm}.

Along with the FCM particle force distributions, the fluctuating stress must also have the appropriate symmetries to satisfy the fluctuation-dissipation theorem when the slip boundaries are present.  In Appendix \ref{sec:AppA}, we show explicitly that this is achieved by having
\begin{eqnarray}
\langle P_{ij}(\mathbf{x})\rangle &=& 0 \label{eq:flucstressslip1} \\
\langle P_{ij}(\mathbf{x})P_{kl}(\mathbf{y})\rangle &=& 2k_BT \Delta_{ijkl} \delta(\mathbf{x} - \mathbf{y}) \nonumber\\
& &+ 2k_BT \Gamma_{ijkl} \delta(\mathbf{x} - \mathbf{Y})
\label{eq:flucstressslip2}
\end{eqnarray}
for $\mathbf{x},\mathbf{y}\in \Omega^\star$ where
\begin{eqnarray}
\Delta_{ijkl}&=&\delta_{ik}\delta_{jl}+\delta_{il}\delta_{jk} \\
\Gamma_{ijkl}&=&\gamma_{ik}\gamma_{jl}+\gamma_{il}\gamma_{jk},
\end{eqnarray}
with $\gamma_{jk} = \delta_{jk} - 2\delta_{3j}\delta_{3k}$, 
and
\begin{equation}
\mathbf{Y} = \mathbf{y} - 2(\mathbf{y}\cdot\bm{\hat{z}}-L_z)\bm{\hat{z}}.
\end{equation}

The symmetrized fluctuating stress Eq. \eqref{eq:flucstressslip1}-\eqref{eq:flucstressslip2} is discretized at each grid point with independent Gaussian random variables whose statistics are 
\begin{equation}
\langle P_{ij}(x_\alpha, x_\beta, x_\gamma)\rangle   =  0  
\end{equation}
\begin{equation}
\def\arraystretch{1.6}
\begin{array}{c}
\langle P_{ij}(x_{\alpha},x_{\beta},x_{\gamma})P_{kl}(x_{\alpha},x_{\beta},x_{\kappa})\rangle  = \\ 
\frac{2k_BT}{(\Delta x)^3}\Delta_{ijkl}\delta_{\gamma \kappa} 
+ \frac{2k_BT}{(\Delta x)^3}\Gamma_{ijkl}\delta_{\gamma \kappa^{im}} 
\end{array}
\end{equation}
where the index $\kappa^{im} = \mod(M -\kappa,M)$. We remind the reader that the indices $\alpha, \beta, \gamma, \kappa$ go from $0$ to $M-1$, where $M$ is the number of discretization points in a given direction.

\subsubsection{Single particle mobility}
\begin{figure}
\includegraphics[width=7.5cm]{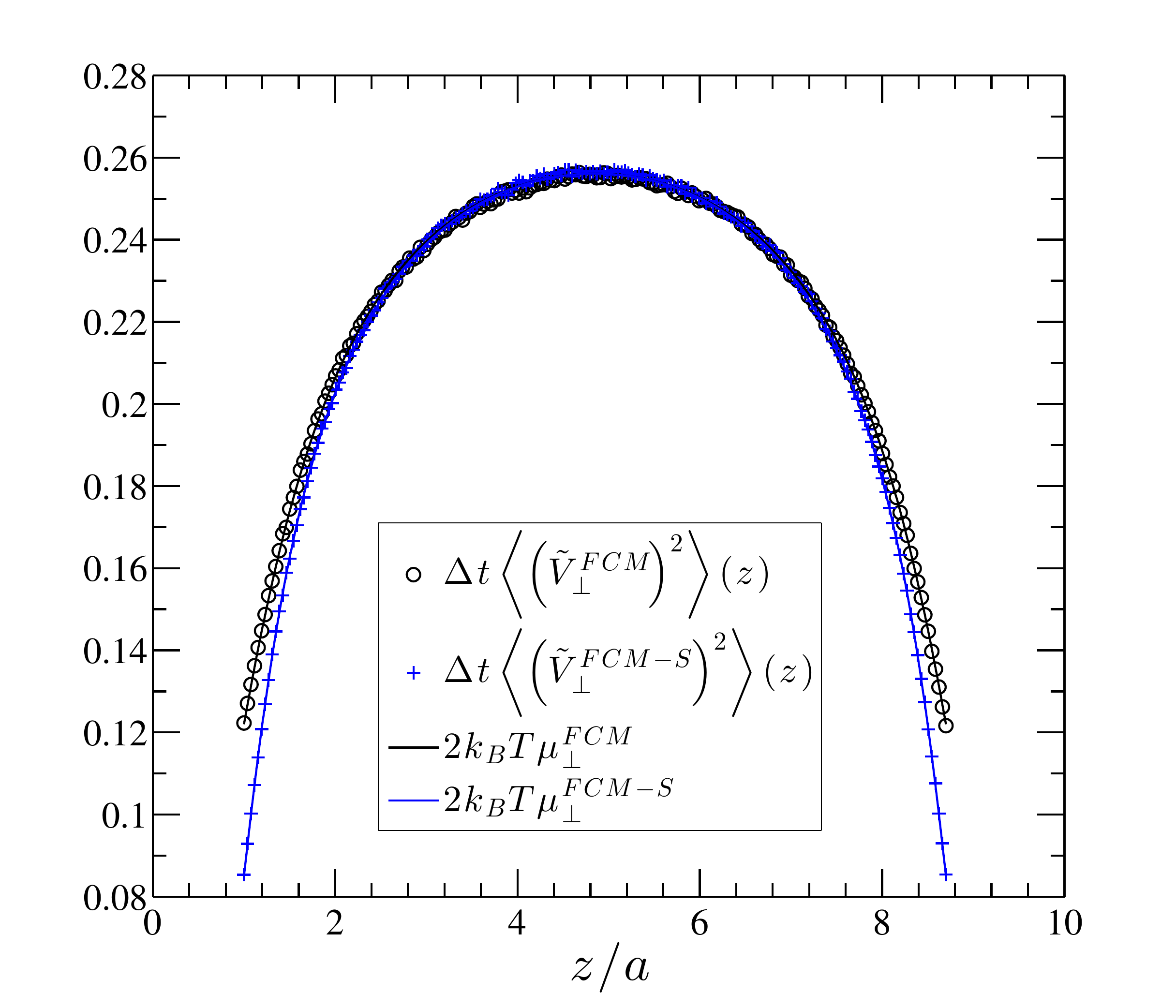}
\caption{A comparison between the wall-normal mobility coefficient and the autocorrelation of the wall-normal particle velocity.
}
\label{fig:Autocorrel_velocity_perp}
\end{figure}

With the slip boundaries present, the mobility matrix for a single particle has the form
\begin{equation}
\mathbf{M}^{VF} = \mu_{\perp}(z) \mathbf{\hat{z}}\mathbf{\hat{z}}^T + \mu_{\parallel}(z)(\mathbf{I} -  \mathbf{\hat{z}}\mathbf{\hat{z}}^T) \label{eq:MVFslip}
\end{equation}
where the mobility coefficients $\mu_{\parallel}(z)$ and $\mu_{\perp}(z)$ depend on the distance from the slip surfaces.  We determine the coefficient $\mu_{\perp}(z)$ for FCM by applying a unit force $\mathbf{F} = \bm{\hat{z}}$ on an isolated sphere and measuring the resulting velocity $V_z(z)$ at $400$ equi-spaced values of $z$ between $z = a$ and $z = L_z-a$. We perform these simulation both with and without the particle stresslets to obtain $\mu^{FCM-S}_{\bot}(z)$ and $\mu^{FCM}_{\bot}(z)$, respectively.  The values of $2k_bT\mu^{FCM-S}_{\bot}(z)$ and $2k_bT\mu^{FCM}_{\bot}(z)$ are provided in Fig. \ref{fig:Autocorrel_velocity_perp}.  We see that both with and without the stresslets, the value of this mobility coefficient decreases as the particle approaches the slip surface.  The addition of the stresslet results in the mobility coefficient depending more strongly on $z$ and further reduces the mobility coefficient by approximately $30\%$ near the boundaries.  We have performed similar computations for $\mu_{\parallel}(z)$ by taking $\mathbf{F} = \bm{\hat{x}}$, but have not included this data for brevity.  We also note that given the symmetries associated with a single slip surface, the exact Stokes flow for a rigid sphere near a slip surface is equivalent to an appropriate two-sphere problem.  A comparison of FCM with these two-sphere problems has been addressed by Lomholt and Maxey\cite{Lomholt2003}, showing that FCM provides accurate results for a wide range of separations.

To ensure that the fluctuation-dissipation theorem for the particles is satisfied when the statistics for the fluctuating stress are given by Eq. \eqref{eq:flucstressslip1}-\eqref{eq:flucstressslip2}, we compute the component of the random particle velocity $\tilde{V}_{\bot}(z)$ normal to the slip surface.  We then can check that its correlations satisfy
\begin{equation}
  \left\langle\tilde{V}_{\bot}^2\right\rangle (z) = \dfrac{2k_BT\mu_{\bot}(z)}{\Delta t}.
  \label{eq:flucdisperp}
\end{equation}
To efficiently compile a large number of samples to accurately determine $\left\langle\tilde{V}_{\bot}^2\right\rangle (z)$, we divide the domain in the $z$-direction into $200$ equispaced parallel planes.   Over each plane, we distribute $50$ particles that do not interact with one another.  For the stresslet case, particular care is taken to ensure that any interactions are removed.  This is done by utilizing directly the stresslet-corrected mobility coefficients measured from simulations of an isolated particle. The procedure is detailed in Appendix \ref{sec:AppC}.

For $10000$ realizations of the fluctuating stress, we compute $\tilde{V}_{\bot}(z)$ for each of the particles.  For each fixed value of $z$, we compute $\left\langle\tilde{V}_{\bot}^2\right\rangle (z)$ by averaging over the realizations and the $50$ particles at that particular value of $z$.  Fig. \ref{fig:Autocorrel_velocity_perp} shows the correlations of the particle normal velocity, $\Delta t\left\langle\tilde{V}_{\bot}^2\right\rangle (z)$, and the normal mobility coefficient in the channel, $2k_BT\mu_{\bot}(z)$.  We see that for both cases, with and without the stresslets, the fluctuation-dissipation relation for the particles, Eq. (\ref{eq:flucdisperp}), is satisfied.  Though not shown, we have also performed the same check for the parallel random velocities and a similar strong agreement was found.

\subsubsection{Equilibrium distribution between two slip surfaces}
\begin{figure}
\includegraphics[width=7.5cm]{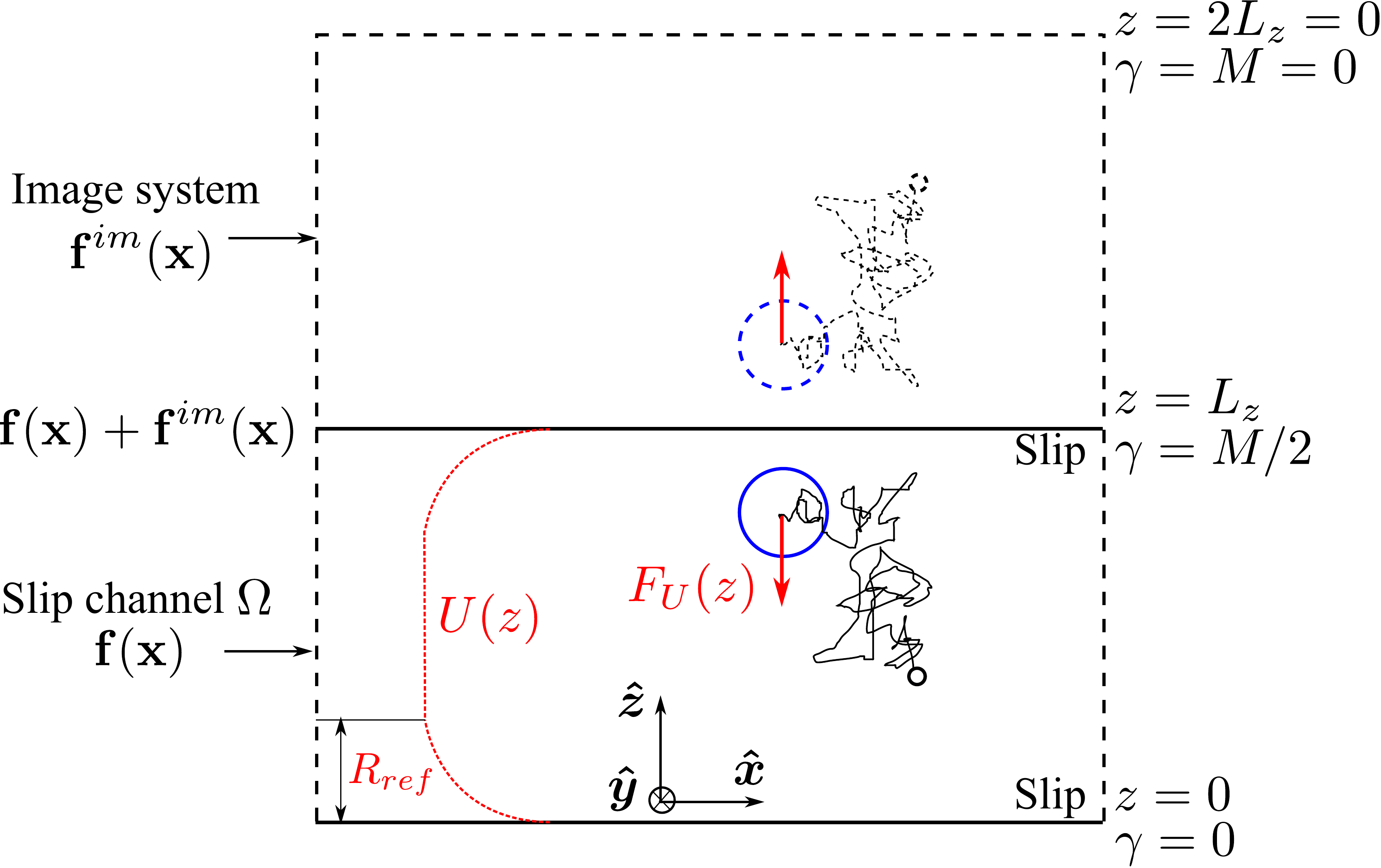}
\caption{Sketch showing a particle (and its image) in the slip channel. The black dashed line represents the periodic boundaries conditions. The index $\gamma$ runs over the $z$-coordinate of the grid. The potential $U(z)$ provides the interaction between the particle with the slip surfaces and is given by Eq. \eqref{eq:SlipChanPot}. The force on the particle is $F_U(z) = -\partial U/\partial z$.  
}
\label{fig:Sketch_channel}
\end{figure}
To demonstrate their RFD approach, Delong \emph{et al.} \cite{Delong2014} performed simulations of one and 100 spherical particles between two no-slip surfaces at $z = 0$ and $z = L_z$ and subject to the potential
\begin{equation}
U(z)=\left\{ \begin{array}{ll}
\dfrac{k}{2}\left(z-R_{ref}\right)^{2} &, z<R_{ref},\\
& \\
\dfrac{k}{2}\left(z-\left(L_{z}-R_{ref}\right)\right)^{2} &, z>L_{z}-R_{ref},\\
 & \\
0 & ,\mbox{ otherwise.}
\end{array}\right.
\label{eq:SlipChanPot}
\end{equation}
When the particle comes within a distance $R_{ref}$ from the wall, it experiences linear, repulsive force whose strength is governed by $k$.  If the time integration scheme correctly accounts for the Brownian drift term, the equilibrium distribution of the particles will be given by the Boltzmann distribution,
\begin{equation}
  P(z) = Z^{-1}\exp\left[-\dfrac{U(z)}{k_BT} \right]
  \label{eq:BoltzDis}
\end{equation}
where $Z = \displaystyle\int_0^{L_z}\exp\left[-\dfrac{U(z)}{k_BT} \right]dz$.  

\begin{figure*}
\centering
\subfloat[]{ \label{fig:PDF_H_no_stresslet} \includegraphics[width=7.5cm]{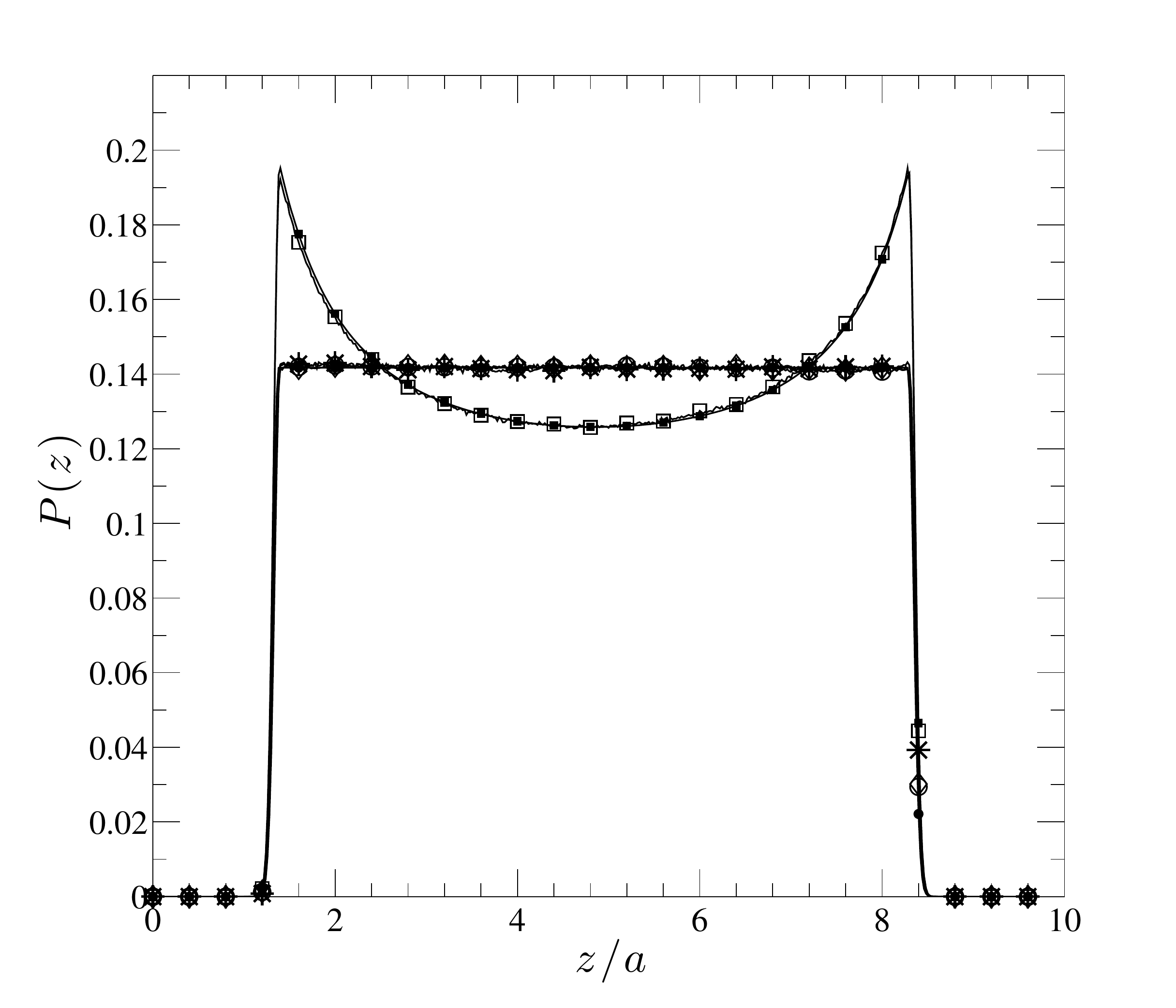}}
\subfloat[]{ \label{fig:PDF_H_stresslet} \includegraphics[width=7.5cm]{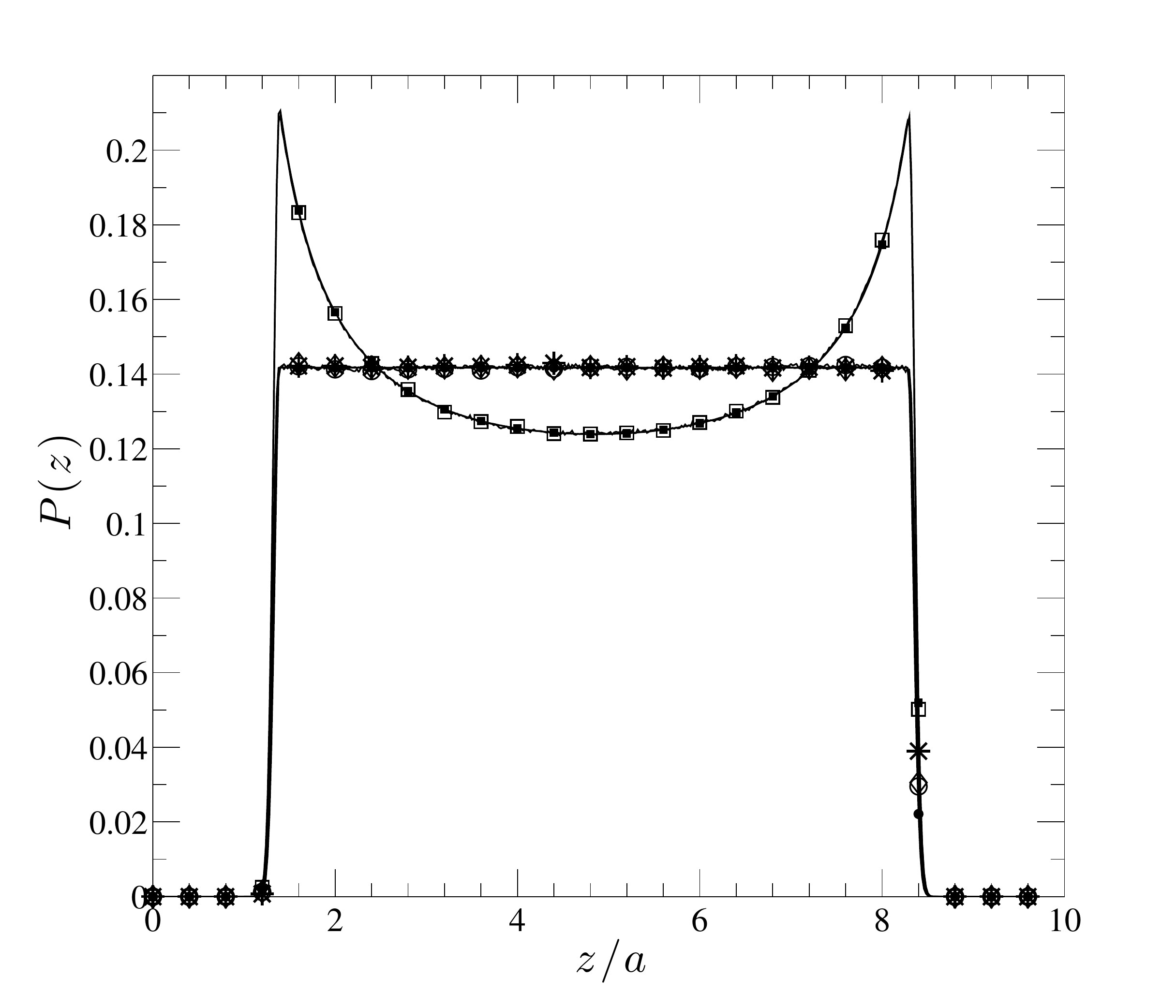}}
\caption{The equilibrium distribution for a Brownian particle subject to the potential $U(z)$.  The distribution from simulations is obtained by averaging $10$ simulations of $500$ non-interacting particles over the time interval $t/t_{D_a} = 6.526 - 391.60 $.
\symbol{\solid}{\bigcircle}{20}{-1.0}{black}{black} : DC without $v^k$, 
\symbol{\solid}{\asterix}{20}{-1.0}{black}{black} : DC with $v^k$, 
\symbol{\solid}{\losange}{20}{-0.3}{black}{black} : central RFD with $\delta = 10^{-6}\Delta x$, 
\symbol{\solid}{\blackcircle}{21}{-1.0}{black}{black} : Gibbs-Boltzmann distribution Eq. \eqref{eq:BoltzDis}, 
\symbol{\solid}{\ssquareb}{20}{0}{black}{black} : Euler-Maruyama Scheme, 
\symbol{\solid}{\blackssquare}{20}{-0.6}{black}{black} : Biased Gibbs-Boltzmann ditribution Eq. \eqref{eq:BiasDis}. 
(a) Without stresslets, 
(b) With stresslet corrections.
} 
\end{figure*}

We consider a similar problem here, now simulating the dynamics of a spherical particle between two slip surfaces, but still subject to the potential $U(z)$.  A sketch of this simulation is provided in Fig. \ref{fig:Sketch_channel}.  Even though we have slip boundary conditions on the walls, the distribution of particles should still be described by Eq. \eqref{eq:BoltzDis}.  We perform these simulations both with and without the stresslets and using both central RFD and the DC.  We also integrate the equations using the Euler-Maruyama scheme, Eq. \eqref{eq:EMFCM}, which should instead yield the biased distribution\cite{Delong2014},
\begin{equation}
  P_B(z) = Z_B^{-1}\exp\left[-\dfrac{U(z)+k_B T\ln(\mu_\bot(z))}{k_BT}\right]
  \label{eq:BiasDis}
\end{equation}
In our simulations, we have set $L_z = 9.7a$, $L_x = L_y = 2L_z$, $R_{ref} = 1.4a$ and $k=24 k_B T/(\Delta x)^2$.  For the RFD simulations we set $\delta = 10^{-6}\Delta x$.  To obtain the equilibrium distribution, each simulations is performed with $N_p = 500$ non-interacting particles and run to time $t/t_{D_a} = 391.60$, where $t_{D_a}= \dfrac{a^2}{k_BT} \bar{\mu}_{\bot}$ is the time required for the particle to diffuse one radius in the $z$-direction based on the $z$-averaged mobility coefficient
\begin{equation}
 \bar{\mu}_{\bot} = \dfrac{1}{L_z-2a}\int_{a}^{L_z-a}\mu_{\bot}(z)dz.
\end{equation}
To determine $P(z)$ from the simulations, a histogram of the particle positions in the $z$-direction is compiled during the time window $t/t_{D_a} = 6.526 - 391.60$.  The results from the simulations without the stresslets are shown in Fig. \ref{fig:PDF_H_no_stresslet}, while those with the stresslets are given in Fig. \ref{fig:PDF_H_stresslet}.  In both figures, we see that central RFD and the DC recover the correct Boltzmann distribution, Eq. (\ref{eq:BoltzDis}).  For completeness, we have also run the DC simulations with $v^k = 0$.  We see that since the potential prevents the particles from getting close to the boundaries, the $v^k = 0$ simulations still provide the correct distribution.  The Euler-Maruyama scheme, however, yields the biased distribution, Eq. (\ref{eq:BiasDis}), as the Brownian drift, $k_BT d\mu_{\perp}/dz$, is non-zero.  We see also that compared to the stresslet-free case, when the stresslets are included the biased distribution exhibits higher peaks close to the boundaries.  This is a direct result of the greater reduction in mobility near the boundaries when the stresslets are included, see Fig. \ref{fig:Autocorrel_velocity_perp}.  One can then expect that using the correct integration scheme becomes even more important in simulations where the singular lubrication interactions with the boundaries are resolved.

\subsubsection{Distribution dynamics}
Along with characterizing the equilibrium distribution, we perform simulations to confirm that fluctuating FCM with the DC also yields the correct distribution dynamics.  For non-interacting particles, the dynamics of the distribution is described by the Smoluchowski equation 
\begin{equation}
 \dfrac{\partial P}{\partial t} = -  \dfrac{\partial }{\partial z} \left[\mu_{\bot} F_{U} P  - k_B T  \mu_{\bot}(z)  \dfrac{\partial P}{\partial z} \right],
\label{eq:adv_diff}
\end{equation}
for the distribution $P(z,t)$ subject to no flux conditions
\begin{equation}
\left.\left[\mu_{\bot} F_{U} P  - k_B T  \mu_{\bot}  \dfrac{\partial P}{\partial z} \right]\right|_{z = 0, L_z} = 0
\label{eq:BC_adv_diff}
\end{equation}
at the slip surfaces ($z=0, z=L_z$).  The advective flux term is proportional to the force $F_{U} = -\partial U / \partial z$ associated with the potential $U$.  We solve Eq. (\ref{eq:adv_diff}) using a first-order finite volume solver where the advective fluxes are calculated using an upwinding scheme and the diffusive terms with a central scheme.   We have validated the solver against analytical solutions of the heat and transport equations.  The mobility coefficient $\mu_{\bot}(z)$ is determined by interpolating the values from our FCM simulations, see Fig. \ref{fig:Autocorrel_velocity_perp}, to the finite volume grid points.

We obtain the distribution dynamics from stresslet-corrected fluctuating FCM simulations by averaging the histogram time dynamics over $20$ independent simulations each with $500$ non-interacting particles.  For these simulations, the initial particle positions are generated by distributing their positions uniformly in the center of the domain in the region $z\in[3L_z/8;5L_z/8]$.  We compare these results with our numerical solution of Eq. \eqref{eq:adv_diff} using the distribution obtained from the fluctuating FCM simulations at $t = 0$ for the initial condition.

\begin{figure}
\centering
\includegraphics[width=7.5cm]{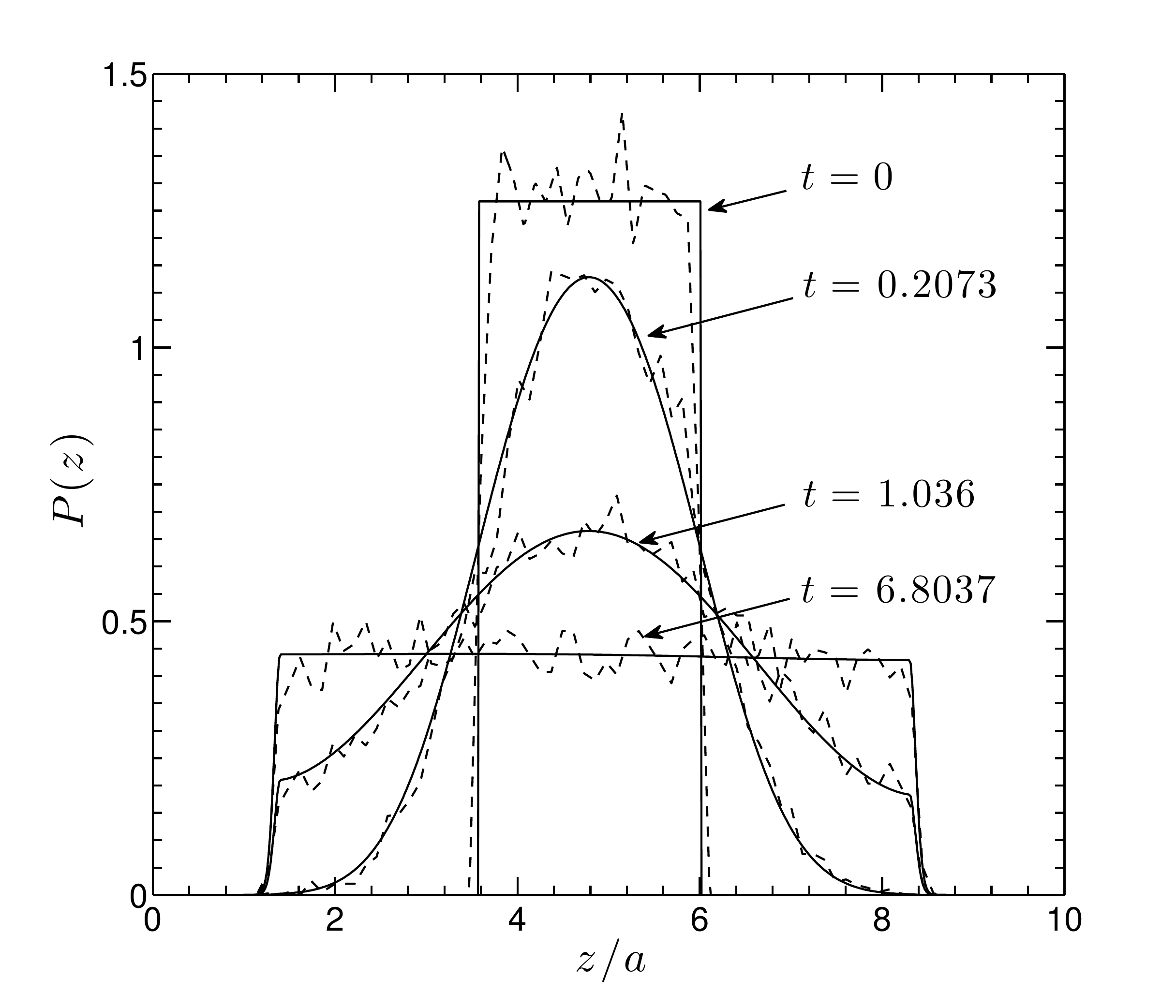}
\caption{The evolution of the particle distribution during the time interval $t/t_{D_a} = 0 - 6.8037$. The results from the particle simulations are achieved by averaging over $20$ simulations, each with $500$ non-interacting particles. The stresslet corrections are included in these simulations.
\symbol{\dashed}{}{11}{0}{black}{black}: fluctuating FCM with the DC, 
\symbol{\solid}{}{12}{0}{black}{black} : solution to the Smoluchowski equation \eqref{eq:adv_diff} - \eqref{eq:BC_adv_diff}. 
}
\label{fig:P_H_time_K_stresslet_20_real_500_part}
\end{figure}

Fig. \ref{fig:P_H_time_K_stresslet_20_real_500_part} shows the dynamics of the distribution obtained from the fluctuating FCM simulations with the DC time integration, as well as the finite volume solution of Eq. \eqref{eq:adv_diff}.  We see that the distributions given by both methods match as they evolve from the initial to the equilibrium state reached at time $t/t_{D_a} = 6.4$.  This confirms further that the DC recovers the dynamics described by the Smoluchowski equation, Eq. \eqref{eq:adv_diff}, as desired.
\subsection{Colloidal gelation and percolation}
\label{sec:interacting}
An important application of the methodology presented in this paper is the large-scale simulation of interacting colloidal particles. 
Depending on the interactions between these particles, colloidal suspensions can take on both solid- and fluid-like properties \cite{Segre2001, Anderson2002}.  When the particles interact via an attractive potential, the suspension can transition from a fluid-like state to that of a solid through the process of gelation \cite{Lu2008}.  This transition depends not only on the potential, but also on the volume fraction occupied by the particles. Recent numerical studies have shown that the dynamics of this aggregation process will also depend on the hydrodynamic interactions between the particles \cite{Yamamoto2008, Furukawa2010, Whitmer2011, Cao2012}.  In this section, we use fluctuating FCM with the DC to simulate of the aggregation and gelation of interacting particles, focusing particularly on the role of hydrodynamic interactions on the resulting structures.  We compare our results with previous studies \cite{Furukawa2010,Cao2012,Delong2014}, demonstrating that fluctuating FCM with the DC provides an effective and computationally efficient approach for studying Brownian suspensions.  

\subsubsection{Collapsing icosahedron}
\label{sec:Icosahedron}
A simple example that highlights the role of hydrodynamic interactions on colloidal aggregation is the collapse of a small cluster of Brownian spherical particles as originally studied by Furukawa and Tanaka\cite{Furukawa2010}.  Specifically, they considered $N_p = 13$ Brownian particles initially at the vertices of a regular icosahedron with edge length $8.08\Delta x$.  The particles interact via a modified Asakura-Oosawa potential 
\begin{equation}
 U_{AO}(r)=\begin{cases}
c\left(D_2^{2}-D_1^{2}\right)r &, r<D\\
c\left(D_2^{2}-1/3r^{2}\right)r &, D_1\leq r<D_2\\
0 &, r\geq D_2,
\end{cases}
\end{equation}
where $r$ is the center-center distance between two particles, as well as a repulsive potential, $U_{SC} = \epsilon_{SC} (2a/r)^{24}$.  The values of the parameters in these potentials are set to $D_1=2.245a$, $D_2 = 2.694a$, $c = 58.5/a^3$, and $\epsilon_{SC} = 10.0$. As the cluster evolved, they monitored the evolution of the radius of gyration
\begin{equation}
R_g(t) = \left[\frac{1}{N_p}\sum\limits_{n=1}^{N_p}\left(\mathbf{Y}^n-\mathbf{Y}^c\right)^2 \right]^{1/2}, 
\end{equation}
where $\mathbf{Y}^c$ is the position of the center of mass.  Their results revealed that the hydrodynamic interactions slow the time of cluster collapse and lead to a final state in which the particles rearrange themselves within the cluster.  Delong \emph{et al.} \cite{Delong2014} used this test case to validate the treatment of hydrodynamic interactions in fluctuating IBM, showing that they matched Brownian Dynamics simulations with hydrodynamics given by the Rotne-Prager-Yamakawa tensor.  

We perform similar simulations of cluster collapse to both show that fluctuating FCM recovers previous results, but also to quantify the effects of the stresslet included in fluctuating FCM.  The simulations are performed using periodic boundary conditions, applying directly the spatial discretization scheme described in Section \ref{sec:SpatDisc}.  In our simulations, we set the magnitude of the repulsive potential to $\epsilon_{SC} = 18.0$, a slightly higher value than the one used by Furukawa and Tanaka\cite{Furukawa2010}.  The values of the other parameters used in our simulations are provided in Table \ref{tab:param_collapse}.  We also note that the inclusion of the particle stresslets necessitates a numerical scheme that accounts for Brownian drift such as the DC or RFD.  If the stresslets were not included, the equations of motion could be integrated using the Euler-Maruyama scheme as the Brownian drift term is zero\cite{Keaveny2014} in an unbounded or periodic domain.  

\begin{table}
\begin{centering}
\begin{tabular}{ccccccccc}
\toprule 
$\Delta U/k_BT$ & $c$  & $\epsilon_{SC}$ & $a/\Delta x$ & $\eta$ & $\Delta t/t_{D_a}$ & $N_{p}$ & $N_g$\\
\colrule
2.39  & $58.5/a^3$ & 18.0 & 3.29 & 1 & $2\cdot10^{-4}$ & 13 & $32^{3}$\\
\botrule
\end{tabular}
\caption{Parameter values for the icosahedron collapse simulations.}
\label{tab:param_collapse}
\end{centering}
\end{table}

\begin{figure*}
\centering
\includegraphics[width=11cm]{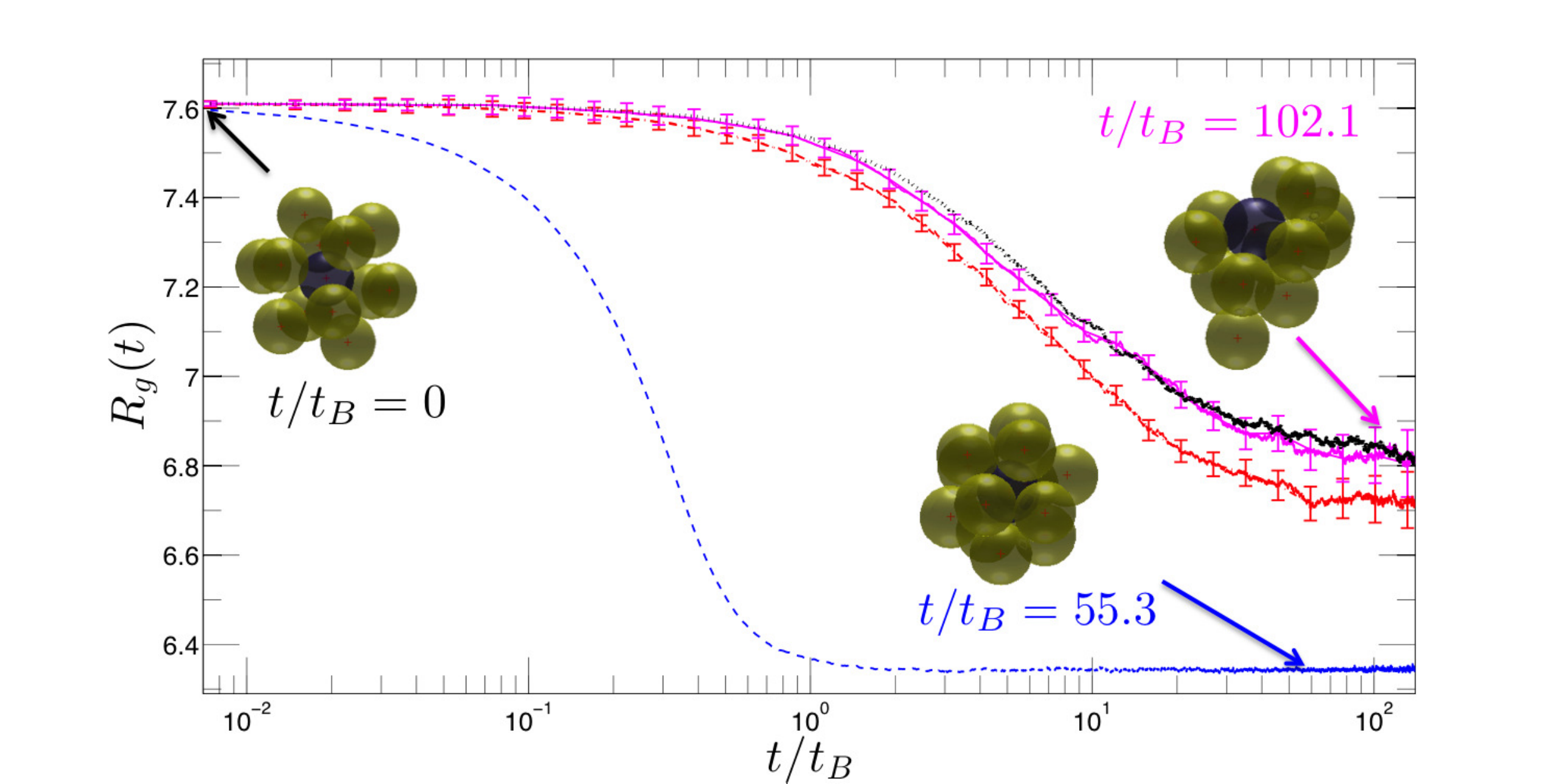}
\caption{The radius of gyration $R_g(t)$ (in units of $\Delta x$) during the time interval $t/t_{D_a} = 0 - 150$. 
The results are obtained by averaging over $150$ independent simulations.
\symbol{\dashed}{}{11}{0}{blue}{black} : no hydrodynamic interactions (HI); 
\symbol{\dashdot}{}{12}{0}{red}{black} : FCM without stresslets, Euler-Maruyama  scheme;
\symbol{\dottedlarge}{}{12}{0}{black}{black} : FCM-S, central RFD;
\symbol{\solid}{}{12}{0}{magenta}{black} :  FCM-S, DC.
Error bars are shown for FCM without stresslets and for FCM-S with the DC.  The inset on the left shows the initial configuration of the particles. The dark purple sphere is initially at the center of the icosahedron. The insets on the right show the spheres at one realization for the simulation with no HI and for FCM-S with the DC, respectively.}
\label{fig:Rg_time}
\end{figure*}

Fig. \ref{fig:Rg_time} shows the time evolution of the average radius of gyration given by fluctuating FCM with and without the particle stresslets.  We also show data from simulations using the same interparticle potentials, but with hydrodynamic interactions ignored completely.  For each case, the averages are obtained from $150$ independent simulations. In addition, when the stresslets are present, we integrate the equations of motion using both central RFD and the DC.  As in Furukawa and Tanaka\cite{Furukawa2010} and Delong \emph{et al.}\cite{Delong2014}, we see from Fig. \ref{fig:Rg_time} that the hydrodynamic interactions impact both the evolution and final value of $R_g$ by slowing down the collapse of the cluster.  Note that in the limit $t\rightarrow \infty$, all the simulations must go to the same final $R_g$ since the equilibrium state does not depend on the dynamics but only on the interaction potential.
Here, we see also that the inclusion of the stresslets leads to a higher value of $R_g$ during the collapse.  The value of $R_g$ with the particle stresslets given by central RFD and the DC are nearly identical.  We note that for each timestep, the central RFD scheme requires the stresslets to be solved for three times.  One central RFD timestep therefore requires approximately $30$ Stokes solves. 
 While providing similar results as the central RFD approach, the DC requires just one iterative solve per time step (about 10 Stokes solves).  This is confirmed from our simulations which show that the average simulation time with the DC is approximately three times less, $T_{sim}^{RFD}/T_{sim}^{DC} \approx 3$, than that for central RFD. Since the $\delta$ parameter is only limited by roundoff errors, forward RFD would provide similar results while saving one stresslet iteration per timestep.  For forward RFD, the simulation time would be double that of the DC.  Again, we note that compared to a completely deterministic simulation, the additional cost per time-step of including Brownian motion for FCM with the DC is just the distribution of the random stresses on the grid, an $O(N_g)$ computation, and one additional Stokes solve, which for our FFT-based solver incurs an $O(N_g \log N_g)$ cost.

\subsubsection{Aggregation and percolation in colloidal suspensions}
\label{sec:Percolation}
As a final test and demonstration of the DC, we perform a series of fluctuating FCM simulations to examine colloidal gelation of a suspension of Brownian particles.  We compare our results with those given by accelerated Stokesian Dynamics (ASD) in Cao \emph{et al.}\cite{Cao2012}.  While we do not incorporate near-field lubrication hydrodynamics as is the case with ASD, we find the fluctuating FCM provides a very similar characterization of the gelation process, and with the DC we find the computations take a fraction of time of the ASD simulations.

As in Cao \emph{et al.}\cite{Cao2012}, we employ an interparticle potential that is a combination of a $(36-18)$ Lennard-Jones-like potential and a repulsive long-range Yukawa potential,
\begin{eqnarray}
 \dfrac{U(r)}{k_BT} & = & A \left[ \left(\dfrac{2a}{r}\right)^{36} -\left(\dfrac{2a}{r}\right)^{18} \right] \nonumber \\
 & & + B\dfrac{\exp\left[-\kappa \left(r-2a\right)\right]}{r}.
\end{eqnarray}
Along with the other parameter values, the exact values of $A$, $B$, and $\kappa$ used in our simulations are provided in Table \ref{tab:param_cao}.  These parameter values match exactly those used by Cao \emph{et al.}\cite{Cao2012}, including the very small time-step that must be taken due to the stiffness of the Lennard-Jones-like potential.  To increase the volume fraction, $\phi$, we increase linearly the number of particles, $N_p$ while keeping the computational domain over which we solve the Stokes equations a fixed size such that $\phi = N_p 4\pi a^3/(3L^3)$.  This is different from Cao \emph{et al.}\cite{Cao2012} where the number of particles was kept fixed as the volume fraction was varied.

\begin{table*}
\begin{centering}
\begin{tabular}{ccccccccc}
\toprule 
$A/k_BT$ & $B/k_BT$ & $\kappa$ & $\Delta t/t_{D_a}$ & $\eta$ & $a$ & $\phi$ & $N_p$ & L\\
\colrule 
60  & $20a$ & $4/a$ & $10^{-4}$ & $1$ & $3.29\Delta x$ & $0.04 - 0.12$ & $558-1674$ & $128\Delta x$\\
\botrule
\end{tabular}
\caption{Simulation parameters for the colloidal gelation simulations}
\label{tab:param_cao}
\end{centering}
\end{table*}

\begin{figure*}
\centering
\subfloat[Number of bonds per particle $N_b/N_p$ as a function of $t$ for different values of $\phi$.]{ \label{fig:Nbonds_part} \includegraphics[width=7cm]{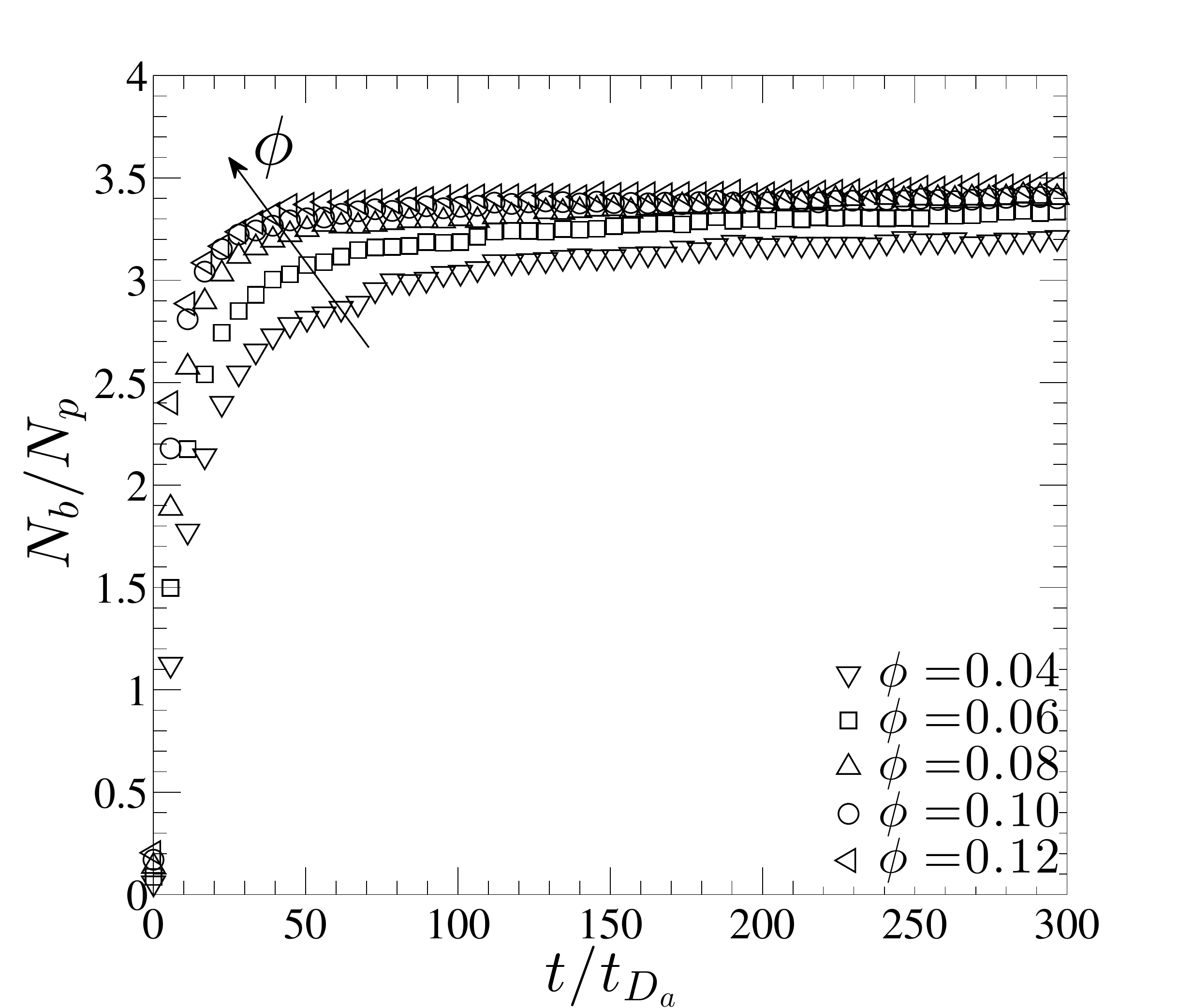}}
\subfloat[Number of clusters in the domain $N_c$ as a function of $t$ for different values of $\phi$.]{ \label{fig:Nclusters} \includegraphics[width=7cm]{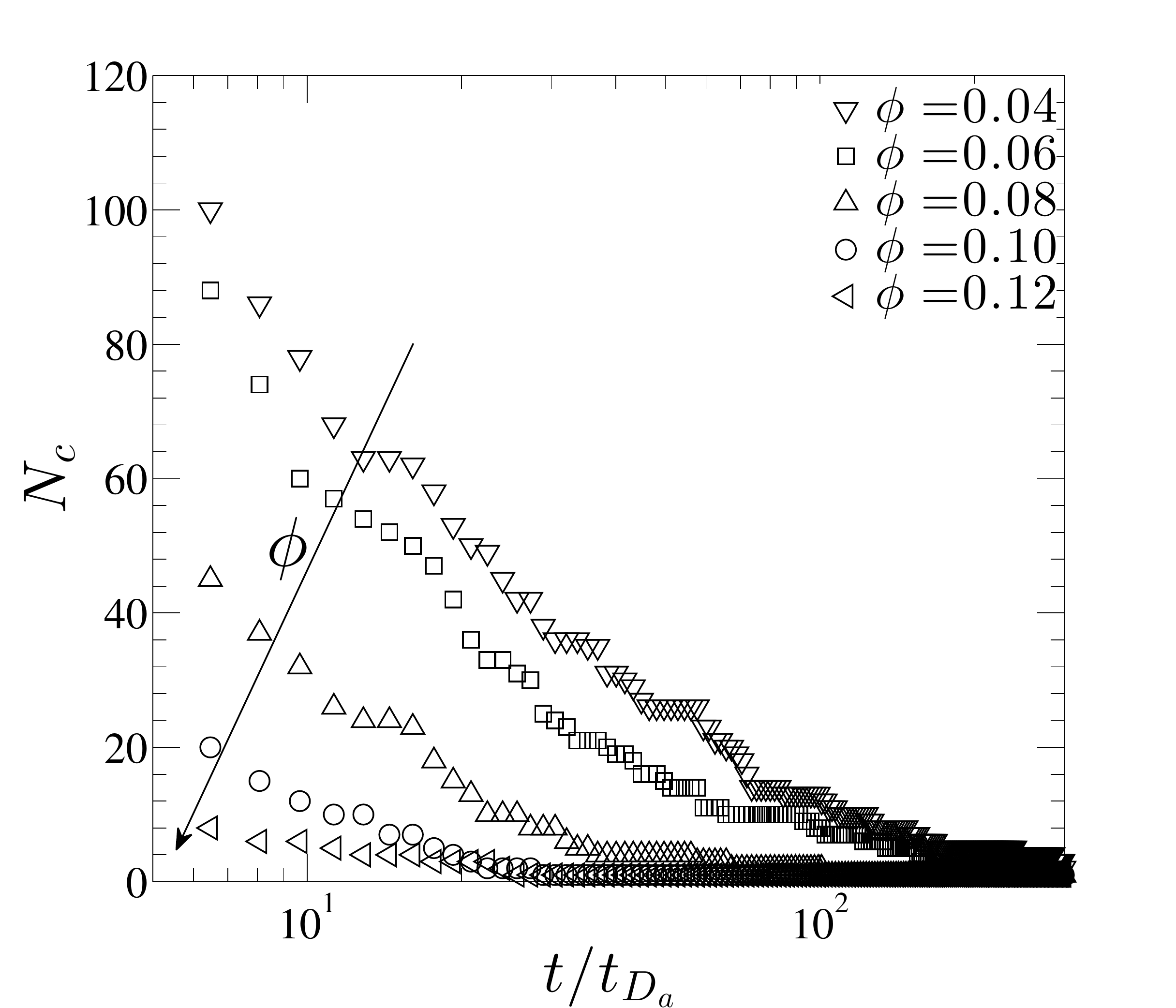}}

\caption{ Time evolution of the aggregation process for $\phi = 0.04 - 0.12$ and $\kappa = 4/a$.
} 
\end{figure*}

As each simulation progresses, we monitor the number of bonds per particle, $N_b/N_p$.  We use the criterion\cite{Cao2012} that a bond between two particles is formed when their center-to-center distance is less than or equal to $2.21a$.  Fig. \ref{fig:Nbonds_part} shows the time evolution of $N_b/N_p$ for different values of $\phi$.  We observe for a fixed time, the number of bonds per particle increases as $\phi$ increases.  We do see, however that as time increases, all simulations approach the same asymptotic value of $N_b/N_p \approx 3.47$.  These observations consistent with the ASD results obtained by Cao \emph{et al.}\cite{Cao2012}.

The aggregation dynamics can also be quantified by the time evolution of the number, $N_c$, of particle clusters.   We determine $N_c(t)$ by first compiling a list of bonded particle pairs at time $t$ and then processing the list using the $k-$clique percolation algorithm implemented in Python by Reid \emph{et al.}\cite{Reid2012}  Fig. \ref{fig:Nclusters} shows $N_c$ as a function of time.  We see that for each value of $\phi$, the number of clusters decreases with time.  We also find that for a fixed time, the number of clusters decreases as $\phi$ increases.  The final structures we observe at $t/t_{D_a}=300$ for different values of $\phi$ are shown in Table \ref{tab:perco}.  For all values of $\phi$ except $\phi = 0.04$, we find that the particles aggregate to form a single structure that spans the entire domain.  For $\phi=0.04$, which is the most dilute case we considered, the final state consists of 2 clusters that are not long enough to connect the opposite sides of the domain.  The tendency to form a single cluster was also noted by Cao \emph{et al.}\cite{Cao2012} and Furukawa and Tanaka\cite{Furukawa2010}.  They also showed that for low volume fractions, hydrodynamic interactions between the particles are needed in order to capture the formation of a single percolated structure. 

\begin{table*}
\begin{center}
\begin{tabular}{cccccc}
\toprule 
 $\phi = $ 0.04 ($N_p = $ 558) & 0.06 (837) & 0.08 (1116) & 0.10 (1395)  & $0.12$ (1674)\\
\colrule
  NP  & P & P & P & P\\
\colrule 
  \includegraphics[width=2.6cm]{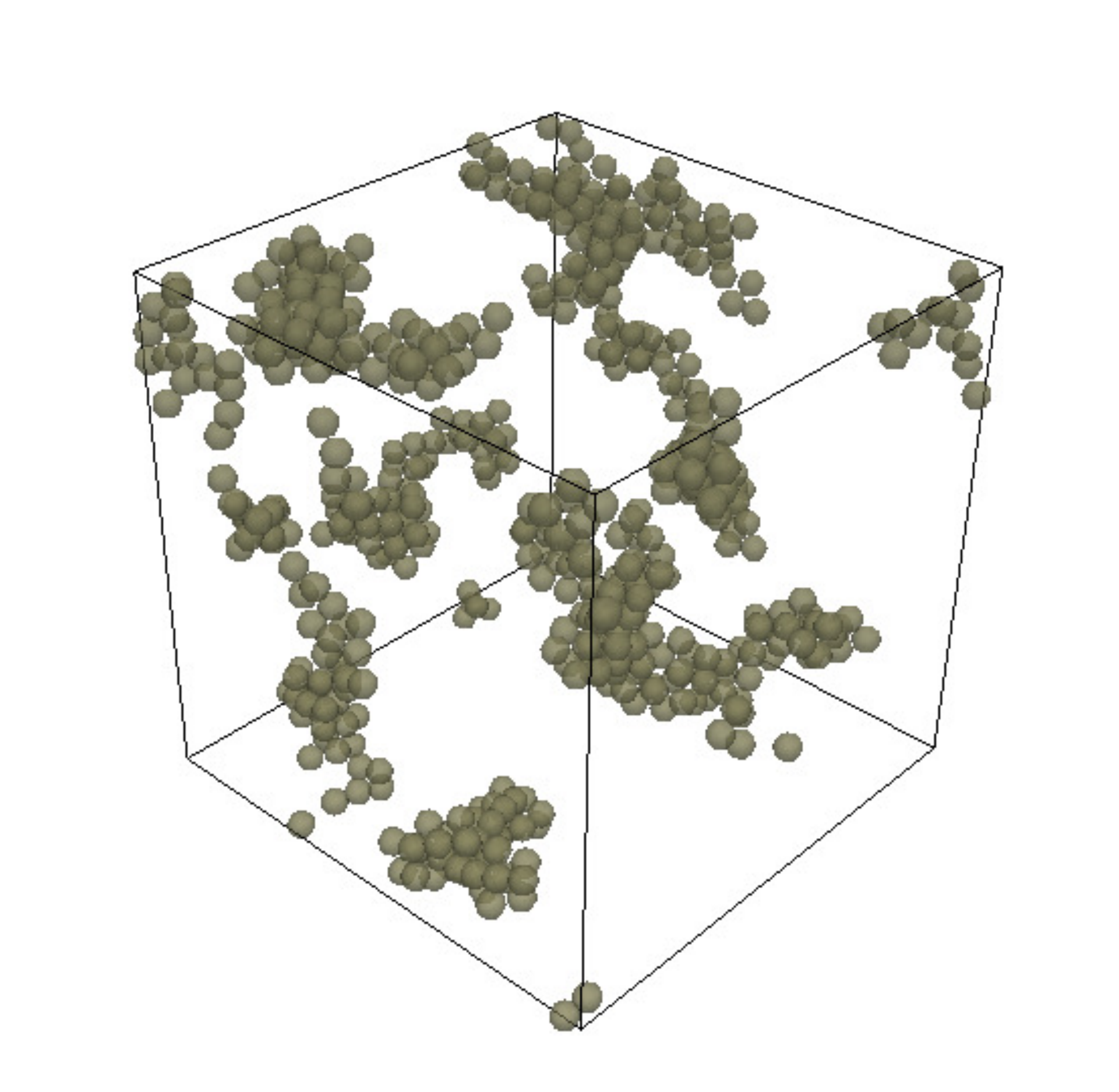} & 
  \includegraphics[width=2.6cm]{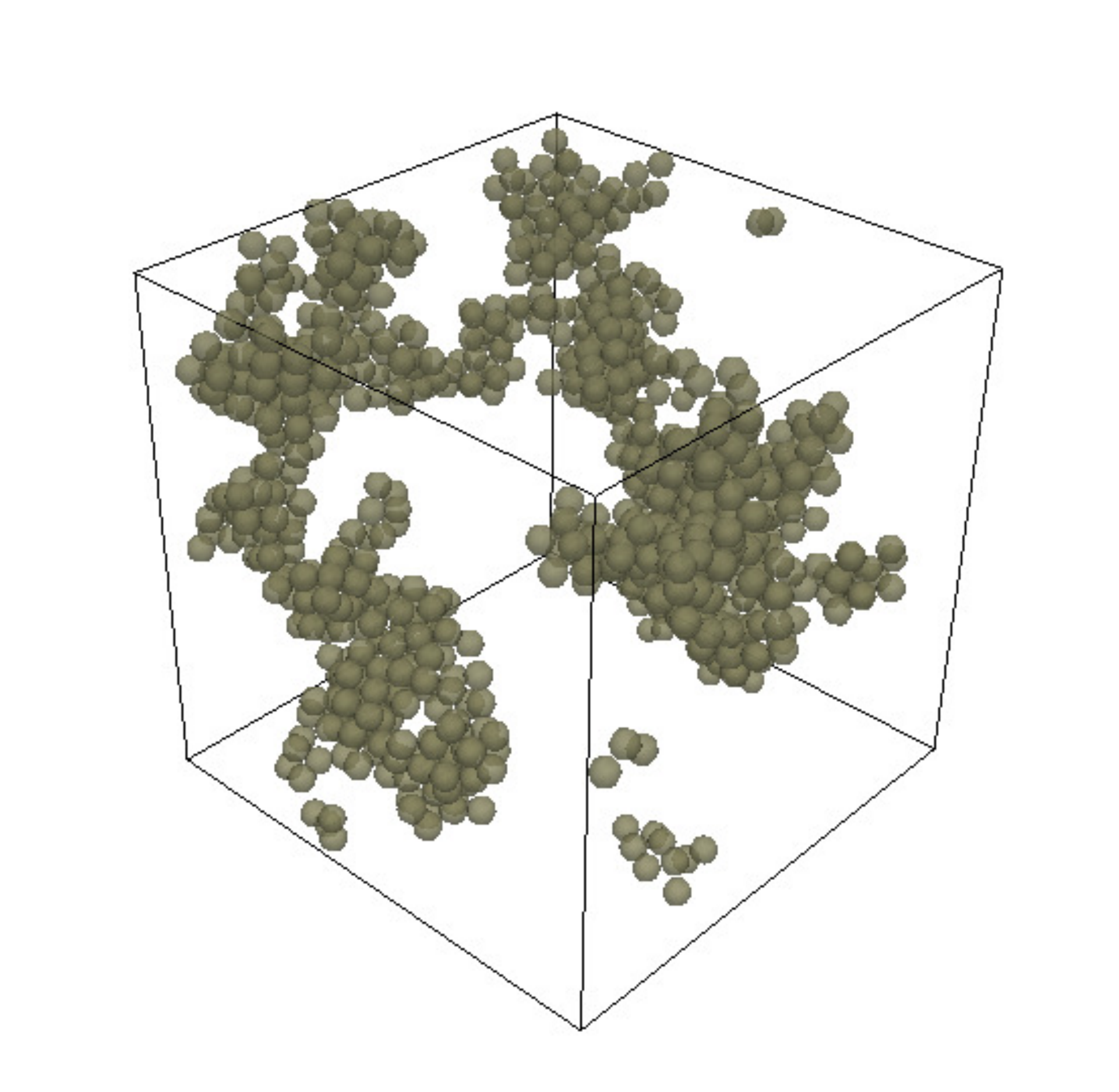} & 
  \includegraphics[width=2.6cm]{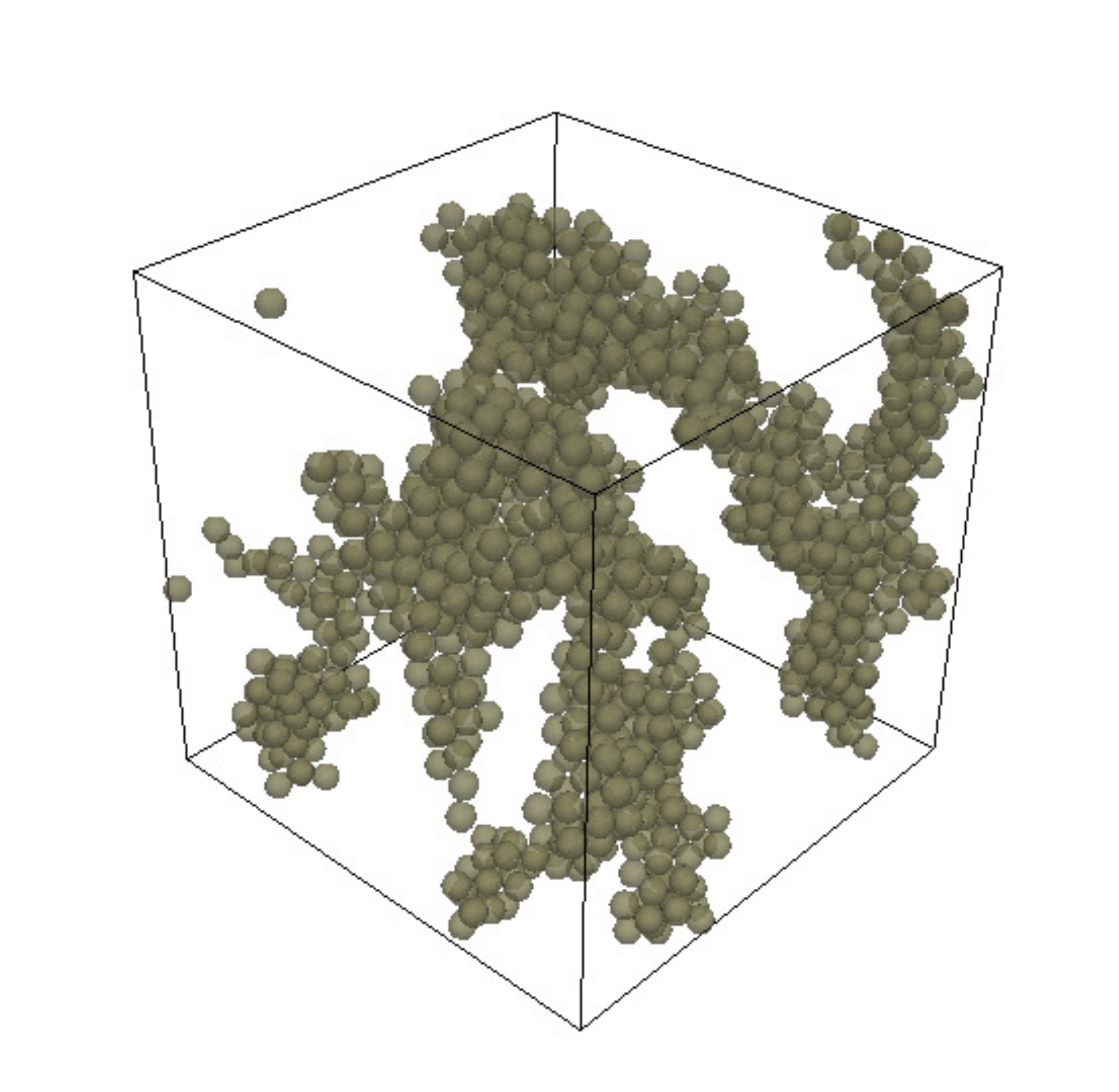} & 
  \includegraphics[width=2.6cm]{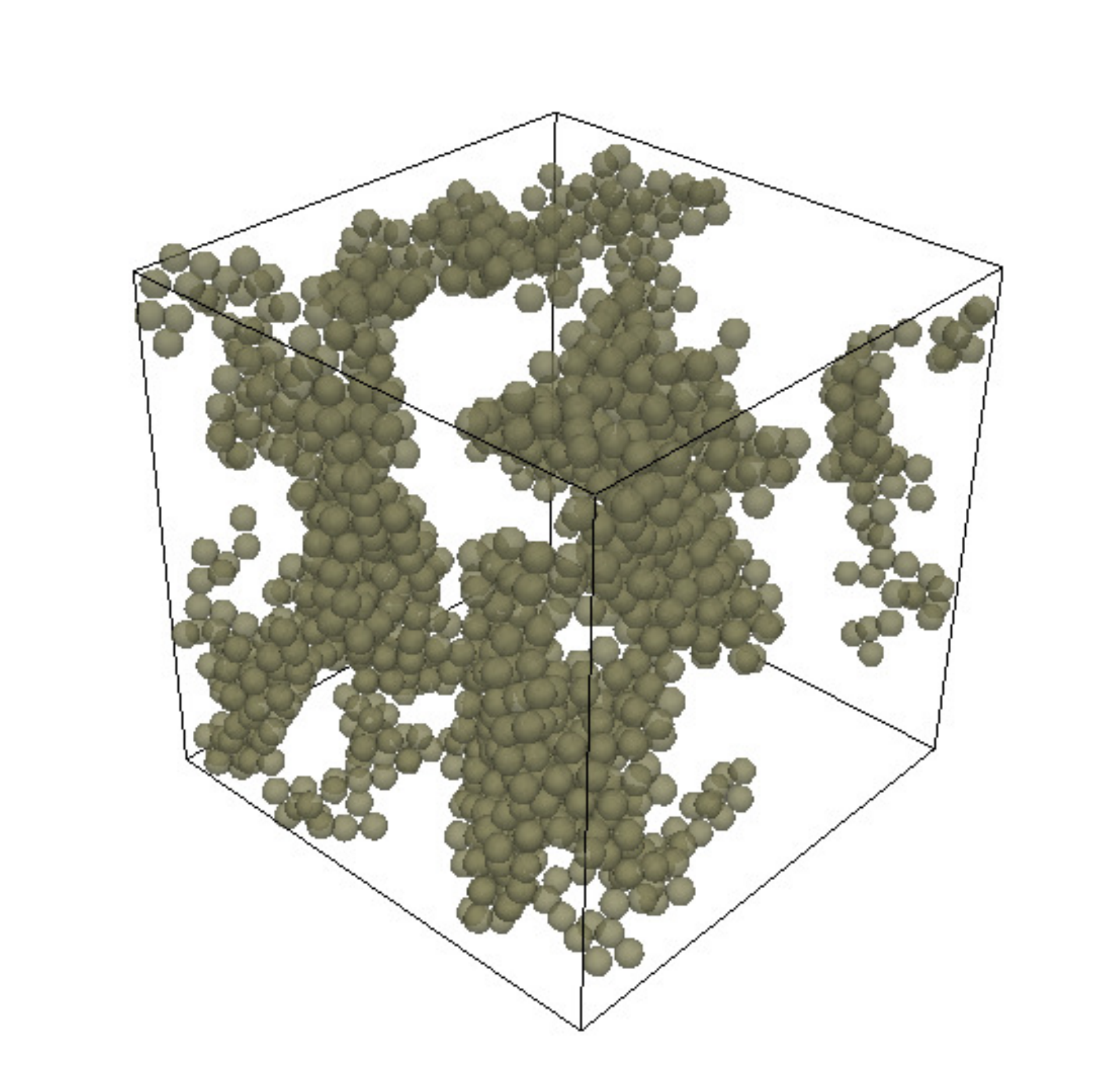} & 
  \includegraphics[width=2.6cm]{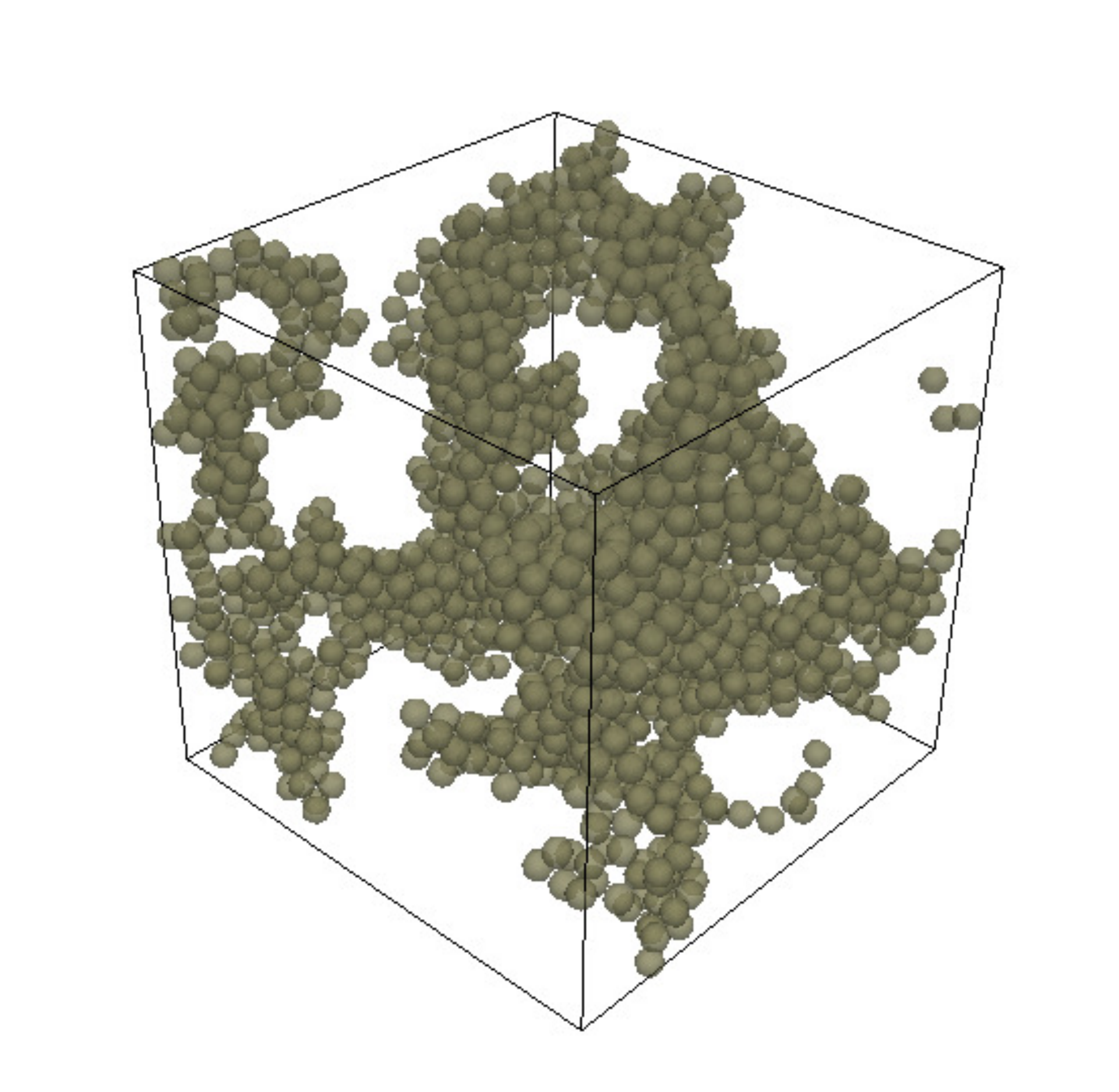}\\
  \colrule 
  $N_c = 2$ & 1 & 1 & 1 & 1 \\
\botrule
\end{tabular}
\caption{State of the suspension at $t/t_{D_a} = 300$ for different values of $\phi$. The numbers in parentheses correspond to the number of particles in the domain $N_p$.  The label ``P'' indicates that the particles aggregated to form a percolated network, while ``NP'' indicates that they have not.  The snapshots show the suspensions at the final time and $N_c$ is the number of clusters in the images (accounting for periodicity).}
\label{tab:perco}
\end{center}
\end{table*}

In addition to achieving very similar results as ASD, fluctuating FCM with the DC integration scheme provides these results at a relatively low computational cost.  The average computational time needed for fluctuating FCM with the stresslets and the DC to reach $t/t_{D_a} = 100$ ($10^6$ time-steps) with $N_p = 1674$ was $2.5$ days. The ASD simulations\cite{Cao2012} required approximately 10 days to reach $t/t_{D_a} = 100$ ($10^6$ timesteps) for half as many particles ($N_p = 874$).  Our approach is $8$ times faster when run on $64$ Intel(r) Ivybridge $2.8$ Ghz cores.  To further test the scalability of fluctuating FCM with DC, we simulated a suspension of $N_p = 3766$ particles at a volume fraction of $\phi = 0.08$ using $64$ Intel(r) Ivybridge $2.8$ Ghz cores. Snapshots from the simulation at different times are provided in Fig. \ref{fig:bigsim}, and we include a movie constructed from our simulations in the supplementary material\cite{SuppMat}.  This simulation required $8$ days to reach $t/t_{D_a} = 145$ ($1.45 \times 10^6$ time steps).  For this larger system size, we again find that the particles do eventually aggregate to form a single structure that percolates the entire domain.

\begin{figure*}
\centering
\subfloat[$t/t_{D_a} = 0$.]{ \label{fig:t0} \includegraphics[width=4.9cm]{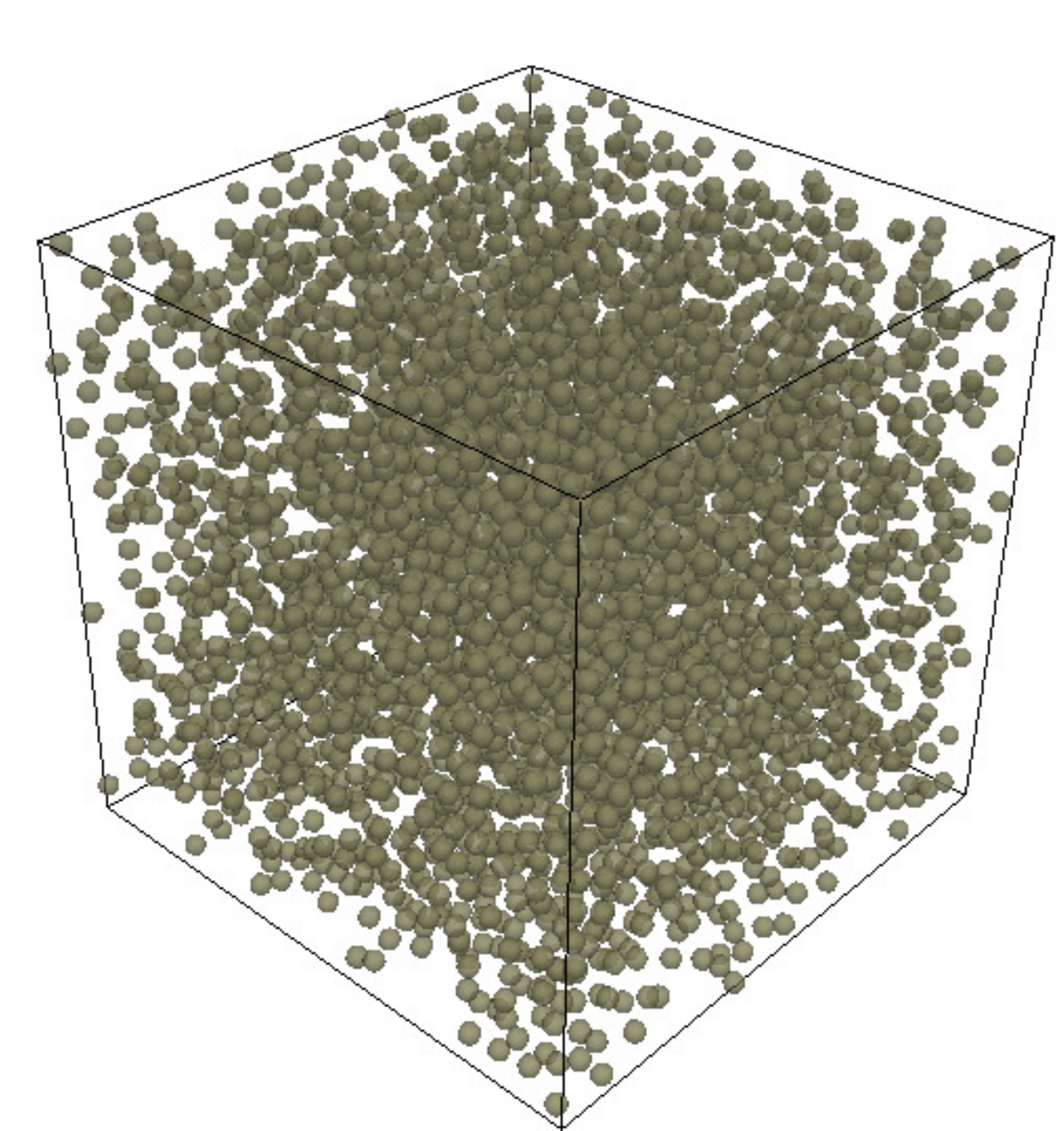}}
\subfloat[$t/t_{D_a} = 8$.]{ \label{fig:t} \includegraphics[width=4.9cm]{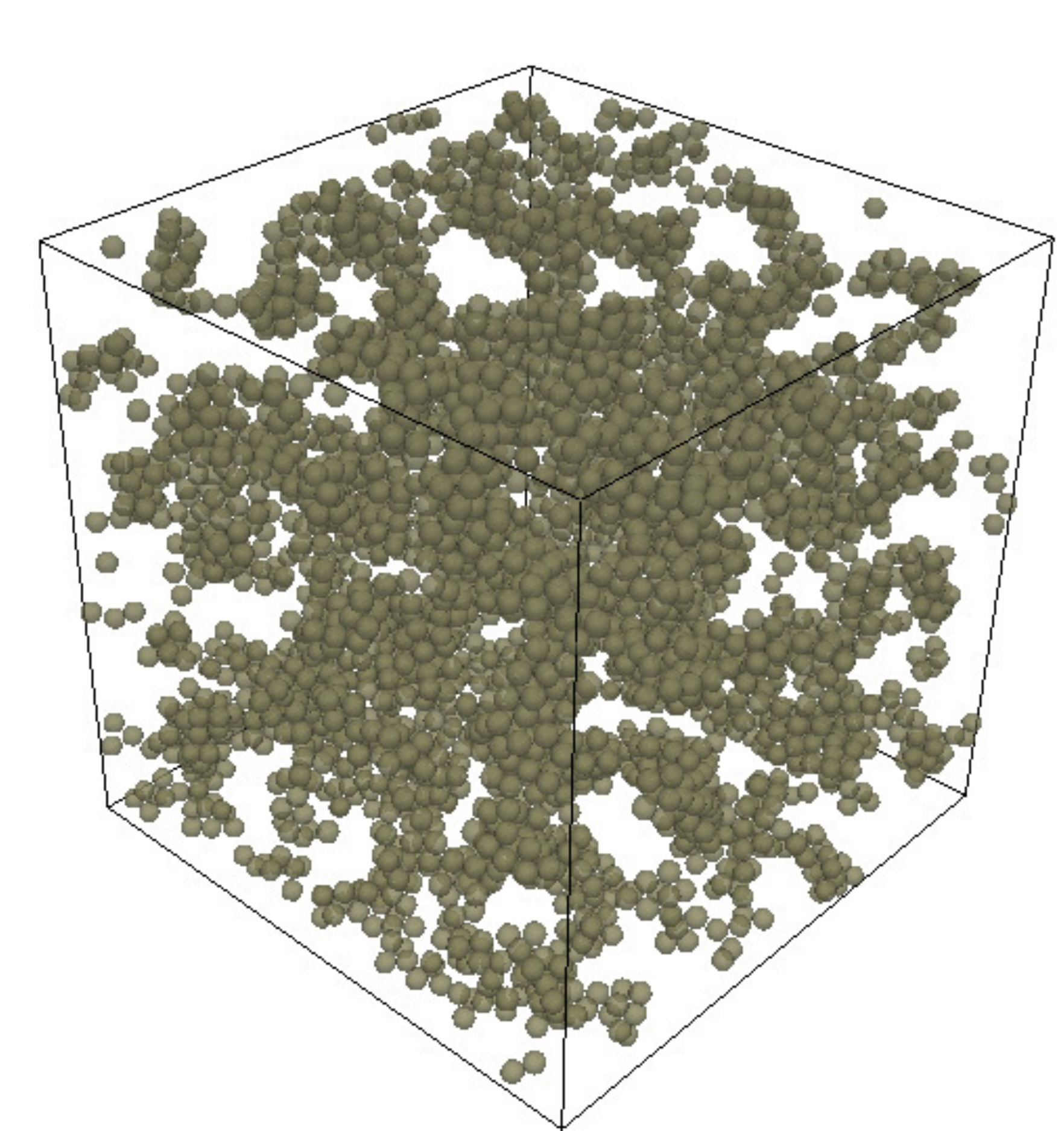}}
\subfloat[$t/t_{D_a} = 145$.]{ \label{fig:t1} \includegraphics[width=4.9cm]{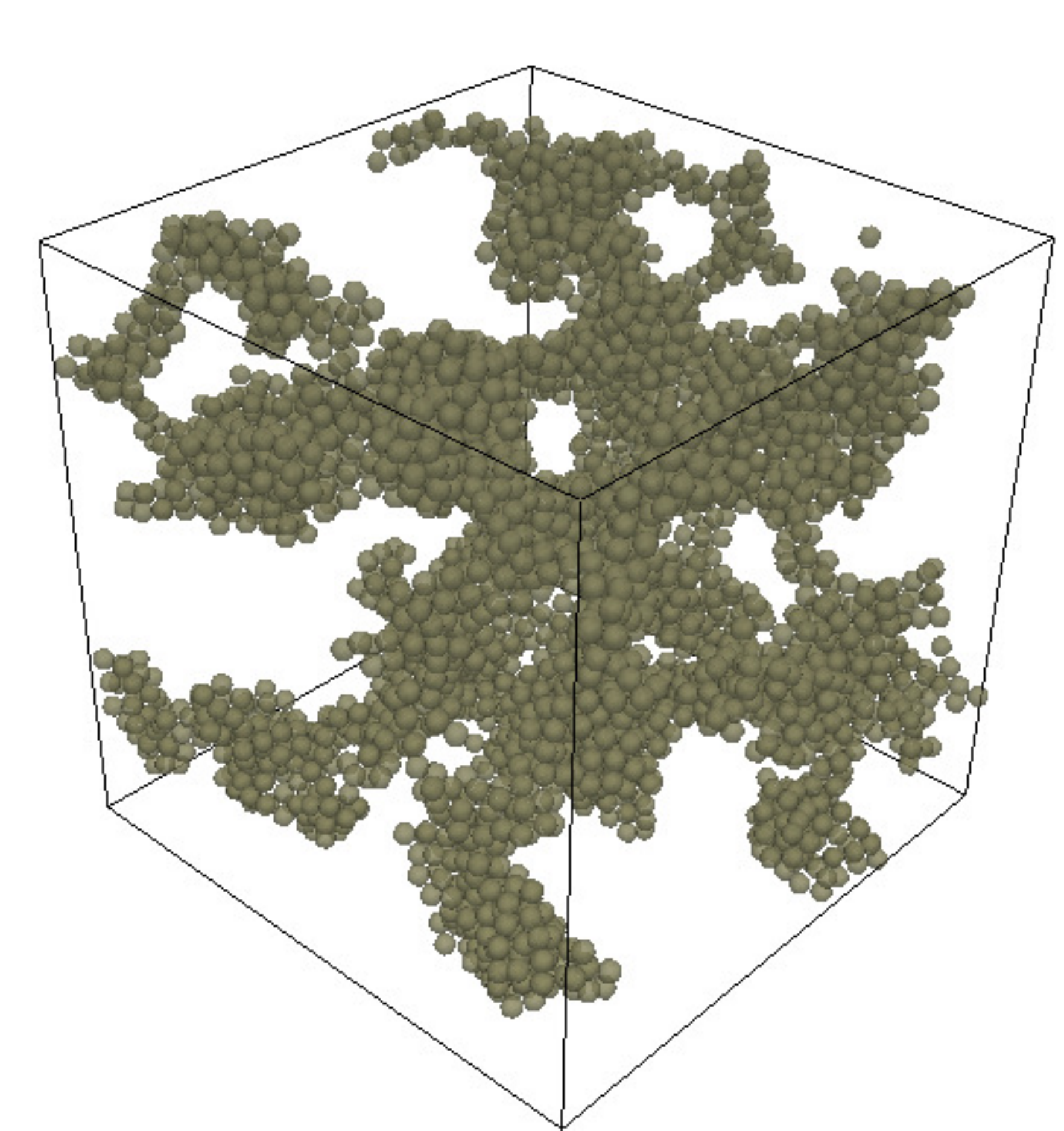}}

\caption{Snapshots of the aggregation process for $\phi = 0.08$ and $N_p = 3766$ (see also supplementary material\cite{SuppMat}).
} \label{fig:bigsim}
\end{figure*}
\section{Conclusions}
In this study, we presented a mid-point time integration scheme we refer to as the drifter-corrector (DC) to efficiently integrate the overdamped equations of motion for hydrodynamically interacting particles when their Brownian motion is computed using fluctuating hydrodynamics.  Methods based on fluctuating hydrodynamics such as the stochastic and fluctuating IBMs and the fluctuating FCM have been shown to drastically reduce the cost of including Brownian motion in simulation by simultaneously computing the deterministic and random particle velocities at each timestep.  

Like Fixman's method and RFD-based schemes, our DC scheme allows for the effects of the Brownian drift term to be included without ever having to compute it directly.  The DC, however, was designed especially for the case where imposed constraints on the rate-of-strain are used to provide a more accurate approximation to the particle hydrodynamic interactions, as is the case with fluctuating FCM.  Imposing these constraints incurs an additional cost for existing schemes.  Using these schemes with fluctuating FCM, we showed that this constraint would at the very least double the cost per timestep, which in practice meant an additional 10 Stokes solves per timestep.  We show that by using the DC with fluctuating FCM, this additional cost can be reduced to a single Stokes solve per time step.  Though we have developed the DC for and tested the scheme with fluctuating FCM, FCM's similarity with other methods, such as the stochastic and fluctuating IBMs, suggests that the DC could be an effective scheme to integrate particle positions for these methods as well.  

Using the DC with fluctuating FCM, we have provided an extensive validation of the scheme, showing that it reproduces the correct equilibrium particle distribution, as well as provides the distribution dynamics in accordance with the Smoluchowski equation.  We have also demonstrated the effectiveness of fluctuating FCM with the DC for colloidal suspension simulations, examining the collapse of a small cluster of particles, as well as the gelation process of a suspension.  In doing so, we were able to both quantify the effect of including higher-order corrections to the hydrodynamic interactions, as well as demonstrate the ability to accurately simulate colloidal suspensions at large-scale.  Indeed, we are currently applying this approach with active particle models \cite{Delmotte2015} to understand recent experimental results\cite{Leptos2009, Kurtuldu2011} on the dynamics of passive Brownian tracers in active particle suspensions.  

The methods described in this work may also be modified and extended to simulate particles with more complicated shapes \cite{Cichocki2015}.  While one approach is to construct rigid \cite{Vazquez2014,Delong2015}, or flexible \cite{Majmudar2012} particles from assemblies of spherical particles, another approach is to modify the shape of the kernel in the projection and volume averaging operators to reflect the shape of the particle itself.  This has been done with FCM for ellipsoidal particles\cite{Liu2009} and, as the underlying framework of the method remains unchanged, fluctuating FCM with the redefined operators should yield particle velocities that satisfy the fluctuation-dissipation theorem.  

A current challenge is to adapt the DC to these particle shapes as the Brownian drift term will also depend on the orientation of the particle. It would be of interest to compare fluctuating FCM results with the experimental measurements of Han \emph{et al.} \cite{Han2006} and the recent simulations of De Corato \emph{et al.} \cite{DeCorato2015}.  We are currently pursuing this line of research to provide a similar set of efficient and accurate simulation techniques to understand the dynamics of suspensions of ellipsoidal Brownian particles.

\section*{Acknowledgements}
We would like to thank Aleksandar Donev and Grigorios Pavliotis for insightful discussions about this work.  BD would like to acknowledge travel support from the STSM COST Action MP1305 grant, the INPT BQR-SMI grant and the MEGEP International Mobility grant. EEK would like to acknowledge travel support from EPSRC Mathematics Platform grant EP/I019111/1.  Simulations were performed on the Calmip supercomputing mesocenter (Univ. of Toulouse) and IDRIS, the centre of high performance computing for the French National Centre for Scientific Research (CNRS).

\appendix
\section{Stress-free boundary conditions with fluctuating hydrodynamics}
\label{sec:AppA}
We have seen in the text that the slip boundary conditions can be imposed by requiring that the fluctuating stress satisfy certain symmetry conditions.  In this appendix, we show analytically that the fluctuating stress with modified statistics yields flow correlations proportional to the appropriate Green's function.

For the case where the fluid is in the positive half-space, i.e. $z > 0$, with the conditions that $u_z = 0$ and $\partial u_{x}/\partial z = \partial u_{y}/\partial z = 0$ at $z = 0$, the Green's function\cite{Lee1979} for this Stokes flow is
\begin{equation}
H_{ij}(\mathbf{x},\mathbf{y}) = G_{ij}(\mathbf{x} - \mathbf{y}) + \gamma_{jk}G_{ik}(\mathbf{x} - \mathbf{Y})
\end{equation}
where 
\begin{eqnarray}
\gamma_{jk} &=& \delta_{jk} - 2\delta_{3j}\delta_{3k}, \\
\mathbf{Y} &=& \mathbf{y} - 2(\mathbf{y}\cdot\bm{\hat{z}})\bm{\hat{z}},\\
G_{ij}(\mathbf{x}) &=& \frac{1}{8 \pi \eta}\left(\frac{\delta_{ij}}{|\mathbf{x}|} + \frac{x_ix_j}{|\mathbf{x}|^3}\right).
\end{eqnarray}
Thus, if there is a point force, $\mathbf{F} = (F_x, F_y, F_z)$, located at the point $\mathbf{y}$ in the flow domain, the boundary conditions can be satisfied by introducing an image point force $\mathbf{F}^{im} = (F_x, F_y, -F_z)$ below the slip surface at $\mathbf{Y}$.

This image system approach to obtain the Green's function readily extends to the case where there are two parallel no flux, slip surfaces at $z = 0$ and $z = L_z$, and periodic boundary conditions at $x = 0$ and $x = L_x$ and $y = 0$ and $y = L_y$.  For our discussion, we consider the specific case where $L_z = L_x/2 = L_y/2 = L/2$, but it can be easily generalized to different domain sizes.  For the case where the fluid domain is given by $z \in [0, L/2)$, $x \in [0, L)$, and $y \in [0, L)$, the Green's function can be expressed as 
\begin{eqnarray}
L_{ij}(\mathbf{x},\mathbf{y}) = \frac{1}{L^3}\sum_{\mathbf{k}\neq 0}\frac{1}{\eta k^2}\left(\delta_{il} - \frac{k_ik_l}{k^2}\right) \nonumber\\
\times\left(\delta_{lj}e^{-i\mathbf{k}\cdot \mathbf{y}}+ \gamma_{lj}e^{-i\mathbf{k}\cdot \mathbf{Y}}\right)e^{i\mathbf{k}\cdot \mathbf{x}}. \label{eq:slipchannelGF}
\end{eqnarray}

To impose the slip condition, we would like then for the fluid flow resulting from the fluctuating stress to have the following statistics
\begin{eqnarray}
\langle\mathbf{\tilde u}(\mathbf{x})\rangle &=& \mathbf{0} \\
\langle\mathbf{\tilde u}(\mathbf{x})\mathbf{\tilde u}^T(\mathbf{y})\rangle &=& 2k_BT \mathbf{L}(\mathbf{x},\mathbf{y}).
\end{eqnarray}
We show here that this is achieved by having 
\begin{eqnarray}
\langle P_{ij}(\mathbf{x})\rangle &=& \mathbf{0} \\
\langle P_{ij}(\mathbf{x})P_{kl}(\mathbf{y})\rangle &=& 2k_BT \Delta_{ijkl} \delta(\mathbf{x} - \mathbf{y}) \nonumber\\
& &+ 2k_BT \Gamma_{ijkl} \delta(\mathbf{x} - \mathbf{Y}).\label{eq:flucstressslip}
\end{eqnarray}
where
\begin{eqnarray}
\Delta_{ijkl}&=&\delta_{ik}\delta_{jl}+\delta_{il}\delta_{jk} \\
\Gamma_{ijkl}&=&\gamma_{ik}\gamma_{jl}+\gamma_{il}\gamma_{jk}.
\end{eqnarray}
We notice that the fluctuating stress is $L$-periodic in each of the three dimensions.  We can therefore use Fourier series to obtain the flow resulting from the fluctuating stress.  We define the Fourier transform as 
\begin{equation}
\mathbf{g}(\mathbf{x}) = \sum_{\mathbf{k}} \mathbf{\hat{g}}(\mathbf{k})e^{i\mathbf{k}\cdot \mathbf{x}}
\end{equation}
and, consequently,
\begin{equation}
\mathbf{\hat{g}}(\mathbf{k}) = L^{-3}\int_{\Omega}\mathbf{\hat{g}}(\mathbf{x})e^{-i\mathbf{k}\cdot \mathbf{x}}d^3\mathbf{x}.
\end{equation}
The flow in Fourier space is given by
\begin{eqnarray}
\hat{u}_i = \frac{1}{\eta k^2}\left(\delta_{ij} - \frac{k_ik_j}{k^2}\right)ik_l \hat{P}_{jl}.
\end{eqnarray}
and, therefore,
\begin{eqnarray}
\langle \hat{u}_i(\mathbf{k})\hat{u}_n(\mathbf{q})\rangle = \frac{-k_lq_r}{\eta^2 k^2q^2}\left(\delta_{ij} - \frac{k_ik_j}{k^2}\right)\nonumber\\
\times\left(\delta_{np} - \frac{q_nq_p}{q^2}\right) \langle \hat{P}_{jl}(\mathbf{k})\hat{P}_{pr}(\mathbf{q})\rangle. \nonumber\\ \label{eq:flowcorr}
\end{eqnarray}
From the definition of the Fourier transform and Eq. \eqref{eq:flucstressslip}, we find that
\begin{eqnarray}
\langle \hat{P}_{ij}(\mathbf{k})\hat{P}_{kl}(\mathbf{q})\rangle & = & 2k_BTL^{-3}\Delta_{ijkl}\delta_{\mathbf{q},\mathbf{-k}} \nonumber \\
& &+ 2k_BTL^{-3}\Gamma_{ijkl}\delta_{\mathbf{q},\mathbf{K}},\nonumber\\
\end{eqnarray}
where $\mathbf{K} = -\bm{\gamma}\mathbf{k}$.  Using this expression in (\ref{eq:flowcorr}) and the fact that $K^2 = k^2$, we find
\begin{eqnarray}
\langle \hat{u}_i(\mathbf{k})\hat{u}_n(\mathbf{q})\rangle &=& \frac{2k_BT}{\eta L^3 k^2}\left(\delta_{in} - \frac{k_ik_n}{k^2}\right)\delta_{\mathbf{q},\mathbf{-k}} \nonumber\\
& & + \frac{2k_BT}{\eta L^3 k^2}\left(\gamma_{in} + \frac{k_iK_n}{k^2}\right)\delta_{\mathbf{q},\mathbf{K}}.\nonumber\\
\end{eqnarray}
In real space, the flow correlations are then
\begin{equation}
\def\arraystretch{1.6}
\begin{array}{l}
\displaystyle \langle \tilde{u}_i(\mathbf{x})\tilde{u}_n(\mathbf{y})\rangle =  \\
\displaystyle\sum_{\mathbf{k}\neq \mathbf{0}}\sum_{\mathbf{q}\neq \mathbf{0}}\frac{2k_BT}{\eta L^3 k^2}\left(\delta_{in} - \frac{k_ik_n}{k^2}\right)\delta_{\mathbf{q},\mathbf{-k}}e^{i\mathbf{k}\cdot\mathbf{x}}e^{i\mathbf{q}\cdot\mathbf{y}}  \\
\displaystyle + \sum_{\mathbf{k}\neq \mathbf{0}}\sum_{\mathbf{q}\neq \mathbf{0}}\frac{2k_BT}{\eta L^3 k^2}\left(\gamma_{in} + \frac{k_iK_n}{k^2}\right)\delta_{\mathbf{q},\mathbf{K}}e^{i\mathbf{k}\cdot\mathbf{x}}e^{i\mathbf{q}\cdot\mathbf{y}}. \\
\end{array}
\end{equation}
Performing the sum over $\mathbf{q}$ for each term, we have
\begin{equation}
\def\arraystretch{1.6}
\begin{array}{l}
\displaystyle \langle \tilde{u}_i(\mathbf{x})\tilde{u}_n(\mathbf{y})\rangle = \\
\displaystyle \sum_{\mathbf{k}\neq \mathbf{0}}\frac{2k_BT}{\eta L^3 k^2}\left(\delta_{in} - \frac{k_ik_n}{k^2}\right)e^{i\mathbf{k}\cdot(\mathbf{x} - \mathbf{y}}) \\
\displaystyle + \sum_{\mathbf{k}\neq \mathbf{0}}\frac{2k_BT}{\eta L^3 k^2}\left(\gamma_{in} + \frac{k_iK_n}{k^2}\right)e^{i\mathbf{k}\cdot\mathbf{x}}e^{i\mathbf{K}\cdot\mathbf{y}}.\nonumber\\
\end{array}
\end{equation}
As $\mathbf{K}\cdot\mathbf{y} = -\mathbf{k}\cdot\mathbf{Y}$, we then have
\begin{equation}
\def\arraystretch{1.6}
\begin{array}{l}
\displaystyle \langle \tilde{u}_i(\mathbf{x})\tilde{u}_n(\mathbf{y})\rangle = \\
\displaystyle \sum_{\mathbf{k}\neq \mathbf{0}}\frac{2k_BT}{\eta L^3 k^2}\left(\delta_{in} - \frac{k_ik_n}{k^2}\right)e^{i\mathbf{k}\cdot(\mathbf{x} - \mathbf{y})} \\
\displaystyle +\sum_{\mathbf{k}\neq \mathbf{0}}\frac{2k_BT}{\eta L^3 k^2}\left(\delta_{in} + \frac{k_ik_n}{k^2}\right)\gamma_{jn}e^{i\mathbf{k}\cdot(\mathbf{x}-\mathbf{Y})}. \\
\end{array}
\end{equation}
Writing this expression slightly differently as, 
\begin{eqnarray}
\langle \tilde{u}_i(\mathbf{x})\tilde{u}_n(\mathbf{y})\rangle = \frac{2k_BT}{L^3}\sum_{\mathbf{k}\neq 0}\frac{1}{\eta k^2}\left(\delta_{il} - \frac{k_ik_l}{k^2}\right)\nonumber\\
\times\left(\delta_{lj}e^{-i\mathbf{k}\cdot \mathbf{y}}+ \gamma_{lj}e^{-i\mathbf{k}\cdot \mathbf{Y}}\right)e^{i\mathbf{k}\cdot \mathbf{x}},\nonumber\\
\end{eqnarray}
we immediately see that 
\begin{equation}
\langle \mathbf{\tilde{u}}(\mathbf{x})\mathbf{\tilde{u}}^T(\mathbf{y})\rangle = 2k_BT \mathbf{L}(\mathbf{x},\mathbf{y}),
\end{equation}
with $\mathbf{L}$ given by Eq. \eqref{eq:slipchannelGF}, as desired.

\section{Error expansion for the DC}\label{sec:AppB}
In this appendix, we present the error analysis showing that the DC is weakly accurate to first-order in time.  To simplify the analysis and make the presentation more compact, we write the scheme using the mobility matrices as 
\begin{eqnarray}
\mathcal{Y}^{k+1/2} &=& \mathcal{Y}^{k} + \frac{\Delta t}{2}\tilde{\mathcal{U}}^{k} \label{eq:tn12}\\
\mathcal{Y}^{k+1} &=& \mathcal{Y}^{k} + \Delta t(1 + v^k)\left[\mathcal{M}^{\mathcal{VF};k+1/2}_{FCM-S}\mathcal{F}^{k+1/2} \right. \nonumber\\
&& + \left. \mathcal{M}^{\mathcal{VT}; k+1/2}_{FCM-S}\mathcal{T}^{k+1/2} + \tilde{\mathcal{V}}^{k+1/2}\right]\label{eq:tn1}
\end{eqnarray}
where 
\begin{eqnarray}
\tilde{\mathcal{U}}^{k} &=& \mathcal{J}^k[\mathbf{\tilde{u}}^k],\\
\tilde{\mathcal{V}}^{k+1/2} &=& \mathcal{J}^{k+1/2}[\mathbf{\tilde{u}}^k] \nonumber\\
&&+ \mathcal{M}^{\mathcal{VS};k+1/2}_{FCM}\mathcal{S}^{k+1/2},\\
\mathcal{S}^{k+1/2}&=& -\mathcal{R}^{\mathcal{ES};k+1/2}_{FCM}\tilde{\mathcal{E}}^{k+1/2},\\
\tilde{\mathcal{E}}^{k+1/2}&=& -\mathcal{K}^{k+1/2}[\mathbf{\tilde{u}}^k],
\end{eqnarray}
and $\mathbf{\tilde{u}}^k$ satisfies Eq. \eqref{eq:FlucStokes} for the realization of the fluctuating stress at time $t_k$.  We now expand the quantities on the right-hand side of Eq. (\ref{eq:tn1}) about $t = t_k$, and using Eq. (\ref{eq:tn12}) we obtain
\begin{widetext}
\begin{eqnarray}
\tilde{\mathcal{V}}_{\alpha}^{k+1/2} &=& \tilde{\mathcal{V}}_{\alpha}^{k} + \frac{\Delta t}{2} \frac{\partial \tilde{\mathcal{V}}_{\alpha}^k}{\partial \mathcal{Y}_\beta} \tilde{\mathcal{U}}_{\beta}^k + \frac{\Delta t^2}{8} \frac{\partial^2 \tilde{\mathcal{V}}_{\alpha}^k}{\partial \mathcal{Y}_\gamma \partial \mathcal{Y}_\beta} \tilde{\mathcal{U}}_{\beta}^k\tilde{\mathcal{U}}_{\gamma}^k + O(\Delta t) \\
\mathcal{F}_{\alpha}^{k+1/2} &=& \mathcal{F}_{\alpha}^{k} + \frac{\Delta t}{2} \frac{\partial \mathcal{F}_{\alpha}^k}{\partial \mathcal{Y}_\beta} \tilde{\mathcal{U}}_{\beta}^k + \frac{\Delta t^2}{8} \frac{\partial^2 \mathcal{F}_{\alpha}^k}{\partial \mathcal{Y}_\gamma \partial \mathcal{Y}_\beta} \tilde{\mathcal{U}}_{\beta}^k\tilde{\mathcal{U}}_{\gamma}^k + O(\Delta t^{3/2}) \\
\mathcal{M}^{\mathcal{VF};k+1/2}_{\alpha \beta} &=& \mathcal{M}^{\mathcal{VF};k}_{\alpha \beta} + \frac{\Delta t}{2} \frac{\partial \mathcal{M}^{\mathcal{VF};k}_{\alpha \beta}}{\partial \mathcal{Y}_\gamma} \tilde{\mathcal{U}}_{\gamma}^k + \frac{\Delta t^2}{8} \frac{\partial^2 \mathcal{M}^{\mathcal{VF};k}_{\alpha \beta}}{\partial \mathcal{Y}_\gamma \partial \mathcal{Y}_\delta} \tilde{\mathcal{U}}_{\delta}^k\tilde{\mathcal{U}}_{\gamma}^k + O(\Delta t^{3/2}).
\end{eqnarray}
\end{widetext}
where we have used Greek letters to index the vectors and mobility matrices and the convention that their repetition implies summation.  The expansions for $\mathcal{T}^{k+1/2}$ and $\mathcal{M}^{\mathcal{VT};k+1/2}_{FCM-S}$ are identical to those for $\mathcal{F}^{k+1/2}$ and $\mathcal{M}^{\mathcal{VF};k+1/2}_{FCM-S}$, respectively.  For clarity, we have dropped the label ``$FCM-S$'' for the mobility matrices involved in the expansion.  We also have from Eq. \eqref{eq:vkexp} that
\begin{equation}
v^k = \frac{\Delta t}{2}\frac{\partial \tilde{\mathcal{U}}^k_{\alpha}}{\partial \mathcal{Y}_\alpha}.
\end{equation}
Inserting the expansions into Eq. \eqref{eq:tn1}, the expansion for the increment $\Delta \mathcal{Y}^k_\alpha = \mathcal{Y}^{k+1}_\alpha - \mathcal{Y}^{k}_\alpha$ is
\begin{widetext}
\begin{eqnarray}
\Delta \mathcal{Y}^k_\alpha &=& \Delta t \tilde{\mathcal{V}}^k_{\alpha} + \Delta t \left[\mathcal{M}^{\mathcal{VF};k}_{\alpha \beta}\mathcal{F}^k_\beta + \mathcal{M}^{\mathcal{VT};k}_{\alpha \beta}\mathcal{T}^k_\beta + \frac{\Delta t}{2}\left(\frac{\partial \tilde{\mathcal{V}}^k_{\alpha}}{\partial \mathcal{Y}_{\beta}} \tilde{\mathcal{U}}^k_{\beta} + \tilde{\mathcal{V}}^k_{\alpha}\frac{\partial \tilde{\mathcal{U}}^k_{\beta}}{\partial \mathcal{Y}_{\beta}}\right)\right] \nonumber\\
&&+\Delta t^2\left[\frac{1}{2}\frac{\partial}{\partial \mathcal{Y}_\gamma}\left(\mathcal{M}^{\mathcal{VF};k}_{\alpha \beta}\mathcal{F}^k_\beta+\mathcal{M}^{\mathcal{VT};k}_{\alpha \beta}\mathcal{T}^k_\beta\right) \tilde{\mathcal{U}}_{\gamma} + \frac{\Delta t}{8}\frac{\partial^2 \tilde{\mathcal{V}}_{\alpha}^k}{\partial \mathcal{Y}_\gamma \partial \mathcal{Y}_\beta}\tilde{\mathcal{U}}_{\beta}^k\tilde{\mathcal{U}}_{\gamma}^k\right] + O(\Delta t^2).
\end{eqnarray}
\end{widetext}
Taking the ensemble average of this expression and using the fact that
\begin{eqnarray}
\frac{\partial}{\partial \mathcal{Y}_{\beta}} \left(\tilde{\mathcal{V}}^k_{\alpha} \tilde{\mathcal{U}}^k_{\beta}\right)&=&\frac{\partial \tilde{\mathcal{V}}^k_{\alpha}}{\partial \mathcal{Y}_{\beta}} \tilde{\mathcal{U}}^k_{\beta} + \tilde{\mathcal{V}}^k_{\alpha}\frac{\partial \tilde{\mathcal{U}}^k_{\beta}}{\partial \mathcal{Y}_{\beta}}
\end{eqnarray}
we have
\begin{eqnarray}
\langle \Delta \mathcal{Y}^k_\alpha \rangle &=& \Delta t\left[\mathcal{M}^{\mathcal{VF};k}_{\alpha \beta}\mathcal{F}^k_\beta + \mathcal{M}^{\mathcal{VT};k}_{\alpha \beta}\mathcal{T}^k_\beta \right.\nonumber\\
&&\left. + \frac{\Delta t}{2}\frac{\partial}{\partial \mathcal{Y}_\beta} \langle\tilde{\mathcal{V}}^k_{\alpha} \tilde{\mathcal{U}}^k_{\beta}\rangle\right] + O(\Delta t^2).\nonumber\\
\end{eqnarray}
Since
\begin{eqnarray}
\langle\tilde{\mathcal{V}}^k_{\alpha} \tilde{\mathcal{U}}^k_{\beta}\rangle &=& \langle\tilde{\mathcal{U}}^k_{\alpha} \tilde{\mathcal{U}}^k_{\beta}\rangle -\mathcal{M}^{\mathcal{VS};k}_{FCM;\alpha \gamma}\mathcal{R}^{\mathcal{ES};k}_{FCM;\gamma \delta}\langle\tilde{\mathcal{E}}^k_{\delta} \tilde{\mathcal{U}}^k_{\beta}\rangle,\nonumber\\
\end{eqnarray}
with the correlations on the right-hand side given\cite{Keaveny2014} by 
\begin{eqnarray}
\langle\tilde{\mathcal{U}}^k_{\alpha} \tilde{\mathcal{U}}^k_{\beta}\rangle &=& \frac{2k_BT}{\Delta t} \mathcal{M}^{\mathcal{VF};k}_{FCM;\alpha \beta}\\
\langle\tilde{\mathcal{E}}^k_{\delta} \tilde{\mathcal{U}}^k_{\beta}\rangle &=& \frac{2k_BT}{\Delta t}\mathcal{M}^{\mathcal{EF};k}_{FCM;\delta \beta}
\end{eqnarray}
we have that 
\begin{equation}
\langle\tilde{\mathcal{V}}^k_{\alpha} \tilde{\mathcal{U}}^k_{\beta}\rangle = \frac{2k_BT}{\Delta t} \mathcal{M}^{\mathcal{VF};k}_{\alpha\beta}. \label{eq:VUcovar}
\end{equation}
We emphasize that the mobility matrix in Eq. \eqref{eq:VUcovar} is the stresslet-corrected mobility matrix, the expression for which is given in Eq. \eqref{eq:FCM-S}.  As a result, we obtain the correct first moment up to first order in time
\begin{eqnarray}
\langle \Delta \mathcal{Y}^k_\alpha \rangle &=& \Delta t \left[\mathcal{M}^{\mathcal{VF};k}_{\alpha \beta}\mathcal{F}^k_\beta + \mathcal{M}^{\mathcal{VT};k}_{\alpha \beta}\mathcal{T}^k_\beta\right]   \nonumber\\
&& + \Delta t k_BT\frac{\partial \mathcal{M}^{\mathcal{VF};k}_{\alpha \beta}}{\partial \mathcal{Y}_\beta} + O(\Delta t^2). \nonumber\\
\end{eqnarray}

The outer product of the increment gives
\begin{eqnarray}
\Delta \mathcal{Y}^k_\alpha \Delta \mathcal{Y}^k_\beta &=& \Delta t^2\tilde{\mathcal{V}}^k_{\alpha}\tilde{\mathcal{V}}^k_{\beta} \nonumber\\
&&+ \Delta t^2\left[\tilde{\mathcal{V}}^k_{\alpha}\tilde{\mathcal{G}}^k_{\beta} + \tilde{\mathcal{V}}^k_{\beta}\tilde{\mathcal{G}}^k_{\alpha}\right] + O(\Delta t^2) \nonumber\\
\end{eqnarray}
where 
\begin{equation}
\tilde{\mathcal{G}}^k_{\alpha} = \mathcal{M}^{\mathcal{VF};k}_{\alpha \beta}\mathcal{F}^k_\beta + \mathcal{M}^{\mathcal{VT};k}_{\alpha \beta}\mathcal{T}^k_\beta + \frac{\Delta t}{2}\frac{\partial}{\partial \mathcal{Y}_\beta} \left(\tilde{\mathcal{V}}^k_{\alpha} \tilde{\mathcal{U}}^k_{\beta}\right).
\end{equation}
Taking the ensemble average of the outer product gives
\begin{eqnarray}
\langle \Delta \mathcal{Y}^k_\alpha \Delta \mathcal{Y}^k_\beta \rangle &=& \Delta t^2\langle \tilde{\mathcal{V}}^k_{\alpha}\tilde{\mathcal{V}}^k_{\beta}\rangle + O(\Delta t^2) \nonumber \\
&=& 2k_BT \Delta t \mathcal{M}^{\mathcal{VF};k}_{\alpha \beta} \nonumber\\
& & + O(\Delta t^2),
\end{eqnarray}
showing that the second moment is also recovered to first order in time.
\subsection{Simplified scheme for periodic, no-slip, and no shear boundary conditions}
From the analysis above, we see that the inclusion of $v^k$ is needed to introduce $\nabla_{\mathcal{Y}}\cdot \tilde{\mathcal{U}}$ in order for the scheme to account for the divergence of the mobility matrix.  If, however, we have that $\nabla_{\mathcal{Y}}\cdot \tilde{\mathcal{U}} = 0$, then we may set $v^k = 0$.  We show here that this is the case when periodic, no-slip, or no flux boundary conditions are imposed at the fluid boundary.  We begin by noticing that the contribution to $\nabla_{\mathcal{Y}}\cdot \tilde{\mathcal{U}}$ from particle $n$ is
\begin{eqnarray}
\frac{\partial \tilde{U}^n_i}{\partial Y^n_i} &=& \int \tilde{u}_i\frac{\partial \Delta(\mathbf{x} - \mathbf{Y}^n)}{\partial Y^n_i} d^3\mathbf{x} \nonumber \\
&=& -\int \tilde{u}_i\frac{\partial \Delta(\mathbf{x} - \mathbf{Y}^n)}{\partial x_i} d^3\mathbf{x} \nonumber \\
&=& -\int \frac{\partial }{\partial x_i}\left(\tilde{u}_i \Delta(\mathbf{x} - \mathbf{Y}^n)\right) d^3\mathbf{x} \nonumber \\
&&+ \int \frac{\partial \tilde{u}_i}{\partial x_i} \Delta(\mathbf{x} - \mathbf{Y}^n) d^3\mathbf{x}. \label{eq:divU}
\end{eqnarray}
where repeated Latin indices imply summation.  The second integral on the right hand side of Eq. \eqref{eq:divU} is zero since $\bm{\nabla}\cdot\mathbf{\tilde{u}} = 0$.  After applying the divergence theorem to the first integral on the right-hand side, we have 
\begin{equation}
\frac{\partial \tilde{U}^n_i}{\partial Y^n_i}  = -\int \tilde{u}_i n_i \Delta(\mathbf{x} - \mathbf{Y}^n) dS,
\end{equation}
where $n_i$ is the normal pointing out of the domain.  Thus, if $\tilde{u}_i n_i=0$ pointwise, which is the case for no-slip and no-flux boundaries, or if periodic boundary conditions are imposed, then this integral is also zero.  As this will be the case for all particles, we then have that
\begin{equation}
\nabla_{\mathcal{Y}}\cdot \tilde{\mathcal{U}} = 0.
\end{equation}
Thus, as a result, we may set $v^k=0$ and the DC scheme will still account for the Brownian drift.

\begin{figure}
\centering
\includegraphics[width=7.5cm]{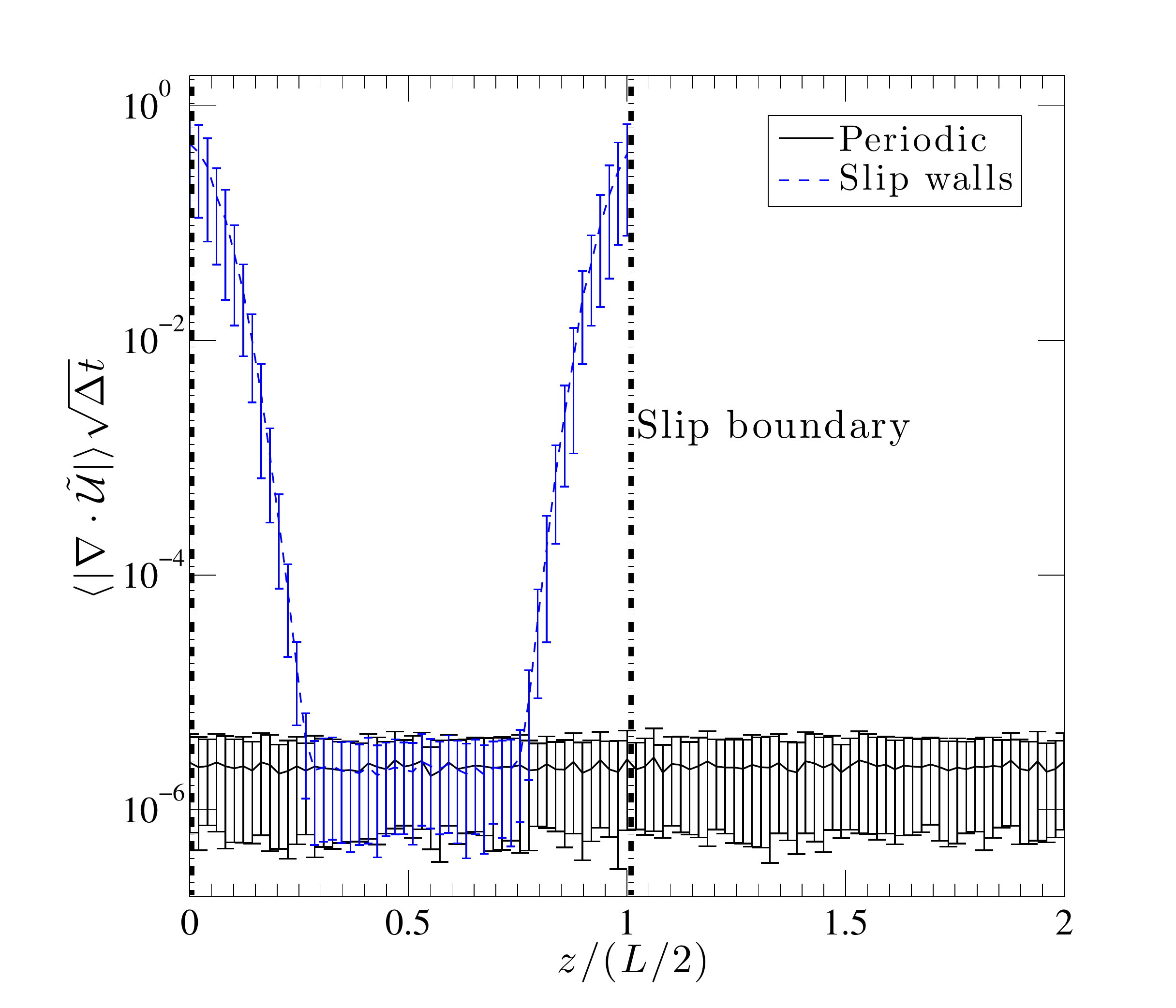}
\caption{$\langle|\nabla\cdot\tilde{\mathcal{U}}(z)|\rangle\sqrt{\Delta t}$ along $z$ in a periodic domain and between two shear free surfaces. The expectation is taken using $200$ realizations of the flow.}
\label{fig:abs_divU}
\end{figure}

As this result relies on using the continous spatial operators, it is important to check that it also holds when the Stokes equations and the projection and volume averaging operators are discretized in space.  In particular, we need to ensure that $\partial \tilde{U}^n_i/\partial Y^n_i$ is nearly zero in the simulations.  Fig. \ref{fig:abs_divU} shows $\langle |\partial \tilde{U}^n_i/\partial Y^n_i|\rangle$ as a function of the $z$-coordinate for a single particle in a periodic domain and between two no-shear surfaces.  For the periodic case, we see that this quantity is $O(10^{-6})$ and does not depend on $z$.  We may therefore safely set $v^k=0$ in our periodic simulations.  For the no-shear boundaries, however, we observe a clear dependence of $\langle |\partial \tilde{U}^n_i/\partial Y^n_i|\rangle$ on $z$ near the boundaries.  Further investigation revealed that this dependence is linked to the sum used to approximate the integral in Eq. \eqref{eq:vkdir} as we found that $\langle |\partial \tilde{U}^n_i/\partial Y^n_i|\rangle$ is well approximated by the error estimate for the quadrature scheme (not shown).  Thus, in our simulations with no-shear boundaries, we retain $v^k$ as part of the time integration.  We note, however, that for the computation of $v^k$, one only needs to consider the particles that overlap the no-shear boundaries.

\section{Energy balance for the force-coupling method}
\label{sec:VisDis}
For a suspension of rigid particles in Stokes flow, the rate of work done by the particles on the fluid is balanced by the viscous dissipation in the fluid and the rate of work done by the external boundary.  Thus,
\begin{eqnarray}
2\eta \int \mathbf{e}:\mathbf{e}d^3\mathbf{x} - \int \mathbf{u}\cdot\bm{\sigma}\cdot \mathbf{\hat{n}} dS &=& \sum_{n}\mathbf{F}^n\cdot\mathbf{V}^n \nonumber\\
&&+ \sum_{n}\bm{\tau}^n\cdot\bm{\Omega}^n\nonumber,\\
\label{eq:VisDisSus}
\end{eqnarray}
where $\mathbf{\hat{n}}$ is the unit normal to the external boundary.
For FCM, we would like to ensure that our definitions of particle motion yield the same energy balance.  Taking the dot product between $\mathbf{u}$ and Eq. \eqref{eq:Stokes} without the fluctuating stress, and integrating over the fluid domain, we find that 
\begin{widetext}
\begin{eqnarray}
2\eta \int \mathbf{e}:\mathbf{e}d^3\mathbf{x} - \int \mathbf{u}\cdot\mathbf{\sigma}\cdot \mathbf{\hat{n}} dS &=& \sum_{n} \mathbf{F}^n\cdot\int \mathbf{u}\Delta_n(\mathbf{x}) d^3\mathbf{x} + \frac{1}{2}\sum_{n} \bm{\tau}^n\cdot \int \mathbf{u} \times \bm{\nabla}\Theta_n(\mathbf{x})d^3\mathbf{x} \nonumber \\
&&+ \frac{1}{2}\sum_{n} \mathbf{S}^n:\int \left(  \mathbf{u}\bm{\nabla}\Theta_n(\mathbf{x})+\left(\mathbf{u}\bm{\nabla}\Theta_n(\mathbf{x})\right)^{T}\right)d^3\mathbf{x}
\label{eq:VisDisFCM}
\end{eqnarray}
\end{widetext}
where we have also integrated by parts to obtain the viscous dissipation and boundary term on the left-hand side.  

We see that in order for Eq. \eqref{eq:VisDisFCM} to comply with the standard energy balance, Eq. \eqref{eq:VisDisSus}, we need to require that the particle velocities are given by Eq. \eqref{eq:FCM_Ju} and the angular velocities are provided by Eq. \eqref{eq:FCM_Qu}.  Additionally, in order for the stresslets to perform no work on the fluid, the local rate of strain given by Eq. \eqref{eq:FCM_Ku} must be constrained to be equal to zero.

\section{Interpolation of the stresslet-corrected mobility coefficients}
\label{sec:AppC}
In order to rapidly generate a large number of statistics to obtain the distribution of a single particle in a slip channel, we determine the motion of the particles using the FCM mobility matrices for the single particle and the volume averages of the fluctuating fluid flow.  This allows the usage of a single realization of the fluctuating fluid flow to determine the velocity of many non-interacting particles.

The velocity of particle $n$ is given by
\begin{equation}
\mathbf{V}_n = \mathbf{M}^{VF}\mathbf{F}_n + \mathbf{\tilde V}_n + \mathbf{C}^{VE}\mathbf{\tilde E}_n  
\end{equation}
where $\mathbf{M}^{VF}$ is given by Eq. \eqref{eq:MVFslip} with the stresslet corrected coefficients,  $\mu^{FCM-S}$, and the matrix $\mathbf{C}^{VE}$ relates the particle velocity and local rate of strain.  The quantities $\mathbf{\tilde V}_n$ and $\mathbf{\tilde E}_n$ are the random velocities and local rates of strain obtain by applying the FCM volume averaging operators to the fluctuating flow field.  

Due to symmetry, $\mathbf{M}^{VF}$ is diagonal (see Eq. \eqref{eq:MVFslip}), and only $C_{z,zz}^{VE}$ and $C_{x,zx}^{VE} = C_{y,zy}^{VE}$ are non-zero.  The values, however, for these non-zero entries depend on the particle's distance from the boundaries.  Thus, to perform the simulations, we compute the non-zero matrix entries at $400$ equi-spaced positions between $z = a$ and $z = L_z-a$ and use linear interpolation to find the values at an arbitrary position along the channel.
\bibliographystyle{plain}
\bibliography{BMbib}
\end{document}